\documentclass[reqno,12pt]{article}
\usepackage{amsfonts,amssymb,amsmath,amscd,bm,bbold,mathrsfs,delarray,subfigure}
\usepackage{hyperref,cite}
\usepackage{xypic}
\usepackage[dvips]{color}
\usepackage[dvips]{graphicx}

\DeclareMathOperator{\ad}{ad}

\DeclareMathOperator{\ch}{ch}
\DeclareMathOperator{\Hom}{Hom}
 
\DeclareMathOperator{\Fun}{Fun}
\DeclareMathOperator{\Map}{Map}

\DeclareMathOperator{\End}{End}

\DeclareMathOperator{\rank}{rank}

\DeclareMathOperator{\tr}{tr}
\DeclareMathOperator{\dd}{d}

\numberwithin{equation}{section}

\newcommand{\sss}{{\hbox{$\sum$}}}

\font\sansserif=cmss12
\font\scriptsansserif=cmss12 at 7 truept
\font\scriptscriptsansserif=cmss10 at 5 truept
\def\sans{\fam=14}
\newcommand{\mathsans}[1]{{{\sans #1}}}

\font\euler=eusm10 at 12 truept
\font\scripteuler=eusm7
\font\scriptscripteuler=eusm5 
\def\eul{\fam=12}
\newcommand{\matheul}[1]{{{\eul #1}}}

\newcommand{\barpartial}{\overline{\partial}}
\newcommand{\barnabla}{\overline{\nabla}}
\newcommand{\bardee}{\overline{D}}
\newcommand{\bardeecpl}{\overline{\mathcal{D}}{}}

\begin{document}

\hrule\vskip.5cm
\hbox to 14.5 truecm{January 2011 \hfil DFUB 11}
\vskip.5cm\hrule
\vskip.7cm
\centerline{\textcolor{blue}{\bf A HETEROTIC SIGMA MODEL }}   
\centerline{\textcolor{blue}{\bf WITH NOVEL TARGET GEOMETRY}}   
\vskip.2cm
\centerline{by}
\vskip.2cm
\centerline{\bf Roberto Zucchini}
\centerline{\it Dipartimento di Fisica, Universit\`a degli Studi di Bologna}
\centerline{\it V. Irnerio 46, I-40126 Bologna, Italy}
\centerline{\it I.N.F.N., sezione di Bologna, Italy}
\centerline{\it E--mail: zucchinir@bo.infn.it}
\vskip.7cm
\hrule
\vskip.6cm
\centerline{\bf Abstract} 
\par\noindent
We construct a $(1,2)$ heterotic sigma model whose target space geometry
consists of a transitive Lie algebroid with complex structure on a Kaehler manifold. 
We show that, under certain geometrical and topological conditions,  
there are two distinguished topological half--twists of the heterotic 
sigma model leading to $A$ and $B$ type half--topological 
models. Each of these models is characterized by the usual topological BRST operator, 
stemming from the heterotic $(0,2)$ supersymmetry, and a second BRST operator anticommuting with the former,
originating from the $(1,0)$ supersymmetry. These BRST operators combined in a certain way
provide each half--topological model with two inequivalent BRST structures and, correspondingly,
two distinct perturbative chiral algebras and chiral rings. The latter are studied in detail and 
characterized geometrically in terms of Lie algebroid cohomology in the quasiclassical limit.

\vfill\eject

\tableofcontents

\vfill\eject

\section{\normalsize \textcolor{blue}{Introduction}}\label{sec:intro}


\hspace{.5cm} The superstring theories which, in the low energy limit, yield 
effective four--dimensional field theories with space--time supersymmetry are described 
by $(0,2)$ superconformal field theories whose right--moving $U(1)$ charges
satisfy suitable integrality conditions. (See \cite{Distler:1987ee}--\cite{Distler:1995mi}
for background and extensive referencing.) 
The subject of $(0,2)$ superconformal field theories is therefore 
of a considerable physical interest. Nevertheless, the amount of research work 
dedicated to these models has been very limited in comparison to the $(2,2)$ models, which 
constitute a special subclass. The reason of this, of course, is that the smaller amount of
world--sheet supersymmetry makes them harder to study. 

Sigma models are nonlinear interacting quantum field theories and, so, are rather difficult
to deal with. In general, they can be described only perturbatively in the large target space 
volume limit. However, as originally showed by Witten in ref. \cite{Witten1}, important aspects 
of supersymmetric sigma models are captured by means of topological twist. 
A topologically twisted supersymmetric sigma model is characterized by the existence 
of a nilpotent BRST charge and the associated operator BRST cohomology. 
On a flat world--sheet, the untwisted parent model and the twisted daughter model are equivalent.
The power of cohomological methods makes then it possible to study rather effectively the sector 
of the former corresponding to the operator BRST cohomology of the latter.
On a non flat world--sheet, the two models are no longer equivalent, but the twisted model
still exists and can be fruitfully studied.  

As found by Witten in ref. \cite{Witten2}, the $(2,2)$ supersymmetric sigma model on a 
Kaehler manifold can be topologically twisted in two inequivalent ways. 
The operator BRST cohomology of the twisted sigma models is known as chiral ring. 
It consists of operator classes constant on the world--sheet and is endowed with a natural 
multiplicative structure. The twisted models are therefore topological field theories. 
The $A$ model chiral ring is isomorphic to the target manifold quantum cohomology,
a deformation of de Rham cohomology related to Gromov--Witten invariants and Floer homology, 
and depends only on the target's Kaehler structure. 
The $B$ model chiral ring is isomorphic to the sheaf cohomology of the exterior algebra 
of the target manifold holomorphic tangent bundle and depends only on the target's complex 
structure. 
For Calabi--Yau manifolds, $A$ and $B$ models are further related by mirror symmetry \cite{Hori:2003ic}.

\vskip.2cm

The $(0,2)$ supersymmetric sigma model on a Kaehler manifold can 
be topologically twisted in just one way. 
The operator BRST cohomology of the twisted sigma model is known as chiral algebra. 
It consists of operator classes varying holomorphically on the world--sheet and 
is endowed with a natural product structure of the operator product expansion type.
The twisted model, therefore, is not a topological field theory 
but a holomorphic field theory.

\vskip.2cm

As remarked again by Witten in ref. \cite{Witten:2005px}, 
the chiral algebra of the twisted $(0,2)$ sigma model is akin to the chiral ring of the twisted 
$(2,2)$ sigma models, but only to a point. 
Like the chiral ring, it is independent from the target manifold's metric, but, 
unlike the chiral ring, it receives complicated perturbative and world--sheet instanton corrections, 
because the model's lack of left--moving supersymmetry. 
In spite of these difficulties, the chiral algebra remains an object of central importance
in the theory. The interest in the subject has been revived in recent years, 
after Witten's discovery that the perturbative chiral algebra 
can be formulated as the cohomology of a sheaf of chiral differential operators
\cite{Malikov:1998dw,Gorbounov:1999}. Almost simultaneously, 
Kapustin \cite{Kapustin:2005pt} reached independently the same conclusion
and, subsequently, Tan \cite{Tan:2006qt}--\cite{Tan:2007bh}
generalized the analysis to heterotic $(0,2)$ supersymmetric sigma models with 
left--moving fermions. 

\vskip.2cm 

The perturbative chiral algebra contains a sector formed by classes with non singular 
operator product expansion, the $(0,2)$ chiral ring, which shares many properties with and is the true 
counterpart of the $(2,2)$ chiral ring, as suggested by its name \cite{Adams:2003zy,
Sharpe:2005fd,Adams:2005tc}. The $(0,2)$ chiral ring is of considerable interest of its own
and lends itself to a nice geometrical and topological characterization
as the $(2,2)$ one. 


In this paper, we construct a heterotic $(1,2)$ supersymmetric sigma model 
whose target space geometry
consists of a transitive Lie algebroid $E$ with complex structure on a Kaehler manifold
$M$. There are several reasons why this is an interesting endeavour, as we now explain

\vskip.15cm

Lie algebroids were first studied by Pradines in the early sixties 
as a vector bundle generalization of Lie algebras \cite{Pradines1}. 
Lie algebroids provide a very general and flexible framework for studying a wide range of 
geometrical structures\cite{Mackenzie1}. Recently, they proved to be essential in the formulation 
of generalized complex geometry \cite{Hitchin1,Gualtieri}, which has attracted much attention 
in the string theory community for its applications to type II flux compactifications \cite{Grana} 
and in the construction of topological sigma models with generalized Kaehler target geometry  
\cite{Kapustin2}\,-- \cite{Zucchini:2006ds}. The study of sigma models with a Lie algebroid target geometry
constitutes therefore a useful theoretical laboratory capable of providing important clues about
the eventual construction of supersymmetric sigma models describing the propagation
of strings in more general flux backgrounds. 

\vskip.15cm

Applications of Lie algebroids to sigma models are plentiful \cite{Kotov:2010wr}.
However, virtually all of them are concerned with smooth Lie algebroids
and the sigma models they yield require gauge fixing and, so, are not immediately suitable
for quantization.
Our experience with sigma models with higher world--sheet supersymmetry 
suggests instead that, if Lie algebroids are to have any role, it is the holomorphic 
rather than the smooth ones which should be involved. 
In our construction, in fact, holomorphic Lie algebroids enter in an essential way. 

\vskip.15cm

As it turns out, the heterotic Lie algebroid sigma model worked out by us 
enjoys a $(0,2)$ world--sheet supersymmetry, as all heterotic sigma models,
and an extra $(1,0)$ supersymmetry originating from the Lie algebroid's geometry. 
Heterotic sigma models with $(1,2)$ supersymmetry were studied long ago
by Hull and Witten in ref. \cite{Hull:1985jv}. However, the origin of the left--moving
supersymmetry in that case was completely different. 
The extra supersymmetry, we hope, should constrain the model enough to make the study
of important structures as the chiral algebra and the chiral ring more manageable
without rendering the model too 'ad hoc' to be interesting.

\vskip.2cm

There are two distinguished topological half--twists of the heterotic Lie algebroid 
sigma model, which lead to an $A$ and a $B$ type half--topological sigma model. Each half--topological 
model is characterized by the usual topological BRST operator, stemming from the heterotic $(0,2)$ 
supersymmetry, and a second BRST operator anticommuting with the former,
originating from the $(1,0)$ supersymmetry. These BRST operators combined in a certain way
provide each half--topological model with two inequivalent BRST structures and, correspondingly,
also two distinct perturbative chiral algebras and chiral rings. Albeit we do not yet fully understand its ultimate 
origin and implications, we consider this to be one of the most salient features of our sigma models.

\vskip.2cm

The first BRST structure answers to the usual one 
of other half--topological heterotic sigma models. The second BRST structure is instead completely novel. 
Though the two structures share a number of properties, the second one is, in the appropriate 
sense explained in the body of the paper, ``more topological'' than the first.
However, in general, the second structure's chiral algebra seemingly does not lend itself to a chiral 
differential operator sheafy description the same way the first structure's does.

\vskip.2cm

The perturbative chiral rings of the two half--topological Lie algebroid sigma models
can be characterized geometrically in the quasiclassical limit, as expected. In particular, 
for the second BRST structure, they are isomorphic to certain Lie algebroid cohomologies.
In this way, the problem of computation of genus $0$ perturbative chiral ring correlators 
can be phrased in Lie algebroid theoretic terms.

\vskip.2cm

The plan of the paper is as follows.
In sect. \ref{sec:liealg}, the theory of smooth Lie algebroids
is reviewed with particular emphasis on the transitive case. 
In sect. \ref{sec:liecmplx}, the theory of Lie algebroid complex and 
holomorphic structures is expounded, providing 
the necessary geometric toolkit for the construction of Lie algebroid sigma models.
In sect. \ref{sec:hetersgm}, the heterotic Lie algebroid sigma model is constructed, 
its supersymmetry is unveiled and its basic quantum aspects described.
In sect. \ref{sec:topsgm}, which is the core part of the paper, 
it is shown that there are two distinguished twists of the
Lie algebroid heterotic sigma model previously constructed, 
the half--topological Lie algebroid sigma models, each of which is 
characterized by two inequivalent BRST structures.
Various topics such as dependence on the target space geometry, perturbative 
chiral algebras and rings and anomalies are analyzed in detail. 
In sect. \ref{sec:cohla}, the chiral rings of the half--topological Lie algebroid 
sigma models are studied in the quasiclassical limit and described 
in terms of  the target Lie algebroid's cohomology.
In sect. \ref{sec:outlook}, we summarize our results and 
indicate future lines of research.

\vfill\eject

\section{\normalsize \textcolor{blue}{Lie algebroids}}\label{sec:liealg}

\subsection{\normalsize \textcolor{blue}{Lie algebroids}}\label{subsec:liealg1}

\hspace{.5cm} Lie algebroids are vector bundles which have the same structural properties as the tangent
bundle of a manifold. On the other hand, Lie algebroids are anchored bundles and, so,
their definition requires a prior independent definition of the tangent bundle.
The latter, therefore, is not merely a particular example of
Lie algebroid. See ref. \cite{Mackenzie1} for an exhaustive treatment of the subject.

A {\it real Lie algebroid} is a smooth real vector bundle $E$ over a manifold $M$ 
equipped with a smooth bundle map $\rho_E:E\mapsto T_M$, called the {\it anchor}, and 
an $\mathbb{R}$--bilinear bracket $[\cdot,\cdot]_E:\Gamma(E)\times \Gamma(E)
\mapsto \Gamma(E)$ with the following properties.

\par\noindent 
~~~1) $[\cdot,\cdot]_E$ is a Lie bracket so that $\Gamma(E)$ is a Lie algebra:
\begin{align}
&[s,t]_E+[t,s]_E=0,
\vphantom{\Big]}
\label{LAA1}
\\
&[s,[t,u]_E]_E+[t,[u,s]_E]_E+[u,[s,t]_E]_E=0,
\vphantom{\Big]}
\label{LAA2}
\end{align}
for $s,t,u\in \Gamma(E)$. 

\par\noindent 
~~~2) $\rho_E$ defines a Lie algebra homomorphism of $\Gamma(E)$ into $\Gamma(T_M)$: 
\begin{equation}
\rho_E[s,t]_E=[\rho_Es,\rho_Et]_{T_M},
\label{LAA3}
\end{equation}
for $s,t\in \Gamma(E)$, where $[\cdot,\cdot]_{T_M}$ is the usual Lie bracket
of vector fields of $M$. 

\par\noindent 
~~~3) The generalized Leibniz rule holds:
\begin{equation}
[s,ft]_E=f[s,t]_E+(\rho_Esf)t,
\label{LAA4}
\end{equation}
for $f\in C^\infty(M)$ and $s,t\in \Gamma(E)$.

The prototype Lie algebroid over $M$ is the tangent bundle $T_M$: the anchor $\rho_{T_M}$
is the identity $1_{T_M}$ and the bracket is the usual Lie bracket $[\cdot,\cdot]_{T_M}$. 
Lie algebroids generalize Lie algebras: a Lie algebra can be viewed as 
a Lie algebroid over the singleton manifold $M=\mathrm{pt}$.

A {\it (base preserving) morphism} of two Lie algebroids $E$, $E'$ over $M$ is a vector bundle 
morphism $\varphi:E\mapsto E'$ such that
\vspace{-.3truecm}
\begin{align}
&\rho_{E'}\varphi=\rho_E,
\vphantom{\Big]}
\label{LAAmorph1}
\\
&\varphi[s,t]_E=[\varphi s,\varphi t]_{E'},
\vphantom{\Big]}
\label{LAAmorph2}
\end{align}
with $s,t\in \Gamma(E)$. 

If $L$, $E$ are two Lie algebroids over $M$ and $L$ is a subbundle 
of $E$, then $L$ is a {\it Lie subalgebroid} of $E$, if the natural injection
$\iota:L\mapsto E$ is a Lie algebroid morphism. 

\subsection{\normalsize \textcolor{blue}{Transitive Lie algebroids}}\label{subsec:liealg2}

\hspace{.5cm} In this paper, we are interested mainly in transitive Lie algebroids. 
A Lie algebroid $E$ is said {\it transitive}, if its anchor $\rho_E$ is surjective. 
When $E$ is transitive, $\ker\rho_E$ is a subbundle of $E$, called the 
{\it adjoint bundle} of $E$, and $\ker\rho_E$ inherits from $E$ a structure of 
Lie algebroid rendering it a Lie subalgebroid of $E$. Since the anchor 
$\rho_{\ker\rho_E}$ of $\ker\rho_E$ 
vanishes, $\ker\rho_E$ is a Lie algebra bundle, that is vector bundle
whose typical fiber is a fixed Lie algebra $\mathfrak{g}$. 

If $E$ is transitive, $\Gamma(\ker\rho_E)$ is a Lie ideal of $\Gamma(E)$. 
The quotient bundle $\tilde E=E/\ker\rho_E$ inherits then from $E$ a structure of Lie algebroid.
The anchor $\rho_{\tilde E}$ of $\tilde E$ defines a Lie algebroid isomorphism of
$\tilde E$ onto $T_M$. 

If $E$ is transitive, then there exists a canonical exact sequence of Lie algebroids
\begin{equation}
\xymatrix{0\ar[r]&\ker\rho_E\ar[r]^{\,\,\,\iota_E}&E\ar[r]^{\rho_E\,\,\,}&T_M\ar[r]&0}
\label{LAAex1}
\end{equation}
called {\it Atiyah sequence}.
A {\it splitting} \footnote{$\vphantom{\bigg[}$  
Splittings are often called connections.
We shall reserve the term connection for a related notion (see below)}
of $E$ is a bundle map $\sigma:T_M \mapsto E$ such that 
\begin{equation}
\rho_E\sigma=1_{T_M}.
\label{LAAex2}
\end{equation}
\eject\noindent

Unlike $\rho_E$, $\sigma$ does not induce in general a morphism of the Lie algebras
$\Gamma(T_M)$ and $\Gamma(E)$. The failure to do so is measured by the {\it curvature} 
of $\sigma$, the field $G_{E\sigma}\in \Gamma(\wedge^2T_M{}^*\otimes\ker\rho_E)$  defined by
\begin{equation}
G_{E\sigma}(x,y)=[\sigma x,\sigma y]_E-\sigma[x,y]_{T_M},
\label{LAAex3}
\end{equation}
with $x,y\in \Gamma(T_M)$. The splitting $\sigma$ is {\it flat}, if $G_{E\sigma}=0$. 

\subsection{\normalsize \textcolor{blue}{Lie algebroid connections}}\label{subsec:liealg3}

\hspace{.5cm} Let $E$ be a Lie algebroid. 
An $E$ Lie algebroid connection on a vector bundle $V$ is an $\mathbb{R}$ bilinear
map $D:\Gamma(E)\times\Gamma(V)\mapsto \Gamma(V)$ satisfying  
\begin{align}
&D_{fs}v=fD_sv, 
\vphantom{\Big]}
\label{LAAconn1} 
\\
&D_s(fv)=fD_sv+(\rho_Esf)v,
\vphantom{\Big]}
\label{LAAconn2}
\end{align}
for any $f\in C^\infty(M)$, $s\in\Gamma(E)$, $v\in\Gamma(V)$. 
The curvature of the connection $D$ is the field $R_D\in\Gamma(\wedge^2 E^*\otimes \End(V))≈ß$ defined by
\begin{equation}
R_D(s,t)v=D_sD_tv-D_tD_sv-D_{[s,t]_E}v,
\label{LAAconn3}
\end{equation}
where $s,t\in\Gamma(E)$, $v\in\Gamma(V)$. $D$ is said {\it flat}, if $R_D=0$.
In that case, $D$ is called a {\it representation} of the Lie algebroid $E$ in the vector bundle $V$. 

When $V=E$, it is also possible to define the 
torsion of $D$, which is the field $T_D\in\Gamma(\wedge^2 E^*\otimes E)$ defined by 
\begin{equation}
T_D(s,t)u=D_st-D_ts-[s,t]_E,
\label{LAAconn4}
\end{equation}
where $s,t\in\Gamma(E)$. $D$ is said {\it torsionless}, if $T_D=0$.

A $T_M$ Lie algebroid connection on $V$ is an ordinary connection on $V$. 
In that case, the definition of curvature and, when $V=T_M$, of torsion
given above reproduce the usual ones.
The notion of Lie algebroid connection is however more general. For instance,
while for a generic $s\in\Gamma(E)$, $D_s$ is a 1st order differential operator on
$\Gamma(V)$, when $s$ is valued in $\ker\rho_E$, $D_s$ is simply a field of
$\Gamma(\End(E))$. 

With any $E$ Lie algebroid connection $D$ on $E$, there is associated canonically an $E$ 
Lie algebroid connection on any of the vector bundles which can be constructed from 
$E$ by dualizing and tensoring in the usual way. We shall denote these connections
with the same symbol $D$. In particular, the associated connection
on $\End(E)$ is defined by the relation $(D_sA)t=D_s(At)-AD_st$ 
with $t\in\Gamma(E)$, for $s\in\Gamma(E)$, $A\in\Gamma(\End(E))$. 

Let $E$ be a transitive Lie algebroid, $V$ be a vector bundle and $D$ be an 
$E$ connection on $V$. If $\sigma$ is a splitting of $E$, then
\begin{equation}
(D_\sigma)_xv=D_{\sigma x}v,
\label{LAAconn5}
\end{equation}
with $x\in\Gamma(T_M)$, $v\in\Gamma(V)$, is an ordinary connection on $V$.
The curvatures of $D$ and $D_\sigma$ are related as
\begin{equation}
R_{D_\sigma}(x,y)v=R_D(\sigma x,\sigma y)v+D_{G_{E\sigma}(x,y)}v,
\label{LAAconn5/1}
\end{equation}
for with $x,y\in\Gamma(T_M)$, $v\in\Gamma(V)$. 

For a transitive Lie algebroid $E$, the expression 
\begin{equation}
\hbox{$D^{\mathrm{ad}}_{Es}z$} 
=[s,z]_E,
\label{LAAconn6}
\end{equation}
with $s\in\Gamma(E)$, $z\in\Gamma(\ker\rho_E)$, defines a canonical $E$ connection on $\ker\rho_E$.
$D^{\mathrm{ad}}_E$ is flat. Conversely, for a generic splitting $\sigma$ of $E$, $D^{\mathrm{ad}}_{E\sigma}$ 
is not flat, being
\begin{equation}
R_{D^{\mathrm{ad}}_{E\sigma}}(x,y)z=[G_{E\sigma}(x,y),z]_{\ker\rho_E},
\label{LAAconn7}
\end{equation}
for $x,y\in\Gamma(T_M)$, $z\in\Gamma(\ker\rho_E)$. 
$D^{\mathrm{ad}}_{E\sigma}$ is flat if $G_{E\sigma}$ is valued in the center
of $\ker\rho_E$, the subbundle of $\ker\rho_E$ corresponding to the center $Z\mathfrak{g}$
of the typical fiber $\mathfrak{g}$ of $\ker\rho_E$.

\subsection{\normalsize \textcolor{blue}{Complex Lie algebroids}}\label{subsec:liealg4}

\hspace{.5cm} For a real vector bundle $V$, let $V^c=V\otimes\mathbb{C}$ be the complexification of $V$.
A {\it complex Lie algebroid} is a smooth complex vector bundle $W$ over a manifold $M$ 
equipped with a smooth bundle map $\rho_W:W\mapsto T_M{}^c$ and
a $\mathbb{C}$--bilinear bracket $[\cdot,\cdot]_W:\Gamma(W)\times \Gamma(W)
\mapsto \Gamma(W)$ such that \eqref{LAA1}--\eqref{LAA4} hold
with $T_M$ replaced by $T_M{}^c$ throughout.
If $E$ is a real Lie algebroid, the complexification 
$E^c$ of $E$ has an obvious induced structure of complex Lie 
algebroid. All the constructions illustrated above
extend without change to complex Lie algebroids.

\vfill\eject 

\section{\normalsize \textcolor{blue}{Complex structures and holomorphic Lie algebroids}}
\label{sec:liecmplx}

\subsection{\normalsize \textcolor{blue}{Holomorphic Lie algebroids}}\label{subsec:liecmplx1}

\hspace{.5cm} Lie algebroid theory can be formulated also in the holomorphic setting
\cite{Weinstein}--\cite{Bruzzo:2010gi}. 
In this case, however, one must take into account the fact that, in general,  
a complex manifold $\mathcal{M}$ admits an infinite dimensional algebra of holomorphic functions 
and, similarly, a holomorphic vector bundle $\mathcal{V}$ over $\mathcal{M}$ 
admits an infinite dimensional space of holomorphic sections only locally on $\mathcal{M}$. For this 
reason, one is forced to work locally on each open set $U$ of $\mathcal{M}$. 
To each relation of the smooth theory, 
there then correspond infinitely many formally analogous relations of the holomorphic theory,
one for each open set $U$ 
and these relations are compatible
with inclusions of open sets $U\subset V$. 
Formally, as is well--known, this can be done employing sheaf theory.
We denote by $\mathcal{O}_{\mathcal{M}}$ the sheaf of holomorphic functions 
of $\mathcal{M}$ and by $\mathcal{O}_{\mathcal{V}}$ the sheaf 
of holomorphic sections of $\mathcal{V}$:
for any open set $U$ of $M$, $\mathcal{O}_{\mathcal{M}}(U)$ is the algebra of holomorphic 
functions on $U$ and $\mathcal{O}_{\mathcal{V}}(U)$ is the space of holomorphic sections of
$\mathcal{V}$ on $U$.

A {\it holomorphic Lie algebroid} is a holomorphic vector bundle $\mathcal{E}$ 
over a complex manifold $\mathcal{M}$ equipped with a holomorphic bundle map 
$\rho_{\mathcal{E}}:\mathcal{E}\mapsto \mathcal{T}_{\mathcal{M}}$ and
a $\mathbb{C}$--bilinear bracket $[\cdot,\cdot]_{\mathcal{E}}:
\mathcal{O}_{\mathcal{E}}\times \mathcal{O}_{\mathcal{E}}
\mapsto \mathcal{O}_{\mathcal{E}}$ commuting with restrictions such that 
\eqref{LAA1}--\eqref{LAA4} are satisfied at the sheaf level (that is  
with $\Gamma(E)$, $\Gamma(T_M)$ and $C^\infty(M)$
replaced throughout by $\mathcal{O}_{\mathcal{E}}(U)$, $\mathcal{O}_{\mathcal{T}_{\mathcal{M}}}(U)$
and $\mathcal{O}_{\mathcal{M}}(U)$, respectively, for all open sets $U$ of $\mathcal{M}$.) 

One can define the notions of transitive holomorphic Lie algebroid and 
holomorphic splitting and those of holomorphic algebroid connections 
as in the smooth theory in obvious fashion. 

\vspace{-.1cm}

\subsection{\normalsize \textcolor{blue}{Lie algebroid complex structures}}\label{subsec:liecmplx2}

\hspace{.5cm} Let $V$ be a real vector bundle over a manifold $M$. 
An {\it almost complex structure} 
\vfil\eject\noindent
of $V$ is a field 
$K\in \Gamma(\End(V))$ satisfying  
\begin{equation}
K^2=-1_V.
\label{CLAx1}
\end{equation}
We denote by $K^c$ the extension of $K$ to the complexification $V^c$ of $V$
and by $V^+$, $\overline{V}{}^+$ the $\pm i$ eigenbundles of $K^c$
\footnote{$\vphantom{\bigg[}$ Usually, the notation $V^{1,0}$, $V^{0,1}$ is used for 
these eigenbundles. We employ the notation $V^+$, $\overline{V}{}^+$ for convenience.}.
When $V=T_M$, we recover the customary notion of 
almost complex structure.

Let $E$ be a real Lie algebroid over $M$ and let $J$ be an almost complex structure of $E$.
$J$ is said a {\it complex structure}, if it satisfies the integrability condition
\begin{equation}
N_{EJ}=0,  
\label{CLAx3}
\end{equation}
where $N_{EJ}\in\Gamma(\wedge^2 E^*\otimes E)$ is the Nijenhuis field of $J$
defined by
\begin{equation}
N_{EJ}(s,t)=[Js,Jt]_E-[s,t]_E-J([Js,t]_E+[s,Jt]_E),
\label{CLAx2}
\end{equation}
with $s,t\in\Gamma(E)$. 
The integrability condition \eqref{CLAx3} is equivalent to 
the closedness of $\Gamma(E^+)$ under the Lie bracket
$[\cdot,\cdot]_{E^c}$. When $E=T_M$, we recover the customary notion of complex 
structure.

Let $E$ be a real Lie algebroid over $M$. A {\it Lie algebroid complex structure} of $E$ is 
a pair $(J,J_M)$, where $J$ and $J_M$ are complex structures of $E$
and $T_M$, respectively, such that the anchor $\rho_E$ of $E$ is complex linear,
\begin{equation}
\rho_EJ=J_M\rho_E.
\label{CLAx4}
\end{equation}
A {\it Lie algebroid with complex structure} is a real Lie algebroid $E$ 
endowed with a Lie algebroid complex structure $(J_E,J_{EM})$.

A {\it Lie algebroid with complex base} is a complex vector bundle $W$ over a 
manifold $M$ together with a complex structure $J_{WM}$ of $T_M$,  
a smooth bundle map $\rho_W:W\mapsto T_M{}^+$ and
a $\mathbb{C}$--bilinear bracket $[\cdot,\cdot]_W:\Gamma(W)\times \Gamma(W)
\mapsto \Gamma(W)$ such that \eqref{LAA1}--\eqref{LAA4} 
\eject\noindent
hold with $E$ and $T_M$ replaced by $W$ and $T_M{}^+$, respectively.
$T_M{}^+$ itself is a complex Lie algebroid with complex base.

If $E$ is a Lie algebroid with complex structure, then 
$E^+$ has an obvious induced structure of 
Lie algebroid with complex base. $\overline{E}{}^+$ also does. 
A lot more happens however. 

The complex structure $J_{EM}$
ensures the existence of a holomorphic coordinate atlas 
on $M$ and, so, renders $M$ a complex manifold, henceforth denoted
by $\mathcal{M}$.
The complex vector bundle $T_M{}^+$ has a canonical holomorphic structure
and is thus a holomorphic vector bundle
$\mathcal{T}_{\mathcal{M}}$ over $\mathcal{M}$, the holomorphic tangent bundle of 
$\mathcal{M}$. Further, the Lie bracket $[\cdot,\cdot]_{T_M{}^+}$ restricts 
to the Lie bracket $[\cdot,\cdot]_{\mathcal{T}_{\mathcal{M}}}$ on 
local holomorphic sections of $\mathcal{T}_{\mathcal{M}}$.
In this way, $\mathcal{T}_{\mathcal{M}}$ gets endowed with a canonical structure 
of holomorphic Lie algebroid.
The natural questions arise about whether the complex 
structure $J_E$ similarly endows the vector bundle $E^+$
with a canonical holomorphic structure making it a holomorphic 
vector bundle $\mathcal{E}$ over $\mathcal{M}$
and whether $\mathcal{E}$ is also canonically a holomorphic Lie algebroid.
Things however are not as simple as for $T_M$, as the following analysis shows. 

\subsection{\normalsize \textcolor{blue}{Lie algebroid holomorphic structures}}\label{subsec:liecmplx3}

\hspace{.5cm} Let $E$ be a Lie algebroid over $M$ 
with complex structure. 
The {\it $\barpartial$ operator}≈ß of $E^+$ is the 
differential operator $\barpartial_{E^+}$ on $\Gamma(E^+)$ defined by
\begin{equation}
\barpartial_{E^+\bar s}t=\Pi_E[\bar s,t]_{E^c}, 
\label{liecmplx1}
\end{equation}
with $s,t\in\Gamma(E^+)$, where $\Pi_E=(1_E-iJ_E)/2\in\Gamma(\End(E^c))$ 
is the projector field on $E^+$. 
$\barpartial_{E^+}$ is a $\overline{E}{}^+$ Lie algebroid connection on $E^+$, that is
\begin{align}
&\barpartial_{E^+\bar f\bar s}t=\bar f\barpartial_{E^+\bar s}t, 
\vphantom{\Big]}
\label{liecmplx2} 
\\
&\barpartial_{E^+\bar s}(ft)=f\barpartial_{E^+\bar s}t+(\bar\rho_{E^+}\bar sf)t,
\vphantom{\Big]}
\label{liecmplx3}
\end{align}
for any $f\in C^\infty_c(M)$, $s,t\in\Gamma(E^+)$ (see eqs. \eqref{LAAconn1}, \eqref{LAAconn2}).
The integrability of the complex structure $J_E$ entails that the 
curvature $R_{\bar\partial_{E^+}}$  of $\barpartial_{E^+}$ vanishes,
\begin{equation}
R_{\bar\partial_{E^+}}(\bar s,\bar t)u=\barpartial_{E^+\bar s}\barpartial_{E^+\bar t}u
-\barpartial_{E^+\bar t}\barpartial_{E^+\bar s}u
-\barpartial_{E^+[\bar s,\bar t]_{\bar E^+}}u=0, 
\label{liecmplx4}
\end{equation}
for $s,t,u\in\Gamma(E^+)$. 

The above construction can be repeated in the case where $E=T_M$. 
As is readily checked, $\barpartial_{T_M{}^+}$ is the usual Cauchy--Riemann operator on 
$\Gamma(T_M{}^+)$. We shall see that $\barpartial_{E^+}$ is a natural generalization of
$\barpartial_{T_M{}^+}$, as suggested by the notation.

The operator $\barpartial_{E^+}$ intertwines naturally with the anchor and the 
Lie bracket of the Lie algebroid $E^+$. Indeed, one has 
\begin{equation}
\barpartial_{T_M{}^+\bar\rho_{E^+}\bar s}(\rho_{E^+}t)=\rho_{E^+}\barpartial_{E^+\bar s}t,
\label{liecmplx6}
\end{equation}
for $s,t\in\Gamma(E^+)$ and, further, 
\begin{equation}
\barpartial_{E^+\bar s}[u,v]_{E^+}=
[\barpartial_{E^+\bar s}u,v]_{E^+}-[\barpartial_{E^+\bar s}v,u]_{E^+}
+\barpartial_{E^+v_{\bar s}}u-\barpartial_{E^+u_{\bar s}}v,
\label{liecmplx7}
\end{equation}
for $s,u,v\in\Gamma(E^+)$, 
where $u_{\bar s}=\overline{\barpartial_{E^+\bar u}s}$. 

Let us call a local section $t$ of $E^+$ {\it $\mathcal{E}$--holomorphic} if 
\begin{equation}
\barpartial_{E^+\bar s}t=0, \qquad s\in\Gamma(E^+).
\label{liecmplx5}
\end{equation}
In general, this equation has only local solutions. 
We denote by $\mathcal{O}_{\mathcal{E}}$ the sheaf of solutions of \eqref{liecmplx5}.
Using \eqref{liecmplx3}, one verifies that, if $f\in \mathcal{O}_{\mathcal{M}}(U)$ 
and $s\in \mathcal{O}_{\mathcal{E}}(U)$, then $fs\in  \mathcal{O}_{\mathcal{E}}(U)$
as well. Thus, $\mathcal{O}_{\mathcal{E}}$ is a sheaf of $\mathcal{O}_{\mathcal{M}}$ modules.

Does the operator $\barpartial_{E^+\bar s}$ defines a holomorphic structure 
on $E^+$ making it a holomorphic vector bundle $\mathcal{E}$ over $\mathcal{M}$
in analogy to $T_M{}^+$?
In that case, is $\mathcal{E}$ naturally also a holomorphic Lie algebroid
again in analogy to $T_M{}^+$?

In order $\barpartial_{E^+\bar s}$ to define a holomorphic structure 
on $E^+$, it is necessary that, locally, there exist $\mathcal{E}$--holomorphic 
frames of $E^+$, that is that the sheaf $\mathcal{O}_{\mathcal{E}}$ is locally free.
This is not the case in general. Let us assume however that it is. 
Then, $E^+$ becomes a holomorphic vector bundle $\mathcal{E}$. 

Eq. \eqref{liecmplx6} is not sufficient to conclude that the anchor 
$\rho_{E^+}$ 
is holomorphic, since in general not every field 
$x\in\Gamma(T_M{}^+)$ is of the form $x=\rho_{E^+}s$
for some field $s\in\Gamma(E^+)$. \eqref{liecmplx6} is anyway compatible
with the holomorphy of $\rho_{E^+}$. So, we can 
assume $\rho_{E^+}$ does have this property. 

Eq. \eqref{liecmplx7} implies that, if $U$ is an open subset of $M$ and 
$u,v\in \mathcal{O}_{\mathcal{E}}(U)$, $[u,v]_{E^+}\in \mathcal{O}_{\mathcal{E}}(U)$ 
as well. Thus, $\mathcal{O}_{\mathcal{E}}$ is closed under the Le brackets
$[\cdot,\cdot]_{E^+}$. 

In summary, in order $E^+$ to become a holomorphic vector bundle
$\mathcal{E}$ over $\mathcal{M}$, we have to assume that the sheaf
$\mathcal{O}_{\mathcal{E}}$ is locally free. Further, in order 
$\mathcal{E}$ to be a holomorphic Lie algebroid, we have to assume 
further that $\rho_{E^+}$ is holomorphic. 
When the Lie algebroid $E$ is transitive,
we are able to make definitely stronger statements.

\subsection{\normalsize \textcolor{blue}{The transitive case}}\label{subsec:liecmplx4}

\hspace{.5cm} Let $E$ be a transitive Lie algebroid and let
$J$ be an almost complex structure of $E$. 
Suppose that $\ker\rho_E$ is invariant under $J$. 
Then, in addition to the primary Nijenhuis field $N_{EJ}$ defined in 
\eqref{CLAx2}, we can define a secondary Nijenhuis field 
$\tilde N_{EJ}\in\Gamma(E^*\otimes\ker\rho_E{}^*\otimes\ker\rho_E{})$ by 
\begin{equation}
\tilde N_{EJ}(s,t)=[Js,Jt]_E+[s,t]_E-J([Js,t]_E-[s,Jt]_E),
\label{transcmplx1}
\end{equation}
with $s\in\Gamma(E)$, $t\in\Gamma(\ker\rho_E)$.
The vanishing of $\tilde N_{EJ}$ is equivalent to the property that 
$[\bar s,t]_{E^c}\in \Gamma(\ker\rho_{E^+})$ for 
$s\in\Gamma(E^+)$, $t\in\Gamma(\ker\rho_{E^+})$, as is easily verified.

Let $E$ be a transitive Lie algebroid with complex structure 
\footnote{$\vphantom{\bigg[}$  We note that, as $\rho_EN_{EJ_E}(s,t)=N_{T_MJ_{EM}}(\rho_Es,\rho_Et)$ 
for $s,t\in\Gamma(E)$ by \eqref{CLAx2} and $\rho_E$ is surjective, 
the integrability of $J_E$ automatically implies that of $J_{EM}$.}.
It is readily seen that the Lie algebroid with complex base
$E^+$ is then also transitive.

As, by \eqref{CLAx4}, $\ker\rho_E$ is invariant under $J_E$,
the secondary Nijenhuis field $\tilde N_{EJ_E}$ is defined. 
It can then be shown that the vanishing of $\tilde N_{EJ_E}$,
\begin{equation}
\tilde N_{EJ_E}=0,
\label{transcmplx2}
\end{equation}
is a necessary and sufficient condition for the sheaf $\mathcal{O}_{\mathcal{E}}$ 
to be locally free. Here is a sketch of the proof. 
$\mathcal{O}_{\mathcal{E}}$ is locally free if and only if 
$E^+$ admits locally an $\mathcal{E}$--holomorphic frame. 
Let $\{e_i\}$ be a local frame of $E^+$. There is an $\mathcal{E}$--holomorphic 
frame of $E^+$ if and only if we can find a local invertible matrix
function $T^i{}_j$ such that $\barpartial_{E^+\bar s}(T^j{}_ie_j)=0$
for $s\in\Gamma(E^+)$. Now, the anchor $\rho_{E^+}:E^+\mapsto T_M{}^+$ 
is surjective. There is thus a bundle map $\sigma^+:T_M{}^+\mapsto E^+$
such that $\rho_{E^+}\sigma^+=1_{E^+}$. The above equation then splits
into two equations: $i$) $\barpartial_{E^+\bar s}(T^j{}_ie_j)=0$ for 
$s\in\Gamma(\ker\rho_{E^+})$ and $ii$) $\barpartial_{E^+\bar \sigma^+\bar x}(T^j{}_ie_j)=0$ 
for $x\in \Gamma(T_M{}^+)$. Eq. $i$ reduces to $\Pi_E[\bar s,e_i]_{E^c}=0$
and thus to \eqref{transcmplx2}. Eq. $ii$ is a genuine differential equation for
$T^i{}_J$. It has a solution provided the appropriate integrability condition 
is satisfied. The condition indeed is as a consequence of the Jacobi identity 
obeyed by the Lie bracket $[\cdot,\cdot]_{E^c}$.

Since $E^+$ is transitive, \eqref{liecmplx6} is sufficient to conclude that 
the anchor $\rho_{E^+}$ is holomorphic, as by the surjectivity of $\rho_{E^+}$, 
every field $x\in\Gamma(T_M{}^+)$ is of the form $x=\rho_{E^+}s$
for some field $s\in\Gamma(E^+)$. 

In summary, if $E$ is a transitive Lie algebroid with complex structure
satisfying \eqref{transcmplx2}, then $\mathcal{O}_{\mathcal{E}}$ is a locally free
sheaf and $\rho_{E^+}$ is holomorphic. $E^+$ has thus a canonical holomorphic structure 
rendering it a transitive holomorphic Lie algebroid $\mathcal{E}$ over
$\mathcal{M}$. Since $\mathcal{O}_{\mathcal{E}}$ is the sheaf of sections of $\mathcal{E}$ 
and is defined by eq. \eqref{liecmplx5}, we can interpret 
$\bar\partial_{E^+}$ as the {\it Cauchy--Riemann operator} on $\Gamma(E^+)$.

\subsection{\normalsize \textcolor{blue}{Induced structures on $\ker\rho_E$ and $\tilde E$}}\label{subsec:liecmplx5}

\hspace{.5cm}  Suppose that $E$ is a transitive Lie algebroid with complex structure
satisfying \eqref{transcmplx2}, as above. 

Since $\ker\rho_E$ is invariant under $J_E$, $J_E$ defines almost complex structures 
on the vector bundles $\ker\rho_E$ and 
$\tilde E=E/\ker\rho_E\simeq T_M$. 
In this way, 
both $\ker\rho_{E}$ and $\tilde E$ are Lie algebroids with complex structure. 
$\ker\rho_{E}$ is not transitive, as its anchor vanishes identically.
$\tilde E$, conversely, is transitive, as its anchor is a vector bundle isomorphism.
Further, $\tilde E$ satisfies \eqref{transcmplx2} trivially. 

Therefore, the above theory does not apply to $\ker\rho_E$ while it does to
$\tilde E$. Nevertheless, the holomorphic structure of $E^+$ induces 
naturally holomorphic structures on $(\ker\rho_E)^+\simeq\ker\rho_{E^+}$ and 
$(\tilde E)^+\simeq \tilde E^+$ and, in the latter case,
the holomorphic structure is precisely the one associated with the structure of
transitive Lie algebroid with complex structure of $\tilde E$.  
In fact, as the anchor $\rho_{\mathcal{E}}$ of $\mathcal{E}$ 
is holomorphic, $\ker\rho_{\mathcal{E}}$ is a holomorphic Lie subalgebroid of $\mathcal{E}$
and, so, a holomorphic Lie algebroid. Consequently, 
$\tilde{\mathcal{E}}=\mathcal{E}/\ker\rho_{\mathcal{E}}\simeq \mathcal{T}_{\mathcal{M}}$
is a holomorphic Lie algebroid.  

\subsection{\normalsize \textcolor{blue}{Standard complex connections}}\label{subsec:liecmplx6}

\hspace{.5cm}  Let $E$ be a transitive Lie algebroid and let $\sigma$ be a splitting of $E$
(cf. subsect. \ref{subsec:liealg2}). We define a field $\omega\in\Gamma(\End(E))$ by
\begin{equation}
\omega=1_E-\sigma\rho_E.
\label{stconn1}
\end{equation}
As $\omega^2=\omega$, $\omega$ is a projector field. In fact, $\omega$  
projects on $\ker\rho_E$. 

Let $E$ be a transitive Lie algebroid and let $(J,J_M)$ be a Lie algebroid
complex 
structure of $E$. A splitting $\sigma$ of $E$ is said {\it complex} with respect to 
$(J,J_M)$ if  
\begin{equation}
\sigma J_M=J\sigma.
\label{stconn2}
\end{equation}
Compare with condition \eqref{CLAx4}. 

Let $E$ be a transitive Lie algebroid with complex structure
satisfying the condition \eqref{transcmplx2}. 
Let $\sigma$ be a complex splitting of $E$ and let $D_M$ be an ordinary connection
on $M$ such that the complex structure $J_{EM}$ is parallel, so that 
\begin{equation}
D_{Mx}J_{EM}=0,
\label{stconn3}
\end{equation}
with $x\in\Gamma(T_M)$. We now set 
\begin{equation}
D_st=\sigma D_{M\rho_Es}(\rho_Et)+[s,\omega t]_E+H_{E\sigma}(\rho_Es,\rho_Et), 
\label{stconn4}
\end{equation}
with $s,t\in\Gamma(T_M)$, where the field $H_{E\sigma}\in\Gamma(T_M{}^{*\otimes 2}\otimes \ker\rho_E)$
is defined by 
\begin{align}
H_{E\sigma}(x,y)&=\frac{1}{4}\Big(G_{E\sigma}(J_{EM}x,J_{EM}y)+G_{E\sigma}(x,y)
\vphantom{\Big]}
\label{stconn4/1}
\\
&\hspace{4.5cm}
+J_E(G_{E\sigma}(J_{EM}x,y)-G_{E\sigma}(x,J_{EM}y)\Big),
\vphantom{\Big]}
\nonumber
\end{align}
with $x,y\in\Gamma(T_M)$, $G_{E\sigma}$ being the curvature of the splitting $\sigma$
(cf. eq. \eqref{LAAex3}). 
Directly from the definition \eqref{stconn4}, we verify that
$D$ is an $E$ Lie algebroid connection on $E$ (cf. eqs. \eqref{LAAconn1}, \eqref{LAAconn2}). 
Note that, when $E=T_M$, $D=D_M$. We call connections of this
form {\it standard}. They will play an important role in the formulation of the Lie 
algebroid heterotic sigma model. The last term in the right hand side of \eqref{stconn4}
may dropped without compromising $D$ being a connection. The reason why it is added
is to make $D$ naturally related to the operator $\barpartial_{E^+}$, as shown momentarily.

The complex structure $J_E$ of $E$ is parallel with respect to $D$, 
\begin{equation}
D_sJ_E=0,
\label{stconn5}
\end{equation} 
with $s\in\Gamma(E)$. Another distinguished property of $D$ is the following. 
We define an ordinary connection on the vector bundle $\Hom(E,T_M)$,
denoted also by $D_M$, 
by the relation $(D_{Mx}\phi)t=D_{Mx}(\phi t)-\phi D_{\sigma x}t$ with 
$t\in\Gamma(E)$, for $x\in\Gamma(T_M)$, $\phi\in\Gamma(\Hom(E,T_M))$. 
We further view the anchor of $E$ as a field $\rho_E\in\Gamma(\Hom(E,T_M))$.
Then, as is easily verified, $\rho_E$ is parallel with respect to $D_M$, 
\begin{equation}
D_{Mx}\rho_E=0,
\label{stconn6}
\end{equation} 
with $x\in\Gamma(T_M)$. 

The torsion $T_D$ of $D$ is given by the formula
\vfil\eject\noindent
\begin{align}
T_D(s,t)&=\sigma T_{D_M}(\rho_Es,\rho_Et)+[s,\omega t]_E-[t,\omega s]_E
\vphantom{\Big[}
\label{stconn7}
\\
&\hspace{3.5cm}
-\omega[s,t]_E+H_{E\sigma}(\rho_Es,\rho_Et)-H_{E\sigma}(\rho_Et,\rho_Es),
\vphantom{\Big[}
\nonumber
\end{align} 
with $s,t\in\Gamma(E)$. We note that, when $T_{D_M}=0$,
$T_D$ is generally non vanishing, but $\ker \rho_E$ valued.

The curvature $R_D$ of $D$ is given by the formula
\begin{align}
R_D(s,t)u&=\sigma R_{D_M}(\rho_Es,\rho_Et)\rho_Eu-H_{E\sigma}(\rho_E[s,t]_E,\rho_Eu)
\vphantom{\Big[}
\label{stconn8}
\\
&
+[s,H_{E\sigma}(\rho_Et,\rho_Eu)]_E-[t,H_{E\sigma}(\rho_Es,\rho_Eu)]_E
\vphantom{\Big[}
\nonumber
\\
&
+H_{E\sigma}(\rho_Es, D_{M\rho_Et}(\rho_Eu))
-H_{E\sigma}(\rho_Et, D_{M\rho_Es}(\rho_Eu)),
\vphantom{\Big[}
\nonumber
\end{align} 
with $s,t,u\in\Gamma(E)$. 

For $s\in\Gamma(E)$ and $t\in\Gamma(\ker\rho_E)$, we have 
\begin{equation}
D_st=[s,t]_E.
\label{stconn9}
\end{equation}
So, upon restriction to $\Gamma(\ker\rho_E)$, $D$ reduces
to the canonical $E$ connection $D^{\mathrm{ad}}_E$ 
of the adjoint bundle $\ker\rho_E$ (cf. eq. \eqref{LAAconn7}).

Since, for $s\in\Gamma(E)$, $t\in\Gamma(\ker\rho_E)$, we have 
$D_st\in\Gamma(\ker\rho_E)$, $D$ induces also an $E$ Lie algebroid connection
on $\tilde E$, where $\tilde E=E/\ker\rho_E$ 
(cf. subsect. \ref{subsec:liealg2}), 
\begin{equation}
D_st=\sigma D_{M\rho_Es}(\rho_Et)\quad \text{mod $\ker\rho_E$},
\label{stconn10}
\end{equation}
with $s\in\Gamma(E)$, $t\in\Gamma(\tilde E)$. As the right hand side of \eqref{stconn10}
vanishes for $s\in\Gamma(\ker\rho_E)$, $D$ induces an $\tilde E$ Lie algebroid connection
on $\tilde E$. Recalling that $\tilde E\simeq T_M$ and that 
$\rho_E:\tilde E\mapsto T_M$ is a Lie algebroid
isomorphism with inverse $\sigma:T_M\mapsto\tilde E$, 
we realize that this connection is essentially the connection
$D_M$ entering in the definition \eqref{stconn4} of $D$.

The $E$ Lie algebroid connection $D$ on $E$ extends by complexification to 
an $E^c$  Lie algebroid connection $D^c$ on $E^c$. 
All the properties of $D$ found above generalize to $D^c$ without change.

By \eqref{stconn5},
$\Gamma(E^+)$ is invariant under $D^c$. In this way, by restriction, we obtain an 
$\overline{E}{}^+$ Lie algebroid connection $\bardeecpl^+$ on $E^+$
(cf. eqs. \eqref{liecmplx2}, \eqref{liecmplx3}). Explicitly, 
\begin{equation}
\bardeecpl^+{}_{\bar s}t=D^c{}_{\bar s}t,
\label{stconn11}
\end{equation}
where $s,t\in\Gamma(E^+)$. 
The remarkable property of this $\bardeecpl^+$ 
is that, if $\bardeecpl_M{}^+=\barpartial_{T_M{}^+}$,
the Cauchy--Riemann operator of $T_M{}^+$, 
then, likewise, 
\begin{equation}
\bardeecpl^+=\barpartial_{E^+},
\label{stconn12}
\end{equation}
the Cauchy--Riemann operator of $E^+$ defining its canonical holomorphic structure 
(cf. eq. \eqref{liecmplx1}). We remark that, 
much as $D_M$ is nor uniquely fixed by the requirement
that $\bardeecpl_M{}^+=\barpartial_{T_M{}^+}$, so $D$ is not the only connection 
enjoying the important property \eqref{stconn12}.

\subsection{\normalsize \textcolor{blue}{Local expressions}}\label{subsec:liecmplx7}

\hspace{.5cm} Let $E$ be a transitive Lie algebroid with complex structure
satisfying \eqref{transcmplx2}. Then, as shown earlier, the base 
$M$ of $E$ is automatically a complex manifold $\mathcal{M}$ and 
$E^+$ has a canonical holomorphic structure making it 
a holomorphic Lie algebroid $\mathcal{E}$ over $\mathcal{M}$.
Consequently, over any sufficiently small open set $U$ of $M$, there exists a holomorphic frame 
$\{\partial_a\}$ of $\mathcal{T}_{\mathcal{M}}$ associated with each set of holomorphic coordinates
$\{z^a\}$ of $\mathcal{M}$ and a holomorphic frame $\{e_i\}$ of $\mathcal{E}$.

Since the anchor $\rho_{\mathcal{E}}$ of $\mathcal{E}$ is holomorphic, we have 
\begin{equation}
\rho_{\mathcal{E}}e_i=\rho^a{}_i\partial_a,
\label{lex1}
\end{equation}
where the anchor structure functions $\rho^a{}_i$ are holomorphic
\begin{equation}
\bar\partial_{\bar a}\rho^b{}_k=0.
\label{lex2}
\end{equation}
As $\mathcal{O}_{\mathcal{E}}(U)$ is closed under the Lie bracket $[\cdot,\cdot]_{\mathcal{E}}$,
we have 
\begin{equation}
[e_j,e_j]_{\mathcal{E}}=f^k{}_{ij}e_k,
\label{lex3}
\end{equation} 
where the bracket structure functions $f^k{}_{ij}$ are holomorphic
\begin{equation}
\bar\partial_{\bar a}f^k{}_{ij}=0.
\label{lex4}
\end{equation} 

The holomorphy of the anchor $\rho_{\mathcal{E}}$ allows us to choose adapted holomorphic
frames. Here and in the following, we use the Latin letters $u,v,w,x,y$ as holomorphic
$\ker\rho_{\mathcal{E}}$ indices and the Latin letters $p,q,r,s,t$ as 
holomorphic $\mathcal{E}/\ker\rho_{\mathcal{E}}$ indices. For an adapted holomorphic frame, \hfil
we have $\{e_i\}=\{e_u\}\cup\{e_p\}$, \hfil where 
$\{e_u\}$ is a holomorphic frame 
of $\ker\rho_{\mathcal{E}}$,
\begin{equation}
\rho_{\mathcal{E}}e_u=0. 
\label{lex6}
\end{equation} 
\eqref{lex6} implies the important relation 
\begin{equation}
\rho^a{}_u=0. 
\label{lex7}
\end{equation}
Since $[s,t]_{\mathcal{E}}\in\mathcal{O}_{\ker\rho_{\mathcal{E}}}(U)$ if either 
$s\in \mathcal{O}_{\ker\rho_{\mathcal{E}}}(U)$ or $t\in \mathcal{O}_{\ker\rho_{\mathcal{E}}}(U)$, 
we have  
\begin{equation}
f^p{}_{uv}=0,\qquad f^p{}_{uq}=0.   
\label{lex8}
\end{equation} 

Let us now view $\mathcal{O}_{\mathcal{E}}(U)$ as a subspace of $\Gamma(E^+|_U)$.
From \eqref{liecmplx1}, it is readily verified that 
$[\bar s,t]_{E^c}=0$ whenever $s,t\in\mathcal{O}_{\mathcal{E}}(U)$. From \eqref{lex3}
and this observation, we have then
\begin{equation}
[e_j,e_j]_{E^c}=f^k{}_{ij}e_k,\qquad [\bar e_{\bar\imath},e_j]_{E^c}=0 \qquad \text{and c. c.}.
\label{lex5}
\end{equation} 
Therefore, only $f^k{}_{ij}$ and c. c. are non vanishing. 

From \eqref{LAA1}--\eqref{LAA4}, the structure functions $\rho^a{}_i$, $f^k{}_{ij}$ 
satisfy
\begin{align}
&f^i{}_{jk}+f^i{}_{kj}=0,
\vphantom{\Big]}
\label{lex9}
\\
&f^i{}_{jm}f^m{}_{kl}+f^i{}_{km}f^m{}_{lj}+f^i{}_{lm}f^m{}_{jk}
\vphantom{\Big]}
\label{lex10}
\\
&\hspace{3cm}+\rho^a{}_j\partial_af^i{}_{kl}+\rho^a{}_k\partial_kf^i{}_{lj}
+\rho^a{}_l\partial_af^i{}_{jk}=0,
\vphantom{\Big]}
\nonumber
\\
&\rho^b{}_i\partial_b\rho^a{}_j-\rho^b{}_j\partial_b\rho^a{}_i-f^k{}_{ij}\rho^a{}_k=0.
\vphantom{\Big]}
\label{lex11}
\end{align}
A host of more explicit relations can be obtained for an adapted frame
by splitting the frame index $i$ as $(u,p)$ and using \eqref{lex7}, \eqref{lex8}. 

In our treatment, the complex Lie algebroid $E$ is always endowed with 
a complex splitting $\sigma$ (cf. eq. \eqref{stconn2}). 
By \eqref{stconn2}, we have 
\begin{equation}
\sigma\partial_a=\sigma^i{}_a e_i\qquad \text{and c. c.}.
\label{lex17}
\end{equation} 
$\sigma$ is characterized by its curvature $G_{E\sigma}$ (cf. eq. \eqref{LAAex3}).
The components $G_{E\sigma}$ are defined by the relations
\begin{equation}
G_{E\sigma}{}^c(\partial_a,\partial_b)=G^i{}_{ab}e_i,\qquad
G_{E\sigma}{}^c(\bar\partial_{\bar a},\partial_b)=G^i{}_{\bar ab}e_i +G^{\bar\imath}{}_{\bar ab}e_{\bar\imath}
\qquad \text{and c. c.}.
\label{lex18}
\end{equation} 
In the right hand side of the first of these relations, a term of the form 
$G^{\bar\imath}{}_{ab}e_{\bar\imath}$ is absent, as this automatically vanishes by the complex 
nature of $\sigma$. 
Applying the definition \eqref{LAAex3}, by a straightforward calculation, we find
\begin{equation}
G^i{}_{ab}=\partial_a\sigma^i{}_b-\partial_b\sigma^i{}_a+f^i{}_{jk}\sigma^j{}_a\sigma^k{}_b,
\qquad G^i{}_{\bar ab}=\bar\partial_{\bar a}\sigma^i{}_b \qquad \text{and c. c.}
\vphantom{\Big]}
\label{lex19}
\end{equation}
Note that, since $G_{E\sigma}$ is $\ker\rho_E$ valued, we have 
\begin{equation}
G^p{}_{ab}=0, \qquad G^p{}_{\bar ab}=0 \qquad \text{and c. c.},
\label{lex20}
\end{equation} 
as may be readily checked. Note also that the vanishing of the components
$G^u{}_{\bar ab}$ is equivalent to the holomorphy of $\sigma$. 

In view of its application to the construction of the Lie algebroid sigma model, 
we provide the local expression of the standard connection $D$ 
(cf. eq. \eqref{stconn4}). We assume that the connection $D_M$ is such that 
$\bardeecpl_M{}^+=\barpartial_{T_M{}^+}$, so that \eqref{stconn12} holds. 
\eqref{stconn12} and the holomorphy condition $\barpartial_{E^+}e_i=0$
entail that
\begin{equation}
D^c{}_{\bar e_{\bar\imath}}e_j=0 \qquad \text{and c. c.}.
\label{lex12}
\end{equation} 
The coefficients of the connection are defined by the relation
\begin{equation}
D^c{}_{e_i}e_j=A^k{}_{ij}e_k \qquad \text{and c. c.},
\label{lex13}
\end{equation} 
which follows from the invariance of $\Gamma(E^+)$ under $D^c$.
The coefficients $A^k{}_{ij}$ can be computed directly 
using \eqref{stconn4}, 
\begin{equation}
A^k{}_{ij}=\sigma^k{}_c\rho^a{}_i\rho^b{}_jA_M{}^c{}_{ab}-\rho^a{}_i\rho^b{}_j\partial_a\sigma^k{}_b
+f^k{}_{il}\omega^l{}_j,
\label{lex14}
\end{equation} 
where the components of $\omega$ are defined 
by $\omega e_i=\omega^j{}_ie_j$ with $\omega^p{}_i=0$, 
in accordance 
with \eqref{stconn1}.
In particular, we have 
\begin{equation}
A^v{}_{iu}=f^v{}_{iu},
\label{lex15}
\end{equation} 
as implied by the property \eqref{stconn9}, and 
\begin{equation}
A^p{}_{ij}=0\qquad \text{for $(i,j)\not=(q,r)$},
\label{lex16}
\end{equation} 
as follows from \eqref{stconn10}. 

The local expressions of the torsion and the curvature of $D^c$ can also be obtained.
We shall not exhibit them however, since they will not be needed in the following. 

\vfill\eject

\section{\normalsize \textcolor{blue}{The heterotic Lie algebroid sigma model}}\label{sec:hetersgm}

\subsection{\normalsize \textcolor{blue}{The heterotic Lie algebroid sigma model: classical theory}}
\label{subsec:hetersgm1}

\hspace{.5cm} As every sigma model, the Lie algebroid 
heterotic sigma model is characterized 
classically by the following data:
$a$) the target space geometry, $b$) the world--sheet geometry, 
$c$) the field content, $d$) the action and $e$) its symmetries.

The target space geometry consists of a transitive Lie algebroid $E$ with complex structure
over a manifold $M$ satisfying \eqref{transcmplx2} 
(cf. subsects. \ref{subsec:liealg2}, \ref{subsec:liecmplx2}). 
$E$ is further furnished with a complex splitting $\sigma$.
Finally, $M$ is endowed with a metric $g$ which is Kaehler with respect its complex structure 
$J_{EM}$. 

The world--sheet is a Riemann surface $\Sigma$ equipped with a spin 
structure $K_\Sigma{}^{1/2}$, a tensor square root of the canonical line bundle
$K_\Sigma$ of $\Sigma$. 

The fields of the model are an embedding field $x\in\Map(\Sigma,M)$, two fermion fields 
$\chi_{\bar \theta}\in \Gamma(\overline{K}_\Sigma{}^{1/2}\otimes x^*\Pi \tilde E^+)$,
$\overline{\chi}_{\bar \theta}\in 
\Gamma(\overline{K}_\Sigma{}^{1/2} \otimes x^*\Pi \overline{\tilde E}{}^+)$, 
two more fermion fields 
$\lambda_\theta\in \Gamma(K_\Sigma{}^{1/2}\otimes x^*\Pi\ker\rho_{E^+})$,
$\lambda^*{}_\theta\in \Gamma(K_\Sigma{}^{1/2}\otimes x^*\Pi\ker\rho_{E^+}{}^*)$
and two boson fields 
$l_{\theta\bar \theta}\in \Gamma(K_\Sigma{}^{1/2}\otimes \overline{K}_\Sigma{}^{1/2} \otimes x^*\ker\rho_{E^+})$,
$l^*{}_{\theta\bar \theta}\in \Gamma(K_\Sigma{}^{1/2}\otimes \overline{K}_\Sigma{}^{1/2} 
\otimes x^*\ker\rho_{E^+}{}^*)$
\footnote{$\vphantom{\bigg[}$ In the string theory literature, the spinor indices
$\theta,\bar\theta$ are generally denoted by $-,+$}. 
Here, $\tilde E^+$ is the quotient vector bundle $E^+/\ker\rho_{E^+}$ 
and $\ker\rho_{E^+}{}^*$ is the dual vector bundle of $\ker\rho_{E^+}$.
$\Pi$ is the parity reversal operator which associates with any vector bundle 
$V$ its odd counterpart $\Pi V$. $x^*V$ denotes the pull--back of a vector bundle $V$ over $M$ 
by the map $x$.

The action of the Lie algebroid 
heterotic sigma model is
\begin{align}
S&=\int_\Sigma d^2z\,\Big[
\frac{1}{2}g_{\bar ab}(x)(\barpartial_{\bar z}x^{\bar a}\partial_z x^b+\partial_z x^{\bar a}\barpartial_{\bar z}x^b)
+ig_{\bar ab}\rho^{\bar a}{}_{\bar p}\rho^b{}_q(x)\overline{\chi}^{\bar p}{}_{\bar\theta}D_z\chi^q{}_{\bar\theta}
\vphantom{\Big]}
\label{hetersgm1} 
\\
&\hspace{1.9cm}+i\lambda^*{}_{u\theta}\bardee_{\bar z}\lambda^u{}_\theta
+f^u{}_{wv}G^w{}_{\bar ab}\rho^{\bar a}{}_{\bar p}\rho^b{}_q(x)\lambda^*{}_{u\theta}\lambda^v{}_\theta
\overline{\chi}^{\bar p}{}_{\bar\theta}\chi^q{}_{\bar\theta}
-l^*{}_{u\theta\bar\theta}l^u{}_{\theta\bar\theta}\Big].
\vphantom{\Big]}
\nonumber
\end{align}
Here, $G^u{}_{\bar ab}$ is the $(1,1)$ component of the curvature $G_{E\sigma}$  of $\sigma$ 
(cf. eq. \eqref{lex19}). 
$D_z$, $\bardee_{\bar z}$ are the components of the connection $x^*D_\sigma$ 
of $x^*E$ yielded via pull--back by 
\vfill\eject\noindent
$x$ of the connection $D_\sigma$
of $E$ associated with a standard $E$ Lie 
algebroid connection $D$ of $E$
and the splitting $\sigma$ as in eq. \eqref{LAAconn5}. 
$D$, in turn, is constructed using $\sigma$ and the customary  Levi--Civita connection 
$\nabla_M$ of $T_M$ according to \eqref{stconn4}. 
The connection $x^*D_\sigma$ extends to $x^*E^c$ by complexification and then
to $x^*E^+$ by restriction, on account of \eqref{stconn5}.  
Since $x^*D_\sigma$ leaves $x^*\ker\rho_{E^+}$ invariant, by \eqref{stconn6},
$x^*D_\sigma$ induces a connection on both $x^*\ker\rho_{E^+}$ and $x^*\tilde E^+$. 
Explicitly, using \eqref{lex14},  we have 
\begin{subequations}
\begin{align}
&D_z\phi^p=\partial_z\phi^p
+(\Gamma^a{}_{cb}\sigma^p{}_a\rho^b{}_q-\partial_c\sigma^p{}_b\rho^b{}_q)(x)\partial_z x^c\phi^q,
\qquad \text{and c. c.,}
\vphantom{\Big]}
\label{hetersgm2a} 
\\
&\bardee_{\bar z}\psi^u=\barpartial_{\bar z}\psi^u
+f^u{}_{iv}\sigma^i{}_a(x)\barpartial_{\bar z}x^a\psi^v,
\vphantom{\Big]}
\label{hetersgm2b} 
\\
&\bardee_{\bar z}\psi^*{}_u=\barpartial_{\bar z}\psi^*{}_u
-f^v{}_{iu}\sigma^i{}_a(x)\barpartial_{\bar z}x^a\psi^*{}_v. 
\vphantom{\Big]}
\label{hetersgm2c} 
\end{align}
\end{subequations}
\!Here and in the following, the abbreviation
``c. c.'' denotes target space complex conjugation only and is thus inert on the
world--sheet.

The action $S$ of the heterotic sigma model has a high amount of symmetry, as we now show.

The $(0,2)$ supersymmetry field variations are given by 
\begin{subequations}
\begin{align}
&\delta_\alpha x^a=i\alpha^{\bar\theta}\rho^a{}_p(x)\chi^p{}_{\bar\theta} \qquad \text{and c. c.,}
\vphantom{\Big]}
\label{hetersgm3a} 
\\
&\delta_\alpha\chi^p{}_{\bar\theta}=-\frac{i}{2}\alpha^{\bar\theta}f^p{}_{qr}(x)\chi^q{}_{\bar\theta}\chi^r{}_{\bar\theta}
-\overline{\alpha}^{\bar\theta}\sigma^p{}_a(x)\barpartial_{\bar z}x^a \qquad \text{and c. c.,}
\vphantom{\Big]}
\label{hetersgm3b} 
\\
&\delta_\alpha\lambda^u{}_\theta=i\alpha^{\bar\theta}\big(-f^u{}_{iv}\sigma^i{}_a\rho^a{}_p(x)
\chi^p{}_{\bar\theta}\lambda^v{}_\theta+l^u{}_{\theta\bar\theta}\big),
\vphantom{\Big]}
\label{hetersgm3c} 
\\
&\delta_\alpha l^u{}_{\theta\bar\theta}=
-i\alpha^{\bar\theta}f^u{}_{iv}\sigma^i{}_a\rho^a{}_p(x)
\chi^p{}_{\bar\theta} l^v{}_{\theta\bar\theta}
\vphantom{\Big]}
\label{hetersgm3d} 
\\
&\hspace{2cm}+i\overline{\alpha}^{\bar\theta}\big(i\bardee_{\bar z}\lambda^u{}_\theta+
f^u{}_{wv}G^w{}_{\bar ab}\rho^{\bar a}{}_{\bar p}\rho^b{}_q(x)\overline{\chi}^{\bar p}{}_{\bar\theta}\chi^q{}_{\bar\theta}
\lambda^v{}_\theta \big),
\vphantom{\Big]}
\nonumber
\\
&\delta_\alpha\lambda^*{}_{u\theta}=i\alpha^{\bar\theta}f^v{}_{iu}\sigma^i{}_a\rho^a{}_p(x)
\chi^p{}_{\bar\theta}\lambda^*{}_{v\theta}+i\overline{\alpha}^{\bar\theta}l^*{}_{u\theta\bar\theta},
\vphantom{\Big]}
\label{hetersgm3e} 
\\
&\delta_\alpha l^*{}_{u\theta\bar\theta}=i\alpha^{\bar\theta}\big(f^v{}_{iu}\sigma^i{}_a\rho^a{}_p(x)
\chi^p{}_{\bar\theta}l^*{}_{v\theta\bar\theta}
\vphantom{\Big]}
\label{hetersgm3f} 
\\
&\hspace{2cm}+i\bardee_{\bar z}\lambda^*{}_{u\theta}
-f^v{}_{wu}G^w{}_{\bar ab}\rho^{\bar a}{}_{\bar p}\rho^b{}_q(x)\overline{\chi}^{\bar p}{}_{\bar\theta}\chi^q{}_{\bar\theta}
\lambda^*{}_{v\theta} \big),
\vphantom{\Big]}
\nonumber
\end{align}
\label{hetersgm3}
\end{subequations} 
\!\!where the spinors 
$\alpha^{\bar\theta}, \overline{\alpha}^{\bar\theta}\in\Gamma(\Pi \overline{K}_\Sigma{}^{-1/2})$
satisfy the antiholomorphy condition 
\eject\noindent
\begin{equation}
\partial_z\alpha^{\bar\theta}=\partial_z\overline{\alpha}^{\bar\theta}=0
\label{hetersgm6}
\end{equation} 
and complex conjugation interchanges $\alpha^{\bar\theta}$, $\overline{\alpha}^{\bar\theta}$
\footnote{$\vphantom{\bigg[}$ In our conventions, if $\xi$, $\eta$ are complex fermion fields,
$\overline{\xi\eta}=\overline{\eta}\overline{\xi}=-\overline{\xi}\overline{\eta}$.}.

The sigma model action $S$ enjoys $(0,2)$ supersymmetry, so that 
\begin{equation}
\delta_\alpha S=0,
\label{hetersgm5}
\end{equation}
provided the splitting $\sigma$ satisfies the condition
\begin{equation}
G^u{}_{ab}=0,      
\label{hetersgm4/1}
\end{equation}
where $G^u{}_{ab}$ is $(2,0)$ component of the curvature $G_{E\sigma}{}^c$
of $\sigma$ (cf. eq. \eqref{lex19}). As a matter of fact, the weaker condition 
$f^v{}_{wu}G^w{}_{ab}=0$ would be sufficient for $(0,2)$ supersymmetry.
However, in the analysis of the cohomological properties of the topological twisted version 
of the model studied in the next section, it will emerge that the
stronger condition \eqref{hetersgm4/1} is required. 
\eqref{hetersgm4/1} will thus be assumed right away. 

The $(1,0)$ supersymmetry field variations are given by 
\begin{subequations}
\begin{align}
&s_\xi x^a=0, \qquad \text{and c. c.,} 
\vphantom{\Big]}
\label{hetersgm10a} 
\\
&s_\xi\chi^p{}_{\bar\theta}=0,\qquad \text{and c. c.,}
\vphantom{\Big]}
\label{hetersgm10b} 
\\
&s_\xi\lambda^u{}_\theta=-\frac{i}{2}\xi^\theta f^u{}_{vw}(x)\lambda^v{}_\theta\lambda^w{}_\theta,
\vphantom{\Big]}
\label{hetersgm10c} 
\\
&s_\xi l^u{}_{\theta\bar\theta}=-i\xi^\theta f^u{}_{vw}(x)\lambda^v{}_\theta l^w{}_{\theta\bar\theta},
\vphantom{\Big]}
\label{hetersgm10d} 
\\
&s_\xi\lambda^*{}_{u\theta}=i\xi^\theta f^w{}_{vu}(x)\lambda^v{}_\theta\lambda^*{}_{w\theta},
\vphantom{\Big]}
\label{hetersgm10e} 
\\
&s_\xi l^*{}_{u\theta\bar \theta}=i\xi^\theta f^w{}_{vu}(x)\lambda^v{}_\theta l^*{}_{w\theta\bar \theta}.
\vphantom{\Big]}
\label{hetersgm10f} 
\end{align}
\label{hetersgm10} 
\end{subequations} 
\!\! where the spinor $\xi^\theta\in\Gamma(\Pi K_\Sigma{}^{-1/2})$
satisfies  the holomorphy condition
\begin{equation}
\barpartial_{\bar z}\xi^\theta=0.
\label{hetersgm11}
\end{equation} 

The sigma model action $S$ enjoys also $(1,0)$ supersymmetry,  
\vfill\eject\noindent
\begin{equation}
s_\xi S=0,
\label{hetersgm12}
\end{equation}
no further restriction on the target space geometry being required.
While the $(0,2)$ supersymmetry is a common property of all heterotic sigma models,
this $(1,0)$ supersymmetry is a special feature of our Lie algebroid sigma model.
The model has therefore a total $(1,2)$ supersymmetry. The $(0,2)$ and $(1,0)$ 
parts of this play however quite different roles in the topological twisted version 
of the model, as we shall see. 

As original noticed in ref. \cite{Witten2}, on a compact Riemann surface 
$\Sigma$, the space of holomorphic sections of the holomorphic line bundles
$K_\Sigma{}^{\,p/2}$ with $p\in\mathbb{Z}$ is only finite dimensional. 
For a world--sheet of this type, conditions \eqref{hetersgm6}, \eqref{hetersgm11} 
have only the trivial solutions
$\alpha^{\bar\theta}=\overline{\alpha}^{\bar\theta}=0$, $\xi^\theta=0$,
and, so, the $(0,2)$ and $(1,0)$ supersymmetries of the sigma 
model are empty (except in genus $1$). 
To have a full infinite dimensional space of solutions
of \eqref{hetersgm6}, \eqref{hetersgm11}, 
it is necessary that $\Sigma$ has punctures. Customarily, one assumes that 
$\Sigma$ is the flat punctured complex plane 
$\mathbb{C}\setminus\{0\}$. In the next section, we shall see
that the Lie algebroid heterotic sigma model, in a topologically twisted form, 
can be formulated on a world-sheet that is a general compact Riemann surface $\Sigma$. 

On a flat world--sheet, in which the supersymmetry parameters 
$\alpha^{\bar\theta}$, $\overline{\alpha}^{\bar\theta}$ and $\xi^\theta$
can be taken constant, the $(1,2)$ supersymmetry algebra
\begin{subequations}
\begin{align}
&[\delta_\alpha,\delta_\beta]=i(\overline{\alpha}^{\bar\theta}\beta^{\bar\theta}
-\overline{\beta}{}^{\bar\theta}\alpha^{\bar\theta})\barpartial_{\bar z},
\vphantom{\Big]}
\label{hetersgm4a} 
\\
&[s_\xi,s_\eta]=0,
\vphantom{\Big]}
\label{hetersgm4b} 
\\
&[\delta_\alpha,s_\xi]\simeq 0
\vphantom{\Big]}
\label{hetersgm4c} 
\end{align}
\label{hetersgm4}
\end{subequations}
\!\!is verified, 
provided the $(0,2)$ supersymmetry condition 
\eqref{hetersgm4/1} is satisfied. 
Here and in the following $\simeq$ denotes equality holding {\it on--shell}.
We remark that \eqref{hetersgm4/1} 
is required only by the validity of \eqref{hetersgm4a}.

The sigma model enjoys also an antiholomorphic $R$ and a holomorphic flavour symmetry, 
under which the sigma model fields
$x$, $\chi_{\bar \theta}$, $\overline{\chi}_{\bar \theta}$, $\lambda_\theta$, $\lambda^*{}_\theta$, 
$l_{\theta\bar \theta}$, $l^*{}_{\theta\bar \theta}$ have $R$/flavour charges 
$(0,0), (-1,0), (1,0), (0,1), (0,-1), (-1,1), (1,-1)$, respectively.
The overall normalization of the two charges is conventional. Different conventions 
are used in \cite{Tan:2006qt}\,--\cite{Tan:2007bh}, \cite{Adams:2005tc}. 

The supersymmetry field variations \eqref{hetersgm3},  \eqref{hetersgm10}
respect the $R$/flavour charges of the sigma model fields 
provided that the supersymmetry parameters $\alpha^{\bar\theta}, \overline{\alpha}^{\bar\theta}$,
$\xi^\theta$ are assigned $R$/flavour charges $(1,0)$, $(-1,0)$, $(0,-1)$ respectively.
It is natural to assume this and we shall do so henceforth. 

\subsection{\normalsize \textcolor{blue}{The Lie algebroid heterotic sigma model: quantum aspects}}
\label{subsec:hetersgm2}

\hspace{.5cm} The Lie algebroid heterotic sigma model is a particular heterotic sigma model
and, so, the usual consistency requirements of the quantum field theory
associated with the latter apply. 

In order the sigma model quantum functional integral to be meaningfully defined, 
the fermion determinants must combine to yield a function on the space of all embeddings
$x$. As is well--known, to this end, the associated determinant 
line bundle $\matheul{L}$ must be trivial. In the heterotic sigma model, this
is achieved if the complex vector bundles $T_M{}^+$, $\ker\rho_{E^+}$
satisfy the condition
\begin{equation}
\ch_2(T_M{}^+)=\ch_2(\ker\rho_{E^+}),
\label{hetersgm35/1}
\end{equation}
where $\ch_2(W)$ denotes the degree $4$ part of the Chern character
of a complex vector bundle $W$.  
Above, the isomorphism $\tilde E^+\simeq T_M{}^+$ has been taken into account.

The $R$ and flavour symmetry are generally anomalous. 
The anomalies manifest themselves as
vacuum background $R$ and flavour charges $\Delta q_R$, $\Delta q_L$.
$\Delta q_R$ is the difference of the numbers of the $\overline{\chi}_{\bar \theta}$,
and the $\chi_{\bar \theta}$ zero--modes. Similarly, $\Delta q_L$
is the difference of the numbers of the $\lambda_\theta$ and the $\lambda^*{}_\theta$
zero--modes. $\Delta q_R$, $\Delta q_L$ are so given by the index formulae
\begin{align}
&\Delta q_R=\int_\Sigma x^*c_1(T_M{}^+),
\vphantom{\Big]}
\label{hetersgm40} 
\\
&\Delta q_L=\int_\Sigma x^*c_1(\ker\rho_{E^+}).
\vphantom{\Big]}
\label{hetersgm41} 
\end{align}
Quantum sigma model correlators can be non zero only if the 
vacuum charges are soaked up by those of the inserted operators. 

Further topological restrictions will arise in the topologically twisted version 
of the sigma model studied in the next section. In that case, 
the fact that the target manifold $M$ and the 
gauge vector bundle $\ker \rho_E$ are encompassed in the Lie algebroid $E$
makes the consistency requirements somewhat more restrictive and, so, 
also more predictive.

\vfil\eject

\section{\normalsize \textcolor{blue}{The half--topological Lie algebroid sigma models}}\label{sec:topsgm}

\subsection{\normalsize \textcolor{blue}{Topological twisting: classical theory}}\label{subsec:topsgm1}

\hspace{.5cm} As is well--known, a supersymmetric field theory on a flat space--time does not remain
supersymmetric on a more general space--time with non trivial topology.
The reason for this is that 
supercharges are constant spinors
and constant spinors exist only on very few space--times. 
A systematic way of constructing supersymmetric field theories on general 
space--times is by topological twist of supersymmetric field theories on
flat ones. The twist is implemented by modifying the space--time covariance of the fields 
depending on their $R$ and flavour symmetry properties in such a way that one or more supercharges 
become scalar and thus can be constant on a general background. 
Twisting is not always possible ore may be possible in more than one way. 
See ref. \cite{Witten2} for a readable exposition of twisting in sigma models

We shall now consider the topological twist of the Lie algebroid heterotic sigma model 
constructed in sect. \ref{sec:hetersgm}. A sigma model field is either the embedding field 
$x\in \Map(\Sigma,M)$ or a section $\phi\in\Gamma(\zeta)$ of some vector bundle $\zeta$ over $\Sigma$.
A twist prescription leaves $x$ unchanged and replaces any other field 
$\phi\in\Gamma(\zeta)$ having $R$/flavour charges $(q_R,q_L)$ with a twisted
counterpart 
$\phi_{\mathrm{tw}}\in\Gamma(\overline{K}_\Sigma{}^{\eta_Rq_R/2}\otimes K_\Sigma{}^{\eta_Lq_L/2}\otimes\zeta)$,
where $\eta_R,~\eta_L$ are integers independent from $\phi$ 
whose values define the type of the twist.

The topological twist of the fields implies a simultaneous twist of
the spinor supersymmetry parameters 
$\alpha^{\bar\theta},\overline{\alpha}^{\bar\theta}\in \Gamma(\overline{K}_\Sigma{}^{-1/2})$,
$\xi^\theta\in\Gamma(K_\Sigma{}^{-1/2})$ of the
$(0,2)$, $(1,0)$ supersymmetry field variations \eqref{hetersgm3}, \eqref{hetersgm10}
so that to ensure the world--sheet covariance of the twisted form of these latter. 
In fact, inspection 
reveals that
the twist of $\alpha^{\bar\theta}, \overline{\alpha}^{\bar\theta}, \xi^\theta$ must be carried out
using the same prescription as that of the fields, where 
$\alpha^{\bar\theta}, \overline{\alpha}^{\bar\theta}, \xi^\theta$ have the 
$R$/flavour charges $(1,0)$, $(-1,0)$, $(0,-1)$ 
found in sect. \ref{sec:hetersgm}, respectively. 

\eject
\vspace{.08cm}

To ensure that one of the two $(0,2)$ supercharges becomes a world--sheet scalar upon twisting, 
as required, we must twist either $\overline{\alpha}^{\bar\theta}$
into $\overline{\alpha}\in\Gamma(\Pi 1_\Sigma)$ or 
$\alpha^{\bar\theta}$ into $\alpha\in\Gamma(\Pi 1_\Sigma)$.
This restricts the choice of $\eta_R$ to $\eta_R=\mp 1$,
as is immediately seen. 
Similarly, to ensure that the $(1,0)$ supercharge becomes a world--sheet scalar upon twisting, 
$\xi^\theta$  must be twisted into $\xi\in\Gamma(\Pi 1_\Sigma)$.
This fixes $\eta_L$ to $\eta_L=-1$. 
The topological supersymmetry
is the residual twisted supersymmetry 
associated with either $\overline{\alpha}$ or $\alpha$ and $\xi$. 

\vspace{.08cm}

We are thus left with the twists defined by $(\eta_R,\eta_L)=(\mp 1,-1)$.
They are called {\it type} $A$ and $B$ twists and they
lead to the type $A$ and $B$ {\it half--topological
Lie algebroid sigma models}, respectively, 
which are studied in this section. We work first at the classical level and postpone 
the analysis of their quantum consistency to the second part of this section. 

\subsection{\normalsize \textcolor{blue}{The type $A$ Lie algebroid sigma model}}\label{subsec:topsgm2}

\hspace{.5cm} The type $A$ Lie algebroid sigma model arises from the implementation of the type
$A$ twist in the heterotic Lie algebroid sigma model (cf. subsect. \ref{subsec:hetersgm1}). 

\vspace{.08cm}

The field content of the $A$ model is obtained from that of the heterotic model by 
turning each field of the latter into its twisted counterpart. 
Proceeding in this way, the embedding field $x\in\Map(\Sigma,M)$ is left unchanged, the fermion fields
$\chi_{\bar \theta}\in \Gamma(\overline{K}_\Sigma{}^{1/2}\otimes x^*\Pi \tilde E^+)$,
$\overline{\chi}_{\bar \theta}\in \Gamma(\overline{K}_\Sigma{}^{1/2} \otimes x^*\Pi \overline{\tilde E}{}^+)$
are replaced by fermion fields 
$\chi_{\bar z}\in\Gamma(\overline{K}_\Sigma\otimes x^*\Pi \tilde E^+)$, 
$\overline{\chi}\in\Gamma(x^*\Pi \overline{\tilde E}{}^+)$,
the fermion fields $\lambda_\theta\in \Gamma(K_\Sigma{}^{1/2}\otimes x^*\Pi\ker\rho_{E^+})$,
$\lambda^*{}_\theta\in \Gamma(K_\Sigma{}^{1/2}\otimes x^*\Pi\ker\rho_{E^+}{}^*)$
turn into fermion fields 
$\lambda\in \Gamma(x^*\Pi\ker\rho_{E^+})$,
$\lambda^*{}_z\in \Gamma(K_\Sigma\otimes x^*\Pi\ker\rho_{E^+}{}^*)$
and the boson fields $l_{\theta\bar \theta}\in 
\Gamma(K_\Sigma{}^{1/2}\otimes \overline{K}_\Sigma{}^{1/2} \otimes x^*\ker\rho_{E^+})$,
$l^*{}_{\theta\bar \theta}\in \Gamma(K_\Sigma{}^{1/2}\otimes \overline{K}_\Sigma{}^{1/2} 
\otimes x^*\ker\rho_{E^+}{}^*)$ yield 
boson fields $l_{\bar z}\in \Gamma(\overline{K}_\Sigma\otimes x^*\ker\rho_{E^+})$,
$l^*{}_z\in \Gamma(K_\Sigma\otimes x^*\ker\rho_{E^+}{}^*)$. 

\vspace{.08cm}

The action of the type $A$ sigma model is obtained from that of the heterotic model
(cf. eq. \eqref{hetersgm1}) by substituting each field of the latter with its twisted counterpart. 
Proceeding in this way, we obtain
\begin{align}
S_A&=\int_\Sigma d^2z\,\Big[
\frac{1}{2}g_{\bar ab}(x)(\barpartial_{\bar z}x^{\bar a}\partial_z x^b+\partial_z x^{\bar a}\barpartial_{\bar z}x^b)
+ig_{\bar ab}\rho^{\bar a}{}_{\bar p}\rho^b{}_q(x)\overline{\chi}^{\bar p}D_z\chi^q{}_{\bar z}
\vphantom{\Big]}
\label{atopsgm1} 
\\
&\hspace{1.9cm}+i\lambda^*{}_{uz}\bardee_{\bar z}\lambda^u
+f^u{}_{wv}G^w{}_{\bar ab}\rho^{\bar a}{}_{\bar p}\rho^b{}_q(x)\lambda^*{}_{uz}\lambda^v
\overline{\chi}^{\bar p}\chi^q{}_{\bar z}-l^*{}_{uz}l^u{}_{\bar z}\Big].
\vphantom{\Big]}
\nonumber
\end{align}

The $(0,2)$ supersymmetry field variations of the $A$ model are obtained from those of the heterotic model 
(cf. eq. \eqref{hetersgm3}) by the simultaneous twist of the sigma model fields and the supersymmetry 
parameters. Under type $A$ twist, the supersymmetry parameters 
$\alpha^{\bar\theta}, \overline{\alpha}^{\bar\theta}\in \Gamma(\overline{K}_\Sigma{}^{-1/2})$ turn into
$\alpha^{\bar z}\in\Gamma(\Pi \overline{K}_\Sigma{}^{-1})$, $\overline{\alpha}\in\Gamma(\Pi 1_\Sigma)$,
respectively. 
The primary type $A$ topological field variations are the truncated $(0,2)$ supersymmetry field 
variations yielded by setting $\alpha^{\bar z}=0$. Proceeding as indicated 
and formally dividing by $\overline{\alpha}$, these read 
\begin{subequations}
\begin{align}
&\delta_A x^a=0, \qquad \qquad \qquad\quad\,\,\,\,
\delta_A x^{\bar a}=i\rho^{\bar a}{}_{\bar p}(x)\overline{\chi}^{\bar p},
\vphantom{\Big]}
\label{atopsgm2a} 
\\
&\delta_A\chi^p{}_{\bar z}=-\sigma^p{}_a(x)\barpartial_{\bar z}x^a,
\qquad
\delta_A \overline{\chi}^{\bar p}=-\frac{i}{2}f^{\bar p}{}_{\bar q\bar r}(x)
\overline{\chi}^{\bar q}\overline{\chi}^{\bar r},
\vphantom{\Big]}
\label{atopsgm2b} 
\\
&\delta_A\lambda^u=0,
\vphantom{\Big]}
\label{atopsgm2c} 
\\
&\delta_A l^u{}_{\bar z}=-\bardee_{\bar z}\lambda^u
+if^u{}_{wv}G^w{}_{\bar ab}\rho^{\bar a}{}_{\bar p}\rho^b{}_q(x)\overline{\chi}^{\bar p}\chi^q{}_{\bar z}\lambda^v,
\vphantom{\Big]}
\label{atopsgm2d} 
\\
&\delta_A\lambda^*{}_{uz}=il^*{}_{uz},
\vphantom{\Big]}
\label{atopsgm2e} 
\\
&\delta_A l^*{}_{uz}=0.
\vphantom{\Big]}
\label{atopsgm2f} 
\end{align}
\label{atopsgm2}
\end{subequations} 
\vspace{-.7cm}

The $(0,2)$ supersymmetry of the heterotic model action, eq. \eqref{hetersgm5}, 
implies that the $A$ model action $S_A$ is invariant under the primary topological symmetry,
\begin{equation}
\delta_A S_A=0.
\label{atopsgm4}
\end{equation}

In analogous fashion, the $(1,0)$ supersymmetry field variations of the $A$ model 
are obtained from those of the heterotic model 
(cf. eq. \eqref{hetersgm10}) by the simultaneous twist of the sigma model fields and the supersymmetry 
parameters. Under type $A$ twist, the $(1,0)$ supersymmetry parameter $\xi^\theta
\in \Gamma(K_\Sigma{}^{-1/2})$ turns into 
$\xi\in\Gamma(\Pi 1_\Sigma)$. The secondary type $A$ topological field variations result in this way.
After formally dividing by $\xi$, they read 
\vspace{-.15truecm}
\begin{subequations}
\begin{align}
&s_A x^a=0, \qquad s_A x^{\bar a}=0,
\vphantom{\Big]}
\label{aintflv1a} 
\\
&s_A\chi^p{}_{\bar z}=0,
\qquad
s_A\overline{\chi}^{\bar p}=0,
\vphantom{\Big]}
\label{aintflv1b} 
\\
&s_A\lambda^u=-\frac{i}{2}f^u{}_{vw}(x)\lambda^v\lambda^w,
\vphantom{\Big]}
\label{aintflv1c} 
\\
&s_A l^u{}_{\bar z}=-if^u{}_{vw}(x)\lambda^vl^w{}_{\bar z},
\vphantom{\Big]}
\label{aintflv1d} 
\\
&s_A\lambda^*{}_{uz}=if^w{}_{vu}(x)\lambda^v\lambda^*{}_{wz},
\vphantom{\Big]}
\label{aintflv1e} 
\\
&s_A l^*{}_{uz}=if^w{}_{vu}(x)\lambda^vl^*{}_{wz}.\hspace{.6cm}
\vphantom{\Big]}
\label{aintflv1f} 
\end{align}
\label{aintflv1} 
\end{subequations} 
\vspace{-.7cm}

The $(1,0)$ supersymmetry of the heterotic model action, eq. \eqref{hetersgm12}, 
implies that the $A$ model action $S_A$ enjoys also the secondary type $A$ topological symmetry,
\begin{equation}
s_A S_A=0.
\label{atopsgm4/1}
\end{equation}

As the heterotic sigma model it comes from, the type $A$ sigma model enjoys 
an $R$ and a flavour symmetry, in which each twisted field has the same $R$ and flavour charge
as its untwisted parent. The $R$/flavour charges of the twisted fields
$x$, $\chi_{\bar z}$, $\overline{\chi}$, $\lambda$, $\lambda^*{}_z$, 
$l_{\bar z}$, $l^*{}_z$ are thus $(0,0), (-1,0), (1,0), (0,1), (0,-1), (-1,1), (1,-1)$,
respectively. 

\vspace{.2truecm}

\subsection{\normalsize \textcolor{blue}{The type $B$ Lie algebroid sigma model}}\label{subsec:topsgm3}

\hspace{.5cm} The type $B$ Lie algebroid sigma model arises from the implementation of the type
$B$ twist in the heterotic Lie algebroid sigma model. 

The field content of the $B$ model is obtained from that of the heterotic model by 
turning each field of the latter into its twisted counterpart. 
Proceeding in this way,  the embedding field $x\in\Map(\Sigma,M)$ is left unchanged, 
the fermion fields
$\chi_{\bar \theta}\in \Gamma(\overline{K}_\Sigma{}^{1/2}\otimes x^*\Pi \tilde E^+)$,
$\overline{\chi}_{\bar \theta}\in 
\Gamma(\overline{K}_\Sigma{}^{1/2} \otimes x^*\Pi \overline{\tilde E}{}^+)$
turn into fermion fields 
$\chi\in \Gamma(x^*\Pi \tilde E^+)$, $\overline{\chi}_{\bar z}\in 
\Gamma(\overline{K}_\Sigma\otimes x^*\Pi \overline{\tilde E}{}^+)$,
the fermion fields $\lambda_\theta\in \Gamma(K_\Sigma{}^{1/2}\otimes x^*\Pi\ker\rho_{E^+})$,
$\lambda^*{}_\theta\in \Gamma(K_\Sigma{}^{1/2}\otimes x^*\Pi\ker\rho_{E^+}{}^*)$
give rise to fermion fields $\lambda\in \Gamma(x^*\Pi\ker\rho_{E^+})$,
$\lambda^*{}_z\in \Gamma(K_\Sigma\otimes x^*\Pi\ker\rho_{E^+}{}^*)$
and the boson fields 
$l_{\theta\bar \theta}\in \Gamma(K_\Sigma{}^{1/2}\otimes \overline{K}_\Sigma{}^{1/2} \otimes x^*\ker\rho_{E^+})$,
$l^*{}_{\theta\bar \theta}\in \Gamma(K_\Sigma{}^{1/2}\otimes \overline{K}_\Sigma{}^{1/2} 
\otimes x^*\ker\rho_{E^+}{}^*)$ yield boson fields
$l\in \Gamma(x^*\ker\rho_{E^+})$,
$l^*{}_{z\bar z}\in \Gamma(K_\Sigma\otimes \overline{K}_\Sigma
\otimes x^*\ker\rho_{E^+}{}^*)$. 

As for the $A$ model, the action of the type $B$ sigma model is obtained from that of the heterotic model
(cf. eq. \eqref{hetersgm1}) by substituting each field of the latter with its twisted counterpart. 
The result is 
\begin{align}
S_B&=\int_\Sigma d^2z\,\Big[
\frac{1}{2}g_{\bar ab}(x)(\barpartial_{\bar z}x^{\bar a}\partial_z x^b+\partial_z x^{\bar a}\barpartial_{\bar z}x^b)
+ig_{\bar ab}\rho^{\bar a}{}_{\bar p}\rho^b{}_q(x)\overline{\chi}^{\bar p}{}_{\bar z}D_z\chi^q
\vphantom{\Big]}
\label{btopsgm1} 
\\
&\hspace{1.9cm}+i\lambda^*{}_{uz}\bardee_{\bar z}\lambda^u
+f^u{}_{wv}G^w{}_{\bar ab}\rho^{\bar a}{}_{\bar p}\rho^b{}_q(x)\lambda^*{}_{uz}\lambda^v
\overline{\chi}^{\bar p}{}_{\bar z}\chi^q
-l^*{}_{uz\bar z}l^u\Big].
\vphantom{\Big]}
\nonumber
\end{align}

Again as for the $A$ model, the $(0,2)$ supersymmetry field variations of the $B$ model are obtained from those 
of the heterotic model (cf. eq. \eqref{hetersgm3}) by the simultaneous twist of the sigma model fields and 
the supersymmetry parameters. Under $B$ twist, the $(0,2)$ supersymmetry parameters 
$\alpha^{\bar\theta}, \overline{\alpha}^{\bar\theta}\in \Gamma(\overline{K}_\Sigma{}^{-1/2})$ become 
$\alpha\in\Gamma(\Pi 1_\Sigma)$, $\overline{\alpha}^{\bar z}\in\Gamma(\Pi \overline{K}_\Sigma{}^{-1})$,
respectively. 
The primary type $B$ topological field variations are the truncated $(0,2)$ supersymmetry field 
variations yielded by setting $\overline{\alpha}^{\bar z}=0$. Proceeding as said 
and formally eliminating $\alpha$, these read 
\begin{subequations}
\begin{align}
&\delta_B x^a=i\rho^a{}_p(x)\chi^p, \qquad \qquad \,\,\,\,\delta_B x^{\bar a}=0, 
\vphantom{\Big]}
\label{btopsgm2a} 
\\
&\delta_B\chi^p=-\frac{i}{2}f^p{}_{qr}(x)\chi^q\chi^r,
\qquad
\delta_B \overline{\chi}^{\bar p}{}_{\bar z}=-\sigma^{\bar p}{}_{\bar a}(x)\barpartial_{\bar z}x^{\bar a},
\vphantom{\Big]}
\label{btopsgm2b} 
\\
&\delta_B\lambda^u=-if^u{}_{iv}\sigma^i{}_a\rho^a{}_p(x)
\chi^p\lambda^v+il^u,
\vphantom{\Big]}
\label{btopsgm2c} 
\\
&\delta_B l^u=-if^u{}_{iv}\sigma^i{}_a\rho^a{}_p(x)\chi^p l^v,
\vphantom{\Big]}
\label{btopsgm2d} 
\\
&\delta_B\lambda^*{}_{uz}=if^v{}_{iu}\sigma^i{}_a\rho^a{}_p(x)
\chi^p\lambda^*{}_{vz},
\vphantom{\Big]}
\label{btopsgm2e} 
\\
&\delta_B l^*{}_{uz\bar z}=if^v{}_{iu}\sigma^i{}_a\rho^a{}_p(x)\chi^p l^*{}_{vz\bar z}
\vphantom{\Big]}
\label{btopsgm2f} 
\\
&\hspace{2cm}-\bardee_{\bar z}\lambda^*{}_{uz}
-if^v{}_{wu}G^w{}_{\bar ab}\rho^{\bar a}{}_{\bar p}\rho^b{}_q(x)\overline{\chi}^{\bar p}{}_{\bar z}\chi^q
\lambda^*{}_{vz}.
\vphantom{\Big]}
\nonumber
\end{align}
\label{btopsgm2}
\end{subequations} 

The $(0,2)$ supersymmetry of the heterotic model action, eq. \eqref{hetersgm5}, 
implies that the $B$ model action $S_B$ is invariant under the primary topological symmetry,
\begin{equation}
\delta_B S_B=0.
\label{btopsgm4}
\end{equation}

In analogous fashion, just as in the $A$ model, the $(1,0)$ supersymmetry field variations of the 
$B$ model are obtained from those of the heterotic model 
(cf. eq. \eqref{hetersgm10}) by the simultaneous twist of the sigma model fields and the supersymmetry 
parameters. Under type $B$ twist, the $(1,0)$ supersymmetry parameter $\xi^\theta
\in \Gamma(K_\Sigma{}^{-1/2})$ turns again into 
$\xi\in\Gamma(\Pi 1_\Sigma)$. The secondary type $B$ topological field variations result in this way.
After formally dividing by $\xi$, they read 
\begin{subequations}
\begin{align}
&s_B x^a=0, \qquad s_B x^{\bar a}=0,
\vphantom{\Big]}
\label{bintflv1a} 
\\
&s_B\chi^p=0,
\qquad
s_B\overline{\chi}^{\bar p}{}_{\bar z}=0,
\vphantom{\Big]}
\label{bintflv1b} 
\\
&s_B\lambda^u=-\frac{i}{2}f^u{}_{vw}(x)\lambda^v\lambda^w, \hspace{.6cm}
\vphantom{\Big]}
\label{bintflv1c} 
\\
&s_B l^u=-if^u{}_{vw}(x)\lambda^vl^w,
\vphantom{\Big]}
\label{bintflv1d} 
\\
&s_B\lambda^*{}_{uz}=if^w{}_{vu}(x)\lambda^v\lambda^*{}_{wz},
\vphantom{\Big]}
\label{bintflv1e} 
\\
&s_B l^*{}_{uz\bar z}=if^w{}_{vu}(x)\lambda^vl^*{}_{wz\bar z}.
\vphantom{\Big]}
\label{bintflv1f} 
\end{align}
\label{bintflv1} 
\end{subequations}  
\vspace{-.7cm}

The $(1,0)$ supersymmetry of the heterotic model action, eq. \eqref{hetersgm12}, 
implies that the $B$ model action $S_B$ enjoys also the secondary type $B$ topological symmetry,
\begin{equation}
s_B S_B=0.
\label{btopsgm4/1}
\end{equation}

As the heterotic sigma model, the type $B$ sigma model enjoys 
an $R$ and a flavour symmetry, in which each twisted field has the same $R$ and flavour charge
as its untwisted parent. 
However, in order to comply later with established cohomological 
conventions, it is convenient to redefine the sign of the $R$ charge. 
The $R$/flavour charges of the twisted fields 
$x$, $\chi$, $\overline{\chi}_{\bar z}$, $\lambda$, $\lambda^*{}_z$, 
$l$, $l^*{}_{z\bar z}$ \hfil are therefore
\eject\noindent
$(0,0), (1,0), (-1,0), (0,1), (0,-1), (1,1), (-1,-1)$,
respectively.

\subsection{\normalsize \textcolor{blue}{Half--topological nature of the twisted sigma models}}
\label{subsec:topsgm4}

\hspace{.5cm} The twisted sigma models illustrated above have a number of features, which make them 
candidate topological field theories. 
In the unified treatment presented below, the label $t$ stands for the types $A,B$.

The algebra of the twisted sigma model fields has a double grading,
where the bidegree of each elementary field is given by its $R$/flavour charges.

The primary and the secondary topological field variation
operators $\delta_t$, $s_t$ are odd, as they flip the statistics of the fields, 
and in fact of bidegree $(1,0)$ and $(0,1)$, respectively. 
The $(1,2)$ supersymmetry algebra \eqref{hetersgm4} entails that 
\begin{subequations}
\begin{align}
&\delta_t{}^2=0, 
\vphantom{\Big]}
\label{topsgm3a} 
\\
&s_t{}^2=0,
\vphantom{\Big]}
\label{topsgm3b} 
\\
&\delta_ts_t+s_t\delta_t\simeq 0,
\vphantom{\Big]}
\label{topsgm3c} 
\end{align}
\label{topsgm3}
\end{subequations}
\!\!provided the $(0,2)$ supersymmetry condition 
\eqref{hetersgm4/1} is satisfied. (Recall that
$\simeq$ denotes equality on--shell.)  
\eqref{hetersgm4/1} is required only by the validity of \eqref{topsgm3a}.

The nilpotency of $\delta_t$ and $s_t$ shows that they are BRST operators. 
In spite of this and the fact that they both originate by
topological twist of the heterotic sigma model, they play quite different roles in the twisted 
sigma models. As we shall now show, $\delta_t$ is a true topological field variation operator,
while $s_t$ is akin to a gauge theoretic Slavnov 
operator. Indeed, $s_t$ represents a non trivial symmetry of the half--to\-po\-logical
sigma models, which, being intimately related to the adjoint bundle $\ker\rho_E$ of $E$, 
we shall call {\it adjoint symmetry}. 

The existence of a nilpotent BRST operator and the associated BRST cohomology 
is a universal feature of all topological sigma models, but it is not by itself sufficient to render 
a sigma model genuinely topological. 
Other requirements must be satisfied: $a$) the sigma model action should be BRST--exact up to 
a topological term and $b$) the stress tensor should be BRST--exact.
Let us check whether our twisted sigma models fulfil these 
taking $\delta_t$ as BRST operator. 

The first requirement is fulfilled. Indeed, $S_t$ has the following structure 
\begin{equation}
S_t=\delta_t\Psi_t+\eta_tI_{\mathrm{top}}, 
\label{btopsgm5}
\end{equation}
where $\Psi_t$ is the gauge fermion
\begin{align}
\Psi_A&=\int_\Sigma d^2z\Big(-g_{\bar ab}\rho^b{}_q(x)\chi^q{}_{\bar z}
\partial_z x^{\bar a}+i\lambda^*{}_{uz}l^u{}_{\bar z}\Big),
\vphantom{\Big]}
\label{btopsgm6a}
\\
\Psi_B&=\int_\Sigma d^2z\Big(-g_{\bar ab}\rho^{\bar a}{}_{\bar p}(x)\overline{\chi}^{\bar p}{}_{\bar z}
\partial_z x^b+il^*{}_{uz\bar z}\lambda^u\Big),
\vphantom{\Big]}
\label{btopsgm6b}
\end{align}
$\eta_A=-1,\eta_B=1$ and $I_{\mathrm{top}}$ is the topological action,
\begin{equation}
I_{\mathrm{top}}=\int_\Sigma d^2z\,
\frac{1}{2}g_{\bar ab}(x)(\partial_z x^{\bar a}\barpartial_{\bar z}x^b-\barpartial_{\bar z}x^{\bar a}\partial_z x^b)
=\int_\Sigma x^*\omega,
\label{btopsgm7}
\end{equation}
$\omega=\frac{i}{2}g_{\bar ab}dz^{\bar a}\wedge dz^b$ being the Kaehler form of $M$. 

The second requirement, conversely, is not fulfilled. The components 
of the stress tensor are in fact found to be
\vspace{-.1cm}
\begin{subequations}
\begin{align}
T_{tzz}&=-g_{\bar ab}(x)\partial_z x^{\bar a}\partial_z x^b-i\lambda^*{}_{uz}D_z\lambda^u,
\vphantom{\Big]}
\label{btopsgm8a} 
\\
T_{t\bar z\bar z}&=-g_{\bar ab}(x)\barpartial_{\bar z}x^{\bar a}\barpartial_{\bar z}x^b
+T'{}_{t\bar z\bar z},
\vphantom{\Big]}
\label{btopsgm8b} 
\\
T_{tz\bar z}&=0,
\vphantom{\Big]}
\label{btopsgm8c} 
\end{align}
\end{subequations}
\!\!where $T'{}_{t\bar z\bar z}$ is given by the expressions
\vspace{-.1cm}
\begin{align}
T'{}_{A\bar z\bar z}&=
ig_{\bar ab}\rho^{\bar a}{}_{\bar p}\rho^b{}_q(x)\bardee_{\bar z}\overline{\chi}^{\bar p}\chi^q{}_{\bar z},
\vphantom{\Big]}
\label{btopsgm8/1a} 
\\
T'{}_{B\bar z\bar z}&=
-ig_{\bar ab}\rho^{\bar a}{}_{\bar p}\rho^b{}_q(x)\overline{\chi}^{\bar p}{}_{\bar z}\bardee_{\bar z}\chi^q
\vphantom{\Big]}
\label{btopsgm8/1b} 
\end{align}
in the two twisted models.
The stress tensor component $T_{zz}$ is $\delta_t$--closed 
\begin{equation}
\delta_t T_{zz}
\simeq 0,
\label{btopsgm9}
\end{equation}
but it is not $\delta_t$--exact.
The stress tensor component $T_{\bar z\bar z}$, instead, is $\delta_t$--exact,
\begin{equation}
T_{t\bar z\bar z}=\delta_tG_{t\bar z\bar z},
\label{btopsgm10}
\end{equation}
where $G_{t\bar z\bar z}$ is given by the expressions
\vspace{-.1cm}
\begin{align}
G_{A\bar z\bar z}&=g_{\bar ab}\rho^b{}_q(x)\chi^q{}_{\bar z}\barpartial_{\bar z} x^{\bar a},
\vphantom{\Big]}
\label{btopsgm10/1a}
\\
G_{B\bar z\bar z}&=g_{\bar ab}\rho^{\bar a}{}_{\bar p}(x)\overline{\chi}^{\bar p}{}_{\bar z}
\barpartial_{\bar z}x^b.
\vphantom{\Big]}
\label{btopsgm10/1b}
\end{align}

We find in this way that our twisted sigma models are not topological field theories.
Since only one of the two non trivial components of the stress tensor is BRST--exact, 
they are {\it half--topological} sigma models \cite{Adams:2005tc}.

What is the role of the BRST operator $s_t$ in all this? Any attempt 
to formulate the twisted sigma models as topological field theories
taking $s_t$ as BRST operator necessarily fails. Since $s_t$ acts trivially
on the embedding and the $\tilde E^+$, $\overline{\tilde E}{}^+$ valued fields, 
there is no way of generating the corresponding terms of the action by letting $s_t$ 
act on a suitable gauge fermion.  Nevertheless, $\delta_t$ and $s_t$ combined 
yield a distinct BRST operator with properties completely analogous to
those of $\delta_t$ above, as we now show. 

For any $u\in\mathbb{C}$, we can construct a BRST operator
\begin{equation}
\delta_t(u)=\delta_t+us_t.
\label{stextr4}
\end{equation}
This is indeed nilpotent on--shell on account of \eqref{topsgm3}, 
\begin{equation}
\delta_t(u)^2\simeq 0.
\label{stextr5}
\end{equation}
One may be tempted to extend the domain of definition of
$\delta_t(u)$ to $u\in\mathbb{CP}$ by setting $\delta'{}_t(u')=u'\delta_t+s_t$.
However, as just noticed, the primary and secondary BRST operators 
play an asymmetrical role in the sigma models, the first one only being the 
counterpart of the usual BRST operator 
of other twisted heterotic sigma models.
So, it seems unlikely that $\delta'{}_t(u')$ would have any significance.

With the BRST operator $\delta_t(u)$, there is associated a BRST cohomology
depending a priori on $u$. Let us see how. Let $f_t$ be the flavour charge. 
As $\delta_t$ and $s_t$ have flavour charges $0$ and $1$, respectively, we have 
$[f_t,\delta_t]=0$ and $[f_t,s_t]=s_t$. So, for $\zeta\in\mathbb{C}$, we have 
\begin{equation}
\exp(\zeta f_t)\delta_t(u)\exp(-\zeta f_t)=\delta_t(u\exp(\zeta)).
\label{stextr6}
\end{equation}
It follows that the $\delta_t(u)$--cohomology is independent from 
$u\in\mathbb{C}\setminus \{0\}$ up to isomorphism. 
Thus, we really have to consider only the BRST
charges $\delta_t(0)$ and, say, $\delta_t(1)$.
Note that $\delta_t(0)=\delta_t$ is the primary BRST charge while $\delta_t(1)=\delta_t+s_t$
is the total BRST charge.

Using \eqref{btopsgm6a}, \eqref{btopsgm6b} and \eqref{aintflv1}, \eqref{bintflv1}, we find that 
\begin{equation}
s_t\Psi_t\simeq 0.
\label{stextr7}
\end{equation}
Further, we have \hphantom{xxxxxxxxxxxxxx}
\begin{equation}
s_tT_{tzz}=0
\label{stextr8}
\end{equation}
and  \hphantom{xxxxxxxxxxxxxxxxxxxxxxxxxxxxxxxxxxxxxxx}
\begin{equation}
s_tG_{t\bar z\bar z}=0,
\label{stextr9}
\end{equation}
as is straightforwardly verified using once more \eqref{aintflv1}, 
\eqref{bintflv1}. By these relations, it is possible to extend the construction carried 
out in the first half of this subsection, replacing $\delta_t$ by $\delta_t+s_t$
throughout. In this way, \eqref{btopsgm5} becomes
\begin{equation}
S_t\simeq(\delta_t+s_t)\Psi_t+\eta_tI_{\mathrm{top}}.
\label{stextr10}
\end{equation}
Likewise, \eqref{btopsgm9} and \eqref{btopsgm10} take the form
\begin{equation}
(\delta_t+s_t)T_{zz}\simeq 0
\label{stextr11}
\end{equation}
and \hphantom{xxxxxxxxxxxxxxxxxxxxxxxxxxxxxxxxxxx} 
\begin{equation}
T_{t\bar z\bar z}=(\delta_t+s_t)G_{t\bar z\bar z}.
\label{stextr12}
\end{equation}
So, $T_{zz}$ and $T_{t\bar z\bar z}$ are also  $\delta_t+s_t$--closed
and $\delta_t+s_t$--exact, respectively. 

The half--topological sigma models have therefore two distinct BRST cohomological structures 
with analogous formal properties, one associated with $\delta_t$, the other associated with 
$\delta_t+s_t$. They will emerge over and over again at various points below
and will constitute the main theme of our analysis.
This feature of the sigma models raises a number of questions.  
Do both these BRST structures have a counterpart in the quantum theory?
If so, which properties do they have and how do they relate to each other?
In the rest of this section, we shall try to answer these queries. 

\subsection{\normalsize \textcolor{blue}{Dependence on target space geometry}}\label{subsec:topsgm5}

\hspace{.5cm}
The target space geometry of the half--topological sigma models
is characterized by several fields: the Lie bracket $[\cdot,\cdot]_E$,
the anchor $\rho_E$, the Lie algebroid 
complex structure $(J_E,J_{EM})$, the complex splitting $\sigma$ and the Kaehler
metric $g$. The cohomological nature of the half--topological sigma models
suggests that, when some of these fields are varied, the action varies by a  
BRST--exact term. 

Consider first a variation $h$ of the metric $g$, the other target space fields remaining 
fixed.
Since the variation must preserve the Hermiticity and the 
Kaehlerness of the metric $g$, the $(2,0)$, $(0,2)$ components of $h$ vanish and
and the $(1,1)$ component satisfies the equation
\begin{equation}
\partial_{\bar a}h_{\bar b c}-\partial_{\bar b}h_{\bar a c}=0 \qquad \text{and c. c..}
\label{topvar1}
\end{equation}
Inspecting the expression of the primary 
field variation
operator $\delta_t$ (cf. eqs. \eqref{atopsgm2}, \eqref{btopsgm2}), 
we observe that it does not depend on the metric. From \eqref{btopsgm5}--\eqref{btopsgm7},
we find then that the resulting variation of the action $S_t$ is given by
\begin{equation}
\Delta_h S_t=\delta_t\Delta_h \Psi_t+\eta_t\Delta_h I_{\mathrm{top}},
\label{topvar2}
\end{equation}
where $\Delta_h \Psi_t$ is the variation of the gauge fermion $\Psi_t$,
\eject\noindent
\begin{align}
\Delta_h \Psi_A&=\int_\Sigma d^2z\Big(-h_{\bar ab}\rho^b{}_q(x)\chi^q{}_{\bar z}\partial_z x^{\bar a}\Big),
\vphantom{\Big]}
\label{topvar3a}
\\
\Delta_h \Psi_B&=\int_\Sigma d^2z\Big(-h_{\bar ab}\rho^{\bar a}{}_{\bar p}(x)
\overline{\chi}^{\bar p}{}_{\bar z}\partial_z x^b\Big)
\vphantom{\Big]}
\label{topvar3b}
\end{align}
and $\Delta_h I_{\mathrm{top}}$ is the variation of the topological term $I_{\mathrm{top}}$,
\begin{equation}
\Delta_h I_{\mathrm{top}}=\int_\Sigma d^2z\,
\frac{1}{2}h_{\bar ab}(x)(\partial_z x^{\bar a}\barpartial_{\bar z}x^b-\barpartial_{\bar z}x^{\bar a}\partial_z x^b)
=\int_\Sigma x^*\Delta_h \omega,
\label{topvar4}
\end{equation}
with $\Delta_h \omega=\frac{i}{2}h_{\bar ab}dz^{\bar a}\wedge dz^b$.
Thus, $\Delta_h S_t$ is $\delta_t$--exact only if $\Delta_h I_{\mathrm{top}}=0$.
This happens when either the $(1,1)$ form $\Delta_h \omega$ is exact
or the embedding field $x$ is homotopically trivial.

Consider next a variation $\tau$ of the splitting $\sigma$, the other target space fields remaining 
fixed. Since the variation must preserve the basic relations \eqref{LAAex2}, \eqref{stconn2}, 
only the components $\tau^u{}_a$ of $\tau$ 
are non vanishing. Further, the $(0,2)$ supersymmetry condition \eqref{hetersgm4/1} requires that
\begin{equation}
\partial_a\tau^u{}_b-\partial_b\tau^u{}_a+f^u{}_{vi}\tau^v{}_a\sigma^i{}_b
+f^u{}_{iv}\sigma^i{}_a\tau^v{}_b=0,
\label{topvar9}
\end{equation}
as follows easily from \eqref{lex19}. 
Inspecting the expression of the primary 
field variation
operator $\delta_t$ (cf. eqs. \eqref{atopsgm2}, \eqref{btopsgm2}), 
we observe that it does depend on the splitting $\sigma$.
Therefore, we cannot expect the resulting variation of the action $S_t$ to be given by an expression
analogous to \eqref{topvar2}. However, using directly the explicit expression of $S_t$ (cf. eqs. 
\eqref{atopsgm1}, \eqref{btopsgm1}) and recalling that of the secondary 
field variation operator $s_t$ (cf. eqs. \eqref{aintflv1}, \eqref{bintflv1} ), we find that
\begin{equation}
\Delta_\tau S_t=s_t\Delta_\tau \Upsilon_t,
\label{topvar5}
\end{equation}
where $\Delta_\tau \Upsilon_t$ is given by
\begin{align}
\Delta_\tau\Upsilon_A&=\int_\Sigma d^2z\lambda^*{}_{uz}\Big(\tau^u{}_a(x)\barpartial_{\bar z}x^a
-i\partial_{\bar a}\tau^u{}_b\rho^{\bar a}{}_{\bar p}\rho^b{}_q(x)\overline{\chi}^{\bar p}\chi^q{}_{\bar z}\Big),
\vphantom{\Big]}
\label{topvar6a}
\\
\Delta_\tau\Upsilon_B&=\int_\Sigma d^2z\lambda^*{}_{uz}\Big(\tau^u{}_a(x)\barpartial_{\bar z}x^a
-i\partial_{\bar a}\tau^u{}_b\rho^{\bar a}{}_{\bar p}\rho^b{}_q(x)\overline{\chi}^{\bar p}{}_{\bar z}\chi^q\Big).
\vphantom{\Big]}
\label{topvar6b}
\end{align}
We discover in this way that, though $\Delta_\tau S_t$ is not $\delta_t$--exact, it is $s_t$--exact. 

All this may look a little bit puzzling. To shed light on this matter, we begin with noticing that,
for a variation $h$ of the metric, one has
\begin{equation}
s_t\Delta_h \Psi_t=0. 
\label{topvar7}
\end{equation}
Therefore, \eqref{topvar2} could be cast as
\begin{equation}
\Delta_h S_t=(\delta_t+s_t)\Delta_h \Psi_t+\eta_t\Delta_h I_{\mathrm{top}}.
\label{topvar8}
\end{equation}
Next, for a variation $\tau$ of the splitting, one has 
\begin{equation}
\delta_t\Delta_\tau\Upsilon_t\simeq 0,
\label{topvar10}
\end{equation}
in virtue of \eqref{topvar9}. Thus, \eqref{topvar2} could be cast as
\begin{equation}
\Delta_\tau S_t\simeq(\delta_t+s_t)\Delta_\tau \Upsilon_t.
\label{topvar11}
\end{equation}
We reach therefore the following conclusions.

Under a variation of the Kaehler metric $g$, the variation of 
the action $S_t$ minus the topological term $I_{\mathrm{top}}$ is both $\delta_t$-- and 
$\delta_t+s_t$--exact. Conversely, under a variation of the splitting
$\sigma$ compatible with the $(0,2)$ supersymmetry condition \eqref{hetersgm4/1}, 
the variation of the action $S_t$ is only $\delta_t+s_t$--exact. 

As far as we can see, the variation of the action $S_t$ resulting from the variation of the 
Lie bracket $[\cdot,\cdot]_E$, the anchor $\rho_E$ and the Lie algebroid complex structure $(J_E,J_{EM})$ 
is neither $\delta_t$-- nor $\delta_t+s_t$--exact. 
$S_t$ depends non trivially on those target space fields.
We shall come back to this point in sect. \ref{sec:outlook}. 

\subsection{\normalsize \textcolor{blue}{Perturbative renormalization and conformal invariance}}
\label{subsec:topsgm6}

\hspace{.5cm} In sigma model perturbation theory, at $1$--loop, the classical action $S_t$ 
of the half--topological sigma models undergoes a renormalization of the form
\begin{equation}
\Delta S_t=\delta_t\Delta\Psi_t,
\label{conftop1}
\end{equation}
where $\Delta\Psi_t$ is the renormalization suffered by the gauge fermion $\Psi_t$,
\vspace{-.02cm}
\begin{align}
\Delta\Psi_A&=\int_\Sigma d^2z\Big(-\kappa_{A1}R_{\bar ab}\rho^b{}_q(x)\chi^q{}_{\bar z}
\partial_z x^{\bar a}+i\kappa_{A2}f^u{}_{wv}g^{\bar ab}G^w{}_{\bar ab}(x)
\lambda^*{}_{uz}l^v{}_{\bar z}\Big),
\vphantom{\Big]}
\label{conftop2}
\\
\Delta\Psi_B&=\int_\Sigma d^2z\Big(-\kappa_{B1}R_{\bar ab}\rho^{\bar a}{}_{\bar p}(x)\overline{\chi}^{\bar p}{}_{\bar z}
\partial_z x^b+i\kappa_{B2}f^u{}_{wv}g^{\bar ab}G^w{}_{\bar ab}(x)
l^*{}_{uz\bar z}\lambda^v\Big),
\vphantom{\Big]}
\label{conftop3}
\end{align} 
$R_{\bar ab}$ being the Ricci tensor of the metric $g_{\bar ab}$. The
$\kappa_{ti}$ are certain cut--off dependent constants. Therefore, 
$\Delta S_t$ is $\delta_t$--exact.  

The renormalization $\Delta\Psi_t$ of the gauge fermion is $s_t$--closed,
\begin{equation}
s_t\Delta \Psi_t\simeq 0,
\label{qadjbrst7}
\end{equation}
as is straightforward  to verify using \eqref{aintflv1}, \eqref{bintflv1}. 
Therefore, the  
renormalization $\Delta S_t$ 
could be also expressed as 
\begin{equation}
\Delta S_t\simeq(\delta_t+s_t)\Delta \Psi_t.
\label{qadjbrst10}
\end{equation}
Consequently, $\Delta S_t$ is $\delta_t+s_t$--exact as well. 

Because of \eqref{btopsgm8c},  the half--topological sigma models are conformally invariant at 
the classical level. The cut--off dependence of the $\kappa_{ti}$ indicates however that 
conformal invariance is generally broken at the quantum level. 
Conformal invariance is preserved quantum mechanically only if $R_{\bar ab}=0$, 
i. e. when $g_{\bar ab}$ is Calabi--Yau, and the splitting $\sigma^i{}_a$
satisfies the appropriate version of the Uhlenbeck--Yau equation, namely 
\begin{equation}
f^u{}_{wv}g^{\bar ab}G^w{}_{\bar ab}=0
\label{conftop4}
\end{equation}
(cf. eq. \eqref{lex19}), as usual. 
Even if these restrictions on the target space geometry do not hold, the renormalization 
of the classical action remains $\delta_t$-- as well as $\delta_t+s_t$--exact . 
We thus expect conformal invariance to continue to hold at the level of BRST cohomology
in the appropriate sense to be defined.

The above conclusion is based on perturbation theory.
Its validity is therefore restricted to the perturbative
regime where the effects of world--sheet instantons are neglected. 
In fact, conformal invariance can be broken by world--sheet instantons 
non perturbatively \cite{Witten:2005px}.
The following analysis assumes conformal invariance and thus 
it holds only perturbatively.

\subsection{\normalsize \textcolor{blue}{BRST charges}}\label{subsec:topsgm7}

\hspace{.5cm} Having clarified the role of the two BRST cohomological structures of the 
half--topological sigma models at the classical and perturbative quantum level, let us 
now consider their quantum operator realization.

In the quantum theory, the field variation operator $\delta_t$ 
becomes a BRST charge $Q_t$ 
\footnote{$\vphantom{\bigg[}$ In the string theory literature, $Q_A$, $Q_B$
are generally denoted by $\overline{Q}_+$, $Q_+$, respectively.} 
characterized by 
the property that, for any local operator $\mathcal{O}$
\begin{equation}
[Q_t,\mathcal{O}]=\delta_t\mathcal{O}, 
\label{qopsbrstc1}
\end{equation}
in the quasiclassical limit, 
$[\cdot,\cdot]$ denoting here and below a graded commutator.
Similarly, the field variation operator $s_t$ 
becomes a BRST charge $\Lambda_t$ such that 
\begin{equation}
[\Lambda_t,\mathcal{O}]=s_t\mathcal{O}
\label{qopsbrstc2}
\end{equation}
again quasiclassically. 

As $\delta_t$ is nilpotent (cf. eq.\eqref{topsgm3a}), so should be the associated charge $Q_t$
\begin{equation}
Q_t{}^2=0.
\label{qopsbrstc3}
\end{equation}
Likewise, the nilpotence of $s_t$ (cf. eq.\eqref{topsgm3b}) leads to the nilpotence of $\Lambda_t$
\begin{equation}
\Lambda_t{}^2=0.
\label{qopsbrstc4}
\end{equation}
As $\delta_t$ and $s_t$ anticommute (cf. eq.\eqref{topsgm3c}), so should their associated
charges,
\begin{equation}
[Q_t,\Lambda_t]=0. \vphantom{\Bigg]}
\label{qopsbrstc5}
\end{equation}

By \eqref{qopsbrstc1}, \eqref{qopsbrstc2},  the BRST charges corresponding to the two 
BRST field variation operators $\delta_t$, $\delta_t+s_t$ of the sigma models at the quantum level
are $Q_t$, $Q_t+\Lambda_t$. By \eqref{qopsbrstc3}--\eqref{qopsbrstc5}, 
they are both nilpotent as required. 

\subsection{\normalsize \textcolor{blue}{Operator BRST cohomology and chiral algebra}}\label{subsec:topsgm8}

\hspace{.5cm} At the quantum level, the half--topological sigma models are characterized by 
their operator BRST cohomology. Since, in the present case, 
we have two BRST structures whose treatment is totally analogous, it is convenient to 
employ a uniform notation. We denote the two topological field variation
operators $\delta_t$, $\delta_t+s_t$ as $\delta^{(1)}{}_t$, $\delta^{(2)}{}_t$
and the associated BRST charges by 
$Q_t$, $Q_t+\Lambda_t$ as $Q^{(1)}{}_t$, $Q^{(2)}{}_t$, respectively.

The basic property of the charge $Q^{(\nu)}{}_t$ is its nilpotence, 
\begin{equation}
Q^{(\nu)}{}_t{}^2=0. 
\label{qopsbrst2}
\end{equation}
Because of it, for any local operator $\mathcal{O}$, we have 
\begin{equation}
[Q^{(\nu)}{}_t, [Q^{(\nu)}{}_t,\mathcal{O}]]=0.
\label{qopsbrst2/1}
\end{equation}
Consequently, with $Q^{(\nu)}{}_t$, there is associated an operator BRST cohomology. 
A local operator $\mathcal{O}$ is $Q^{(\nu)}{}_t$--closed if
\begin{equation}
[Q^{(\nu)}{}_t,\mathcal{O}]=0.
\label{qopsbrst3}
\end{equation}
A $Q^{(\nu)}{}_t$--closed local operator $\mathcal{O}$ is $Q^{(\nu)}{}_t$--exact if 
\begin{equation}
\mathcal{O}=[Q^{(\nu)}{}_t,\mathcal{X}].
\label{qopsbrst4}
\end{equation}
for some local operator $\mathcal{X}$. The {\it operator BRST cohomology}
is just the operator $Q^{(\nu)}{}_t$--cohomology, that is the quotient of the 
algebra of $Q^{(\nu)}{}_t$--closed local operators by the ideal of 
$Q^{(\nu)}{}_t$--exact ones. 

The {\it topological operators} $\mathcal{O}$ are precisely the $Q^{(\nu)}{}_t$--closed ones.
The quantum correlators of topological operators depend only on their 
$Q^{(\nu)}{}_t$--co\-homology classes, as $\langle [Q^{(\nu)}{}_t,\mathcal{X}]\rangle=0$ 
for any operator $\mathcal{X}$. Thus, inside correlators, we have  
\begin{equation}
\mathcal{O}\approx \mathcal{O}+[Q^{(\nu)}{}_t,\mathcal{X}].
\label{qopsbrst4x}
\end{equation}

We have seen above that the classical half--topological sigma models are 
conformally invariant;  $T_{z\bar z}$ vanishes (cf. eq. \eqref{btopsgm8c}). Further, 
$T_{zz}$ is $\delta^{(\nu)}{}_t$--closed (cf. eqs. \eqref{btopsgm9}, \eqref{stextr11}) 
and $T_{\bar z\bar z}$ is $\delta^{(\nu)}{}_t$--exact
(cf. eqs. \eqref{btopsgm10}, \eqref{stextr12}). 
These properties do not automatically extend to the quantum sigma models.

The $\delta^{(\nu)}{}_t$--exactness of the renormalization of the classical 
action (cf. eqs. \eqref{conftop1}, \eqref{qadjbrst10}) indicates that conformal invariance 
holds in the quantum theory at the level of $Q^{(\nu)}{}_t$--cohomology. 
This statement needs 
to be qualified
\cite{Witten:2005px,Kapustin:2005pt,Adams:2005tc}. 

In a quantum conformal field theory with central charge $c$, it is more natural to employ, in place of 
the covariant stress tensor $T$ used so far, the conformal stress tensor $T^c$. The components 
of the two are related as
\begin{subequations}
\begin{align}
&T^c{}_{zz}=T_{zz}-\frac{c}{12\pi}R_{zz},
\vphantom{\Big]}
\label{qopsbrst5}
\\ 
&T^c{}_{\bar z\bar z}=T_{\bar z\bar z}-\frac{c}{12\pi}R_{\bar z\bar z},
\vphantom{\Big]}
\label{qopsbrs6}
\vphantom{\Big]}
\\
&T^c{}_{z\bar z}=T_{z\bar z}-\frac{c}{12\pi}R_{z\bar z},
\label{qopsbrs7}
\end{align}
\end{subequations}
where $R_{zz}$, $R_{\bar z\bar z}$ and $R_{z\bar z}$ are the components of the projective connection 
and the Ricci tensor of the world--sheet metric $h_{z\bar z}$
\footnote{$\vphantom{\bigg[}$ We recall that 
$R_{zz}=\partial_z\Gamma^z{}_{zz}-\frac{1}{2}(\Gamma^z{}_{zz})^2$,
$R_{\bar z\bar z}=\barpartial_{\bar z}\Gamma^{\bar z}{}_{\bar z\bar z}
-\frac{1}{2}(\Gamma^{\bar z}{}_{\bar z\bar z})^2$ and
$R_{z\bar z}=-\barpartial_{\bar z}\Gamma^z{}_{zz}=
-\partial_z\Gamma^{\bar z}{}_{\bar z\bar z}$
where $\Gamma^z{}_{zz}=\partial_z\ln h_{z\bar z}$, 
$\Gamma^{\bar z}{}_{\bar z\bar z}=\barpartial_{\bar z}\ln h_{z\bar z}$
are the components of the Levi--Civita connection. $R_{z\bar z}$ can be expressed as 
$R_{z\bar z}=\frac{1}{2}Rh_{z\bar z}$, where $R$ is the Ricci scalar.}.
In what follows, we assume that $h_{z\bar z}$
has constant Ricci scalar 
\footnote{$\vphantom{\bigg[}$ 
According to the uniformization theorem,
this is always possible. Upon doing so, we have 
$\barpartial_{\bar z}R_{zz}=0$, $\partial_zR_{\bar z\bar z}=0$.}.
Conformal invariance implies that $T^c{}_{z\bar z}=0$ 
and that 
\vfil\eject\noindent
$\barpartial_{\bar z}T^c{}_{zz}=0$, 
$\partial_zT^c{}_{\bar z\bar z}=0$. 

In the half--topological sigma models, which are conformally invariant 
only at the level of $Q^{(\nu)}{}_t$--cohomology, weaker conditions hold.
The sigma models are still characterized by their central charge $c_t$, 
as any ordinary conformal field theory. 
However, $T^c{}_{tz\bar z}$ vanishes only up to a $Q^{(\nu)}{}_t$--exact term.  So, we have 
\begin{equation}
T^c{}_{tz\bar z}=[Q^{(\nu)}{}_t,G^{(\nu)}{}_{tz\bar z}], \vphantom{\bigg[} 
\label{qopsbrst8}
\end{equation}
for some operator $G^{(\nu)}{}_{tz\bar z}$. Similarly, $T^c{}_{tzz}$ is holomorphic and 
$T^c{}_{t\bar z\bar z}$ is antiholomorphic only up to a $Q^{(\nu)}{}_t$--exact term.
We find indeed the relations
\begin{equation}
\barpartial_{\bar z}T^c{}_{tzz}=-[Q^{(\nu)}{}_t,\nabla_zG^{(\nu)}{}_{tz\bar z}], 
\qquad \partial_zT^c{}_{t\bar z\bar z}=-[Q^{(\nu)}{}_t,\barnabla_{\bar z}G^{(\nu)}{}_{tz\bar z}], 
\vphantom{\Bigg]}
\label{qopsbrst9}
\end{equation}
which follow from \eqref{qopsbrst8} and the covariant stress energy conservation equations, 
$\barnabla_{\bar z}T_{tzz}+\nabla_zT_{tz\bar z}=0$,
$\nabla_zT_{t\bar z\bar z}+\barnabla_{\bar z}T_{tz\bar z}=0$.

In the classical theory, the component $T_{t\bar z\bar z}$ of the stress tensor is 
$\delta^{(\nu)}{}_t$--exact (cf. eqs. \eqref{btopsgm10}, \eqref{stextr12}). In the quantum theory,  
an analogous property holds. 
The component $T^c{}_{t\bar z\bar z}$ of the stress tensor is  
$Q^{(\nu)}{}_t$--exact up to a central term. More precisely, we have
\begin{equation}
T^c{}_{t\bar z\bar z}+\frac{c_t}{12\pi}R_{\bar z\bar z}=[Q^{(\nu)}{}_t,G^{(\nu)}{}_{t\bar z\bar z}], 
\label{qopsbrst10}
\end{equation}
for some operator $G^{(\nu)}{}_{t\bar z\bar z}$ 
\footnote{$\vphantom{\bigg[}$  
The central term $\frac{c_t}{12\pi}R_{\bar z\bar z}$ in the left hand side 
ensures that the latter transforms as a quadratic differential 
under a change of world--sheet coordinates as the left hand side does.}.
The validity of \eqref{qopsbrst10} can be argued as follows.
We first note that the left hand side of \eqref{qopsbrst10} is just the 
component $T_{t\bar z\bar z}$ of the covariant stress tensor. 
We then recall that an infinitesimal deformation of the complex structure of the world--sheet 
$\Sigma$, corresponding to the infinitesimal Beltrami differentials
$(\mu^z{}_{\bar z},\overline{\mu}{}^{\bar z}{}_z)$, results in an insertion of the 
form $\int_\Sigma d^2z(\mu^z{}_{\bar z}T_{tzz}+\overline{\mu}{}^{\bar z}{}_zT_{t\bar z\bar z})$ inside correlators.
Now, in the quantum theory, the action undergoes a $1$--loop renormalization 
$\Delta S_t$ of the form \eqref{conftop1}, \eqref{qadjbrst10}, 
which keeps it $\delta^{(\nu)}{}_t$--exact up to a topological term.
Under the deformation, the Beltrami differential component 
$\overline{\mu}{}^{\bar z}{}_z$ enters only in the variation of $\Delta\Psi_t$
but not in that of $\delta^{(\nu)}{}_t$, as is easy  to realize inspecting the expressions of 
$\delta^{(\nu)}{}_t$ (cf. eqs. \eqref{atopsgm2}, \eqref{btopsgm2}, \eqref{aintflv1}, \eqref{bintflv1})  
and of $\Delta\Psi_t$ (cf. eqs. \eqref{conftop2}, \eqref{conftop3}). $\overline{\mu}{}^{\bar z}{}_z$ 
therefore couples to a $\delta^{(\nu)}{}_t$--exact 
object indicating that 
\eqref{qopsbrst10} must hold quantum mechanically. 

In the classical theory, the component $T_{tzz}$ of the stress tensor is 
$\delta^{(\nu)}{}_t$--closed (cf. eqs. \eqref{btopsgm9}, \eqref{stextr11}). 
However, we cannot expect this property to extend at the quantum level
as for $T_{t\bar z\bar z}$. 
$T^c{}_{tzz}$ is not $Q^{(\nu)}{}_t$--closed 
in general.
This can be seen by an argument similar to the one expounded in the previous paragraph. 
Under a deformation of the world--sheet complex structure, the Beltrami differential 
component $\mu^z{}_{\bar z}$ enters in the variation of $\delta^{(\nu)}{}_t$ 
in the $1$--loop renormalization $\Delta S_t$ 
of the action and, so, it does not necessarily couple to a $\delta^{(\nu)}{}_t$--closed object.
If, however, this does nevertheless happens, 
then $T^c{}_{tzz}$ will be $Q^{(\nu)}{}_t$--closed. This is the case in particular when
$\Delta S_t$ vanishes, i. e. when $g_{\bar ab}$ is Calabi--Yau and the splitting 
$\sigma$ satisfies the Uhlenbeck--Yau equation \eqref{conftop4}
and the half--topological sigma models are strictly conformal quantum mechanically.
Weaker conditions on the target space geometry may however suffice. 

Suppose now that the world--sheet $\Sigma$ is the standard flat punctured complex plane 
$\mathbb{C}\setminus\{0\}$. Since, by \eqref{qopsbrst9}, $T^c{}_{tzz}$,  
$T^c{}_{t\bar z\bar z}$ are (anti)holomorphic up to a $Q^{(\nu)}{}_t$--exact
term, we can expand $T^c{}_{t\bar z\bar z}$, $T^c{}_{t\bar z\bar z}$ 
in Laurent series in the usual way, $T^c{}_{tzz}=\sum_n L_{tn} z^{-n-2}$, 
$T^c{}_{t\bar z\bar z}=\sum_n \overline{L}_{tn} \bar z^{-n-2}$.
The Laurent modes $L_{tn}$, $\overline{L}_{tn}$ however are defined 
only up to a $Q^{(\nu)}{}_t$--exact ambiguity.

As $T^c{}_{t\bar z\bar z}$ is $Q^{(\nu)}{}_t$--exact up to a central term, by
\eqref{qopsbrst10}, the modes $\overline{L}_{tn}$ also are. 
Their adjoint action thus descends on the 
operator $Q^{(\nu)}{}_t$--cohomology and is in fact trivial \footnote{$\vphantom{\bigg[}$ 
If $T$ is a $Q^{(\nu)}{}_t$--closed operator, then the adjoint action of $T$ is defined 
on the operator 
$\vphantom{\bigg[}Q^{(\nu)}{}_t$--cohomology. Indeed, as $[Q^{(\nu)}{}_t,T]=0$,
$[Q^{(\nu)}{}_t,[T,\mathcal{O}]]=0$ for any operator $\mathcal{O}$ such that 
$[Q^{(\nu)}{}_t,\mathcal{O}]=0$ and $[T,\mathcal{O}]=\pm [Q^{(\nu)}{}_t,[T,\mathcal{X}]]$ 
whenever $\mathcal{O}=[Q^{(\nu)}{}_t,\mathcal{X}]$. 
Further, if $T$ is a $Q^{(\nu)}{}_t$--exact operator,
then the adjoint action of $T$ on the operator $Q^{(\nu)}{}_t$--cohomology is trivial.
Indeed, $T=[Q^{(\nu)}{}_t,V]$ for some operator $V$ and, so,
$[T, \mathcal{O}]=[Q^{(\nu)}{}_t,[V,\mathcal{O}]]$ for any operator $\mathcal{O}$ such that 
$[Q^{(\nu)}{}_t,\mathcal{O}]=0$.}.
Conversely, as $T^c{}_{tzz}$ is not $Q^{(\nu)}{}_t$--closed in general, 
the modes $L_{tn}$ also are not. 
Consequently, their adjoint action is not defined on the operator $Q^{(\nu)}{}_t$--cohomology. 
However, translation invariance guarantees that
$\partial_z$ commutes with $Q^{(\nu)}{}_t$. As $L_{t-1}=\partial_z$, we have $[Q^{(\nu)}{}_t,L_{t-1}]=0$. 
Moreover, as $Q^{(\nu)}{}_t$ is a scalar, it commutes with the spin operator
$S_t$. Since $S_t=L_{t0}-\overline{L}_{t0}$ and $\overline{L}_{t0}$ is $Q^{(\nu)}{}_t$--exact,
we have $[Q^{(\nu)}{}_t,L_{t0}]=0$. Hence, the adjoint action of $L_{t-1}$ and  
$L_{t0}$ still descends on the operator $Q^{(\nu)}{}_t$--cohomology \cite{Adams:2005tc}. 
When $T^c{}_{tzz}$ happens to be $Q^{(\nu)}{}_t$--closed, all the modes $L_{tn}$ also are 
and their adjoint action is thus defined on the operator $Q^{(\nu)}{}_t$--cohomology.

As observed in \cite{Witten:2005px}, these properties have important
consequences. If $\mathcal{O}$ is any $Q^{(\nu)}{}_t$--closed operator,
then $[\overline{L}_{t0},\mathcal{O}]$ and 
$\barpartial_{\bar z}\mathcal{O}=[\overline{L}_{t-1},\mathcal{O}]$
are $Q^{(\nu)}{}_t$--exact. Therefore, the operator $Q^{(\nu)}{}_t$--cohomology consists 
of antiholomorphic scaling dimension $0$ holomorphic operator classes. 
So, if  $\mathcal{O}_\alpha$ is a basis of $Q^{(\nu)}{}_t$--closed operators
modulo $Q^{(\nu)}{}_t$--exact ones representing the $Q^{(\nu)}{}_t$--cohomology, 
the coefficients $c_{\alpha\beta}{}^\gamma(z)$ entering in
the operator product expansion
\begin{equation}
\mathcal{O}_\alpha(z)\mathcal{O}_\beta(0)
\approx\sss_\gamma c_{\alpha\beta}{}^\gamma(z)\mathcal{O}_\gamma(0)
\label{qopsbrst12x}
\end{equation}
are in fact all meromorphic. 
The holomorphic operator $Q^{(\nu)}{}_t$--cohomology, equip\-ped with the 
operator product expansion structure \eqref{qopsbrst12x}, constitutes 
the {\it chiral algebra} $\mathcal{A}^{(\nu)}{}_t$ of the half--topological sigma model.

Since  $L_{t-1}$ and  $L_{t0}$ act on the operator $Q^{(\nu)}{}_t$--cohomology, 
$\mathcal{A}^{(\nu)}{}_t$ carries a natural action
of $\partial_z$ and is further graded according to 
holomorphic scaling dimension. 

The spectrum of holomorphic scaling dimensions $h$ of 
$\mathcal{A}^{(\nu)}{}_t$ consists of non negative integers. This 
can be shown as follows.
At the classical level, operator 
cohomology classes of $\mathcal{A}^{(\nu)}{}_t$
with non vanishing $h$ can be constructed using exclusively
the field $\lambda^*{}_{uz}$ and the $\partial_z$ derivatives of the
fields $x^a$, $x^{\bar a}$ and $\lambda^*{}_{uz}$. (The $\partial_z$ 
derivatives  of the fields $\chi^p$, $\overline{\chi}^{\bar p}$ as
well as the fields $l^u$, $l^*{}_u$ can be eliminated using the field 
equations.) The resulting classes have non negative integer $h$. 
Their spin is therefore $s=h$, since $s=h-\overline{h}$, where 
$\overline{h}$  is the antiholomorphic scaling dimension of the 
classes, which always vanishes. At the quantum level, there are
consequently no perturbative quantum corrections of 
the values of $h$, since the spin $s=h$ must remain integer.

For dimensional reasons, the coefficients $c_{\alpha\beta}{}^\gamma(z)$
entering in the operator product expansion \eqref{qopsbrst12x} have the form
\begin{equation}
c_{\alpha\beta}{}^\gamma(z)
=\frac{f_{\alpha\beta}{}^\gamma}{z^{h_\alpha+h_\beta-h_\gamma}}, 
\label{qopsbrst12y}
\end{equation}
where $f_{\alpha\beta}{}^\gamma$ is a constants and $h_\iota$ is the
scaling dimension of $\mathcal{O}_\iota$.

Since $h_\iota\geq 0$, the operator classes $\mathcal{O}_\alpha$
with $h_\alpha=0$ have a non singular operator product expansion. 
When inserted in the same point they therefore 
form a ring $\mathcal{R}^{(\nu)}{}_t$, the {\it chiral ring} of the half-topological model sigma model,
the heterotic counterpart of the chiral ring of $(2,2)$ theories.
Correlators of chiral ring classes are
holomorphic and hence constant when the world--sheet $\Sigma$ is compact.

$\mathcal{A}^{(\nu)}{}_t$ is graded also by $R$ and flavour charge.
In the operator product expansion \eqref{qopsbrst12x}, 
the $R$ and flavour charges of the two sides must match. 
Then, in \eqref{qopsbrst12y}, $f_{\alpha\beta}{}^\gamma=0$ unless
$q_{R\alpha}+q_{R\beta}-q_{R\gamma}=q_{L\alpha}+q_{L\beta}-q_{L\gamma}=0$, 
where $q_{R\iota}$, $q_{L\iota}$  are the $R$/flavour charges of $\mathcal{O}_\iota$. 

\subsection{\normalsize \textcolor{blue}{Topological twist and anomalies}}\label{subsec:topsgm9}

\hspace{.5cm} At the quantum level, the half--topological sigma models worked out above must satisfy
the same consistency requirement as the parent heterotic sigma model, namely that the fermion 
determinants combine to yield a function on the space of all embeddings
$x$ or, equivalently, that the associated determinant line bundle $\matheul{L}$ 
be trivial (cf. subsect. \ref{subsec:hetersgm2}). In the half--topological sigma models, this
is achieved if the complex vector bundles $T_M{}^+$, $\ker\rho_{E^+}$
satisfy the condition \eqref{hetersgm35/1} relating their Chern characters and a further 
condition, viz 
\begin{equation}
\frac{1}{2}c_1(T_\Sigma{}^+)\big((2\bar s-1)c_1(T_M{}^+)-(2s-1)c_1(\ker\rho_{E^+})\big)=0
\label{qtopsgm1}
\end{equation}
\footnote{$\vphantom{\bigg[}$ Here, $T_\Sigma{}^+$, $T_M{}^+$, $\ker\rho_{E^+}$
are shorthands for the box product vector bundles over $\Sigma\times M$ $T_\Sigma{}^+\boxtimes 1_M$,
$1_\Sigma\boxtimes T_M{}^+$, $1_\Sigma\boxtimes \ker\rho_{E^+}$, respectively.}, 
where $\bar s$ and $1-s$ are the spins of the fermion fields $\chi$ and $\lambda$
(that is, from the 
world--sheet point of view, $\chi$ and $\lambda$ are sections of  
$\overline{K}_\Sigma{}^{\bar s}$ and $K_\Sigma{}^{1-s}$, respectively). 
We note that \eqref{qtopsgm1} is automatically verified 
in the untwisted case, where $s,\bar s=1/2$, but not so in general in the twisted ones,
where $s,\bar s\not=1/2$. Now, $s=1$, $\bar s=1$, for the $A$ twist, 
and $s=1$, $\bar s=0$, for the $B$ twist.
Then, \eqref{qtopsgm1} becomes
\begin{equation}
-\frac{1}{2}c_1(T_\Sigma{}^+)\big(\eta_tc_1(T_M{}^+)+c_1(\ker\rho_{E^+})\big)=0,
\label{qtopsgm2}
\end{equation}
where $\eta_A=-1,\eta_B=1$, as before. 

In order \eqref{qtopsgm2} to be satisfied, it suffices that 
either $c_1(T_\Sigma{}^+)=0$  
or $\eta_tc_1(T_M{}^+)+c_1(\ker\rho_{E^+})=0$. Assume that the latter condition obtains. 
Then, on account of \eqref{hetersgm35/1}, 
the 1st and 2nd Chern classes of the vector bundles $T_M{}^+$, $\ker\rho_{E^+}$ satisfy
\begin{align}
&-\eta_tc_1(T_M{}^+)=c_1(\ker\rho_{E^+}),
\vphantom{\Big]}
\label{hetersgm35} 
\\
&c_2(T_M{}^+)=c_2(\ker\rho_{E^+}).
\vphantom{\Big]}
\label{hetersgm36} 
\end{align}
Let us see how these topological restrictions affect the Lie algebroid $E^+$.

Since the Lie algebroid $E^+$ is transitive, it fits into the Atiyah sequence
\begin{equation}
\xymatrix{0\ar[r]&\ker\rho_{E^+}\ar[r]^{\,\,\,\,\,\iota_{E^+}}&E^+\ar[r]^{\rho_{E^+}\,\,\,\,\,}&T_M{}^+\ar[r]&0}
\label{hetersgm37}
\end{equation}
(cf. eq. \eqref{LAAex1}), which implies the relation
$c(E^+)=c(\ker\rho_{E^+})c(T_M{}^+)$ among total Chern classes.
Using \eqref{hetersgm35}, \eqref{hetersgm36}, we find 
\eject\noindent
\begin{align}
&c_1(E^+)=(1-\eta_t)c_1(T_M{}^+),
\vphantom{\Big]}
\label{hetersgm38} 
\\
&c_2(E^+)=2c_2(T_M{}^+)-\eta_tc_1(T_M{}^+)^2. 
\vphantom{\Big]}
\label{hetersgm39} 
\end{align}
By \eqref{hetersgm38}, the cohomology class $c_1(E^+)$ is even for the $A$ model and vanishes
for the $B$ model. If $E=T_M$, then necessarily $c_i(E^+)=0$, 
$i=1,2$. In that case, further, $M$ is Calabi--Yau.

At the de Rham cohomology level, condition \eqref{hetersgm35} is required by the consistency of the 
twist of the heterotic sigma model. As we recalled at the beginning 
of this section, 
twisting amounts to altering the covariance of the sigma model fields depending on their $R$ 
and flavour charges. In the sigma model, twisting  is implemented by coupling the world--sheet 
spin connection to the appropriate linear combination $J_{\mathrm{tw}}$ of the 
$R$ and a flavour currents $J_R$, $J_L$. With $s$ and $\bar s$ defined as earlier, we have 
\begin{equation}
J_{\mathrm{tw}}=(2\bar s-1)J_R+(1-2s)J_L.
\label{qtopsgm3/1}
\end{equation}
In our case, $J_{\mathrm{tw}}$ turns out to be
\begin{equation}
J_t=-\frac{1}{2}(\eta_tJ_R+J_L).
\label{qtopsgm3}
\end{equation}
To ensure that the quantum twisted fields have the same covariance as their 
classical counterparts, the current $J_t$ must be non-anomalous, i. e.
the corresponding  vacuum background charge must vanish,
\begin{equation}
\Delta q_t=-\frac{1}{2}(\eta_t\Delta q_R+\Delta q_L)=0.
\label{qtopsgm4}
\end{equation}
Using \eqref{hetersgm40}, \eqref{hetersgm41}, 
\eqref{qtopsgm4} can be cast compactly as
\begin{equation}
-\frac{1}{2}\int_\Sigma \big(\eta_tx^*c_1(T_M{}^+)+x^*c_1(\ker\rho_{E^+})\big)=0.
\label{qtopsgm5}
\end{equation}
\eqref{hetersgm35} is a sufficient condition for \eqref{qtopsgm5} to be satisfied.

While we require that the twisted general covariance is non-anomalous, the 
\eject\noindent
global 
$R$ and flavour symmetries still could, and generally will, suffer an anomaly after twisting.
The vacuum background $R$ and flavour charges, with $s$ and $\bar s$ defined as above, 
are given by 
\begin{align}
&\Delta q_{R\bar s}=
d(1-2\bar s)(\ell-1)+\int_\Sigma x^*c_1(T_M{}^+),
\vphantom{\Big]}
\label{qtopsgm6}
\\ 
&\Delta q_{Ls}=
r(1-2s)(\ell-1)+\int_\Sigma x^*c_1(\ker\rho_{E^+}),
\vphantom{\Big]}
\label{qtopsgm7} 
\end{align}
where $\ell$ is the genus of $\Sigma$ and $d=\rank T_M{}^+$, $r=\rank\ker\rho_{E^+}$.  
In the untwisted case, where $s,\bar s=1/2$, 
we recover \eqref{hetersgm40}, \eqref{hetersgm41}. 
In the twisted cases, taking into account that we have conventionally
redefined the sign of the $R$ charge in the $B$ model, we find the expressions 
\vspace{-.1cm}
\begin{align}
&\Delta q_{Rt}=-
d(\ell-1)-\eta_t\Delta q_R,
\vphantom{\Big]}
\label{qtopsgm41}
\\
&\Delta q_{Lt}=-
r(\ell-1)+\Delta q_L,
\vphantom{\Big]}
\label{qtopsgm40} 
\end{align}
where $\Delta q_R$, $\Delta q_L$ given by \eqref{hetersgm40}, \eqref{hetersgm41}. 
Again, quantum sigma model correlators can be non zero only if these 
vacuum charges are soaked up by those of the inserted operators.

\subsection{\normalsize \textcolor{blue}{Algebro--geometric conditions}}\label{subsec:topsgm10}

\hspace{.5cm} There are certain algebro--geometric conditions which must be added
to the topological anomaly cancellation conditions \eqref{hetersgm35/1} and \eqref{qtopsgm2}
of the half--topolog\-ical sigma models.
They ensure the existence of a natural trace on the relevant chiral ring 
which, in turn, is required for the consistent computation of 
correlators of chiral ring operator classes. We refer the reader to ref. \cite{Sharpe:2005fd} 
(see also ref. \cite{Adams:2005tc}) 
for a detailed discussion of this point. Here, we shall limit ourselves to outline the argument.

Let us consider for definiteness the string tree level $\ell=0$. In the perturbative 
regime, assumed in our analysis, world--sheet instantons do not contribute and 
the sigma model path integral localizes at constant embeddings $x\in\Map(\Sigma,M)$. 
The moduli space of constant embeddings is just the target space $M$. 
Thus, the quantum correlator $\langle\mathcal{O}\rangle$ of any chiral ring element 
$\mathcal{O}$ should be expressible as an integral over $M$ of a suitable 
top form $\omega_{\mathcal{O}}$ depending on $\mathcal{O}$.
\begin{equation}
\langle\mathcal{O}\rangle=\int_M\omega_{\mathcal{O}}. 
\label{hetersgm50}
\end{equation}
To ensure the convergence of the integral, we require $M$ to be compact.
Clearly, $\mathcal{O}$ determines only the de Rham cohomology class 
$[\omega_{\mathcal{O}}]_{\mathrm{dR}}$ of $\omega_{\mathcal{O}}$. 

From \eqref{qtopsgm41}, \eqref{qtopsgm40}, in the situation considered, 
the vacuum background $R$ and flavor charges are $\Delta q_{Rt}=d$ and $\Delta q_{Lt}=r$.
Therefore, the correlator $\langle\mathcal{O}\rangle$ can be non vanishing only 
if the $R$ and flavour charges of $\mathcal{O}$ have the values $d$ and $r$, respectively.
Correspondingly, the class 
$[\omega_{\mathcal{O}}]_{\mathrm{dR}}$ is trivial unless 
this condition is met.

The algebro--geometric conditions mentioned at beginning of this subsection
are sufficient conditions for the existence of the appropriate correspondence
$\mathcal{O}\mapsto [\omega_{\mathcal{O}}]_{\mathrm{dR}}$. They 
take the form of isomorphism relations between certain (anti)holomorphic line bundles
(denoted below by $\cong$). 

For the type $A$ sigma model, there is just one condition, viz
\begin{equation}
\wedge^{r}\ker\rho_{\mathcal{E}}{}^*\cong\wedge^{d}\mathcal{T}_{\mathcal{M}}{}^*.
\label{hetersgm36/1}
\end{equation}

For the type $B$ sigma model, the statement of the conditions is slightly more involved.
To begin with, we recall that the target space geometry involves   
a splitting $\sigma$ of $E$ satisfying \eqref{stconn2} and whose curvature 
$G_{E\sigma}{}^c$ has vanishing $(2,0)$ component (cf. eq. \eqref{hetersgm4/1}). 
Let $\mathcal{D}^{\mathrm{ad}}_{E\sigma}$ be the ordinary $(1,0)$ connection of $\ker\rho_{E^+}$
obtained from the connection $D^{\mathrm{ad}}_{E\sigma}$ by complexification and restriction 
to $\Gamma(\ker\rho_{E^+})$ (cf. subsect. \ref{subsec:liealg3}). 
By relation \eqref{LAAconn7}, being $G_{E\sigma}{}^{2,0}=0$, 
$\mathcal{D}^{\mathrm{ad}}_{E\sigma}$ has vanishing $(2,0)$ curvature.
$\mathcal{D}^{\mathrm{ad}}_{E\sigma}$ therefore defines an antiholomorphic structure on the complex 
vector bundle $\ker\rho_{E^+}{}$ making it an antiholomorphic vector bundle 
$\mathcal{K}_{E^+\sigma}$ over $\mathcal{M}$. 
The first condition then reads 
\begin{equation}
\wedge^r\mathcal{K}_{E^+\sigma}{}^*\cong 1_{\mathcal{M}}. \vphantom{\Bigg]}
\label{hetersgm36/2a}
\end{equation}
The second condition requires that $\mathcal{M}$ is a Calabi--Yau manifold,
\begin{equation}
\wedge^{d}\mathcal{T}_{\mathcal{M}}{}^*\cong 1_{\mathcal{M}}. \vphantom{\Bigg]}
\label{hetersgm36/2}
\end{equation}

For the type $A$ sigma model, \eqref{hetersgm36/1} implies that  
\begin{equation}
c_1(T_M{}^+)=c_1(\ker\rho_{E^+}). \vphantom{\Bigg]}
\label{hetersgm36/3}
\end{equation}
\eqref{hetersgm36/3} is just \eqref{hetersgm35}. 
For the type $B$ sigma model, \eqref{hetersgm36/2a}, \eqref{hetersgm36/2} together lead to 
\begin{equation}
c_1(T_M{}^+)=c_1(\ker\rho_{E^+})=0. \vphantom{\Bigg]}
\label{hetersgm36/4}
\end{equation}
\eqref{hetersgm36/4} is  compatible with but more restrictive than 
\eqref{hetersgm35}.

For the second BRST structure of the half--twisted sigma models, there is a further condition:
the typical fiber $\mathfrak{g}^+$ of the Lie algebra bundle $\ker \rho_{E^+}$ 
must be {\it unimodular}, that is $\tr\ad X=0$ for any $X\in \mathfrak{g}^+$.
Explicitly, this can be phrased as a simple condition on the structure functions
$f^u{}_{vw}$, namely
\begin{equation}
f^v{}_{uv}=0.
\label{hetersgm36/5}
\end{equation}
For the $B$ model, the unimodularity condition has another important consequence:
unlike $\mathcal{K}_{E^+\sigma}$, 
the antiholomorphic line bundle $\wedge^r\mathcal{K}_{E^+\sigma}{}^*$ is actually independent
from the splitting $\sigma$
\footnote{$\vphantom{\bigg[}$
The $(1,0)$ connection induced by $\mathcal{D}^{\mathrm{ad}}_{E\sigma}$
on $\wedge^{r}\ker_{E^+}$ is given locally by $\partial_a-f^v{}_{iv}\sigma^i{}_a$
Under a variation $\tau$ of $\sigma$ at fixed $\rho_E$
(cf. subsect. \ref{subsec:topsgm5}), the variation of the induced connection is
$-f^v{}_{uv}\tau^u{}_a$, which vanishes if \eqref{hetersgm36/5} holds. It follows that
the antiholomorphic structure of $\mathcal{K}_{E^+\sigma}$ is indeed independent
from $\sigma$.}. Hence, so is also condition \eqref{hetersgm36/2a}. 

The justification of the restrictions listed above will be provided 
in sect. \ref{sec:cohla} below, once the structure of the 
sigma model chiral rings will have been unveiled.

\vfill\eject

\section{\normalsize \textcolor{blue}{Lie algebroid cohomology and chiral ring}}\label{sec:cohla}

\subsection{\normalsize \textcolor{blue}{Lie algebroid cohomology}}
\label{subsec:cohla2}


\hspace{.5cm} In this section, we shall unveil the relation between the cohomology 
of certain Lie algebroids with complex base belonging to the target space geometry of the
heterotic Lie algebroid sigma model and the BRST cohomology defining the chiral rings 
of the 
half--twisted sigma models in the perturbative quasiclassical regime. 
This will provide an elegant geometric interpretation of those rings. 

A Lie algebroid $E$ over $M$ is characterized by its cohomology $H_{LA}{}^*(E)$. 
This is defined as follows. 

Let $E[1]$ be the parity shifted form of $E$. In each trivializing neighborhood, $E[1]$ 
is coordinatized by  base coordinates $x^A$ of degree $0$ and fiber coordinates $\xi^I$ 
of degree $1$. The algebra of functions of $E[1]$, $\Fun(E[1])$, is then graded according 
to the polynomial degree in the $\xi^I$.

There exists a canonical degree $1$ vector field $Q_E$ on $E[1]$ 
given locally by 
\begin{equation}
Q_E=\rho^A{}_I\xi^I\partial_{xA}-\frac{1}{2}f^I{}_{JK}\xi^J\xi^K\partial_{\xi I},
\label{cohla1}
\end{equation}
where $\rho^A{}_I$, $f^I{}_{JK}$ are the anchor and bracket structure functions of $E$
with respect to a local frame $\{e_I\}$
\footnote{$\vphantom{\bigg]}$ The structure functions are defined as usual
by the relations $\rho_E e_I=\rho^A{}_I\partial_A$ and $[e_I,e_J]_E=f^k{}_{IJ}e_K$ 
and obey relations formally identical to \eqref{lex9}--\eqref{lex11}.}. 
From \eqref{LAA1}--\eqref{LAA3}, it follows that $Q_E$ is homological, 
\begin{equation}
Q_E{}^2=0.
\label{cohla2}
\end{equation}

The Lie algebroid cohomology $H_{LA}{}^*(E)$ is then the cohomology of the complex 
$(\Fun(E[1]),Q_E)$. When $E=T_M$, it reduces to the customary de Rham cohomology
$H_{d_M}{}^*(M)$ of $M$, as is immediate to see. So, it can be considered a natural 
generalization of the latter.

The Lie algebroid cohomology $H_{LA}{}^*(W)$ of a complex Lie algebroid $W$ 
is de-
\eject\noindent
fined in the same fashion.  If $E$ is a Lie algebroid, then 
$H_{LA}{}^*(E^c)$ is the complexification of $H_{LA}{}^*(E)$.

\subsection{\normalsize \textcolor{blue}{Complex and holomorphic Lie algebroid cohomology}}\label{subsec:cohla3}

\hspace{.5cm} Let $W$ be a Lie algebroid with complex base $M$ (cf. subsect. \ref{subsec:liecmplx2}). 
$W$ is characterized by a Lie algebroid cohomology $H_{LA}{}^*(W)$
defined in an analogous fashion. One considers the parity shifted 
vector bundle $W[1]$ and coordinatizes it by degree $0$ base coordinates
$z^a, z^{\bar a}$ and degree $1$ fiber coordinates $\xi^i$. $\Fun(W[1])$ is then graded 
according to the polynomial degree in the $\xi^i$. Next, one constructs a canonical degree
$1$ holomorphic vector field on $W[1]$, viz
\begin{equation}
Q_W=\rho^a{}_i\xi^i\partial_{za}-\frac{1}{2}f^i{}_{jk}\xi^j\xi^k\partial_{\xi i},
\label{cohla15}
\end{equation}
with the anchor and bracket structure functions $\rho^a{}_i$, $f^i{}_{jk}$ 
taken with respect to a local 
frame $\{e_i\}$ of $W$.
Then, due to \eqref{LAA1}--\eqref{LAA3}, $Q_W$ is homological,
\begin{equation}
Q_W{}^2=0.
\label{cohla16}
\end{equation}
$H_{LA}{}^*(W)$ is the cohomology of the complex $(\Fun(W[1]),Q_W)$. 
When $W=T_M{}^+$, we obtain the holomorphic Dolbeault cohomology of $M$, 
$H_{\partial_M}{}^{*,0}(M)$. 

Let $\mathcal{E}$ be a holomorphic Lie algebroid on a complex manifold $\mathcal{M}$
(cf. subsect. \ref{subsec:liecmplx1}).
Then, with $\mathcal{E}$, there is associated a holomorphic Lie algebroid cohomology
$H_{LA}{}^*(\mathcal{E})$ as follows. Instead of a complex, we have a sheaf of complexes
$(\Fun(\mathcal{E}[1]),\mathcal{Q}_{\mathcal{E}})$, where
$\Fun(\mathcal{E}[1])$ is the sheaf such that, for any open set $U$ of $M$,
$\Fun(\mathcal{E}[1])(U)$ is the algebra of holomorphic functions on
$\mathcal{E}|_U[1]$ and $\mathcal{Q}_{\mathcal{E}}$ is 
defined according to  \eqref{cohla15} using a holomorphic frame $\{e_i\}$. 
The Lie algebroid cohomology $H_{LA}{}^*(\mathcal{E})$ is the sheaf associated with the presheaf
$H'_{LA}{}^*(\mathcal{E})$ such that, for any open set $U$ of $M$,
$H'_{LA}{}^*(\mathcal{E})(U)$ is the cohomology of $(\Fun(\mathcal{E}[1])(U),\mathcal{Q}_{\mathcal{E}})$.

Suppose that $E$ is a transitive Lie algebroid with complex structure 
satisfying \eqref{transcmplx2}, the case mainly treated in this paper.
Then, $E^+$ is in particular a Lie algebroid with complex base and, as such, it is 
equipped with its cohomology $H_{LA}{}^*(E^+)$. 

Using \eqref{cohla15}, we can extend $Q_{E^+}$ to a homological vector field on $E^c[1]$.
However, while $Q_{E^+}$ does not depend on the framing $\{e_i\}$ of $E^+$ used,
the resulting extension does. There is nevertheless a canonical choice of the extension.
Since $E^+$ has a canonical holomorphic structure making it a holomorphic Lie 
algebroid $\mathcal{E}$, we can use the associated holomorphic framing $\{e_i\}$.
We denote the resulting extension by $\mathcal{Q}_E$
\footnote{ $\vphantom{\bigg]}$ 
A change of the smooth frame $\{e_i\}$ induces a change of fiber coordinates $\xi^i$, in which the new fiber 
coordinates depend holomorphically on the old ones but generally non holomorphically on the base 
coordinates. The derivatives $\partial_a$ in the two frames differ so by terms 
proportional to $\partial_{\bar\imath}$. Therefore, the right hand side of 
\eqref{cohla15} is covariant only up to terms of the latter form.
These are inert on $\Fun(E^+[1])$ but not on $\Fun(E^c[1])$. 
If, however, we restrict to holomorphic frames, the $\partial_{\bar\imath}$ terms vanish
and the right hand side of \eqref{cohla15} is covariant. \label{foot:1}}. 
The homological vector field $Q_{E^c}$ of $E^c$ splits as
\vspace{-.1cm}
\begin{equation}
Q_{E^c}=\mathcal{Q}_E+\overline{\mathcal{Q}}_E,
\label{cohla17}
\end{equation}
as is easy to check. 
The homological vector field $\mathcal{Q}_{\mathcal{E}}$ of $\mathcal{E}$ is simply 
the restriction of $\mathcal{Q}_E$ to $\Fun(\mathcal{E}[1])$. 

From now on, we assume that $E$ is a transitive Lie algebroid with complex structure 
satisfying \eqref{transcmplx2}. Further, we pick a splitting $\sigma$ of $E$ satisfying condition
\eqref{stconn2} and whose curvature $G_{E\sigma}{}^c$ has vanishing $(2,0)$ component 
(cf. eq. \eqref{hetersgm4/1}). 

\subsection{\normalsize \textcolor{blue}{The Lie algebroid ${\hat{E}}^+$ and its cohomology}}\label{subsec:cohla4}

\hspace{.5cm} Now, we are going to construct a distinguished Lie subalgebroid  
$\widehat{E}^+$ of $E^c$, which is a Lie algebroid with complex base and whose 
cohomology turns out to be intimately related to the BRST cohomology defining the 
type $A$ half--topological sigma model chiral ring.

Consider the complex vector bundle $\widehat{E}^+{}_\sigma=\overline{T}_M{}^+\oplus\ker\rho_{E^+}$.
We endow $\widehat{E}^+{}_\sigma$ with the anchor defined by 
\hphantom{xxxxxxxxxxxxxxxxxxxxxx} 
\begin{equation}
\rho_{\hat E^+{}_\sigma}(\overline{x}\oplus u)=\overline{x},
\label{acohom1}
\end{equation}
with $x\in\Gamma(T_M{}^+)$, $u\in\Gamma(\ker\rho_{E^+})$, and with the 
Lie bracket 
defined by 
\begin{equation}
[\overline{x}\oplus u,\overline{y}\oplus v]_{\hat E^+{}_\sigma}
=\overline{[x,y]_{T_M{}^+}}\oplus\big([\sigma^c\overline{x},v]_{E^c}
-[\sigma^c\overline{y},u]_{E^c}+[u,v]_{E^c}\big), 
\label{acohom2}
\end{equation}
with $x,y\in\Gamma(T_M{}^+)$, $u,v\in\Gamma(\ker\rho_{E^+})$.
Condition \eqref{transcmplx2} ensures that the right hand side of 
\eqref{acohom2} belongs to $\Gamma(\widehat{E}^+{}_\sigma)$.
Condition \eqref{hetersgm4/1} is required by the fulfillment of the Jacobi identity. 
$\widehat{E}^+{}_\sigma$ acquires in this way a structure of Lie algebroid with complex base, as is 
readily verified.

The bundle map $\widehat{\varpi}_\sigma:\widehat{E}^+{}_\sigma\rightarrow E^c$
defined by $\widehat{\varpi}_\sigma(\overline{x}\oplus u)=\sigma^c\overline{x}+u$
for $x\in\Gamma(T_M{}^+)$, $u\in\Gamma(\ker\rho_{E^+})$ is a 
monomorphism of complex Lie algebroids (cf. subsect. \ref{subsec:liealg1}). 
Let $\widehat{E}^+$ be the image of $\widehat{\varpi}_\sigma$ 
in $E^c$. Then, $\widehat{E}^+$ is a complex Lie subalgebroid of $E^c$ and
a Lie algebroid with complex base. 
Further, 
$\widehat{\varpi}_\sigma:\widehat{E}^+{}_\sigma\rightarrow \widehat{E}^+$
is an isomorphism of Lie algebroids with complex base. From now on, we identify $\widehat{E}^+{}_\sigma$
and $\widehat{E}^+$ leaving the isomorphism $\widehat{\varpi}_\sigma$ understood.

Let us compute the homological vector field $Q_{\hat E^+}$. 
As $E^+$ has a canonical holomorphic structure, there is a holomorphic
framing $\{e_i\}$. Then, $\{\partial_{\bar a}\}\cup\{e_u\}$ constitutes a distinguished 
framing of $\widehat{E}^+$. Let $\zeta^{\bar a}$, $c^u$ be the corresponding fiber coordinates
of $\widehat{E}^+[1]$. Then
\begin{equation}
Q_{\hat E^+}=\zeta^{\bar a}\partial_{z\bar a}-\frac{1}{2}f^u{}_{vw}c^vc^w\partial_{cu}.
\label{acohom3}
\end{equation}
It is straightforward to check that $Q_{\hat E^+}$ is nilpotent
\begin{equation}
Q_{\hat E^+}{}^2=0,
\label{acohom4}
\end{equation}
as required.

Next, we observe that $Q_{\hat E^+}$ decomposes as
\begin{equation}
Q_{\hat E^+}=\tilde Q_{\hat E^+}+\Lambda_{\hat E^+},
\label{acohom5}
\end{equation}
where $\tilde Q_{\hat E^+}$, $\Lambda_{\hat E^+}$ are given by 
\vspace{-.15truecm}
\begin{align}
&\tilde Q_{\hat E^+}=\zeta^{\bar a}\partial_{z\bar a},
\vphantom{\Big]}
\label{acohom6}
\\
&\Lambda_{\hat E^+}=-\frac{1}{2}f^u{}_{vw}c^vc^w\partial_{cu}.
\vphantom{\Big]}
\label{acohom7}
\end{align}
$\tilde Q_{\hat E^+}$, $\Lambda_{\hat E^+}$ are nilpotent and anticommute
\vspace{-.15cm}
\begin{align}
&\tilde Q_{\hat E^+}{}^2=0,
\vphantom{\Big]}
\label{acohom8}
\\
&\Lambda_{\hat E^+}{}^2=0,
\vphantom{\Big]}
\label{acohom9}
\\
&\tilde Q_{\hat E^+}\Lambda_{\hat E^+}+\Lambda_{\hat E^+}\tilde Q_{\hat E^+}=0.
\vphantom{\Big]}
\label{acohom10}
\end{align}

By \eqref{acohom8}, $(\Fun(\widehat{E}^+[1]),\tilde Q_{\hat E^+})$ is a complex.
The algebra $\Fun(\widehat{E}^+[1])$ is isomorphic to the algebra of
$\wedge^*\ker_{\mathcal{E}}{}^*$--valued $(0,*)$ forms $\Omega^{0,*}(M,\wedge^*\ker_{\mathcal{E}}{}^*)$.
Under the isomorphism, 
$\tilde Q_{\hat E^+}$ is identified with the standard Dolbeault operator
$\barpartial_{\wedge^*\ker_{\mathcal{E}}{}^*}$. The cohomology of 
$(\Fun(\widehat{E}^+[1]),\tilde Q_{\hat E^+})$ is thus isomorphic
to the Dolbeault cohomology $H_{\bar\partial}{}^*(\wedge^*\ker_{\mathcal{E}}{}^*)$, that is 
the sheaf cohomology $H^*(\mathcal{O}_{\wedge^*\ker_{\mathcal{E}}{}^*})$. 

We note that $\Fun(\widehat{E}^+[1])$ is  bigraded according to the polynomial degree
in $\zeta^{\bar a}$, $c^u$ and that, 
by \eqref{acohom8}--\eqref{acohom10}, $(\Fun(\widehat{E}^+[1]),\tilde Q_{\hat E^+},\Lambda_{\hat E^+})$ is 
a double complex. By \eqref{acohom5}, the total cohomology of this latter is the cohomology 
of the complex $(\Fun(\widehat{E}^+[1]),Q_{\hat E^+})$, that is the Lie algebroid cohomology
$H_{LA}{}^*(\widehat{E}^+)$. This can therefore be computed, at least in principle, 
using standard spectral sequence methods
\footnote{ $\vphantom{\bigg]}$ 
$\ker\rho_{E^+}$ is a Lie algebroid with vanishing anchor.  It has 
a canonical holomorphic structure induced by that of $E^+$, with which there are associated a 
holomorphic framing $\{e_u\}$ and fiber coordinates $c^u$ of $\ker\rho_{E^+}[1]$. 
The homological vector field $Q_{\ker\rho_{E^+}}$ is given by 
$Q_{\ker\rho_{E^+}}=-\frac{1}{2}f^w{}_{uv}c^uc^v\partial_{cw}$. 
$Q_{\ker\rho_{E^+}}$ is therefore the fiberwise Chevalley--Eilenberg differential of the 
Lie algebra bundle $\ker\rho_{E^+}$ and, so, $H_{LA}{}^*(\ker\rho_{E^+})$ 
is the space of sections of a vector bundle whose typical fiber 
the Chevalley--Eilenberg cohomology $H_{CE}{}^*(\mathfrak{g}^+)$ of the typical 
fiber $\mathfrak{g}^+$ of $\ker\rho_{E^+}$. 
 As is apparent, $Q_{\ker\rho_{E^+}}$
is formally identical to the homological vector field $\Lambda_E$. This furnishes
an interpretation of the latter.}. 
\vspace{.1cm}

\vfill\eject

\subsection{\normalsize \textcolor{blue}{
Lie algebroid cohomology of  ${\hat{E}}^+$ and $A$ model chiral ring}}\label{subsec:cohla5}

\hspace{.5cm} Now, we are ready to unveil the relation between the Lie algebroid 
cohomology of $\widehat{E}^+$ and the $A$ model chiral ring.

As explained in subsect. \ref{subsec:topsgm8}, the chiral ring $\mathcal{R}^{(\nu)}{}_A$ of the 
$\nu$--th BRST structure of the type $A$ half--topological sigma model consists of 
operator $Q^{(\nu)}{}_A$--cohomology classes of vanishing scaling dimensions $h,\bar h$. 
Each class is represented by a scalar local operator $\mathcal{O}$ 
\footnote{$\vphantom{\bigg]}$ Here and below, the adjective ``scalar'' means
``world--sheet scalar''.}
such that $[Q^{(\nu)}{}_A,\mathcal{O}]=0$ defined modulo an operator of the form 
$[Q^{(\nu)}{}_A,\mathcal{X}]$ with $\mathcal{X}$ an arbitrary scalar local operator
\footnote{~Since the BRST charge $Q^{(\nu)}{}_A$ is itself  a 
scalar, 
it cannot change the world--sheet covariance of the local operators $\mathcal{F}$ which it acts upon.
Therefore, $\mathcal{X}$ must be also be a 
scalar.}. Let $\mathsans{F}_A$ denote the algebra of the $A$ model 
scalar local operators. Then, $(\mathsans{F}_A,[Q^{(\nu)}{}_A,\cdot\,])$ is a complex
and the chiral ring $\mathcal{R}^{(\nu)}{}_A$ is its cohomology.

The operators $\mathcal{F}\in \mathsans{F}_A$ are all of the form
\begin{equation}
\mathcal{F}=\sum_{m,n}\frac{1}{m!n!}\phi_{\bar p_1\ldots \bar p_m;u_1\ldots u_n}(x)\overline{\chi}^{\bar p_1}
\cdots \overline{\chi}^{\bar p_m}\lambda^{u_1}\cdots \lambda^{u_n}.
\label{acohla26}
\end{equation}
The fields $\chi^p{}_{\bar z}$, $\lambda^*{}_{uz}$
as well as all the $\partial_z$, $\barpartial_{\bar z}$ derivatives of fields cannot contribute 
to $\mathcal{F}$ because of its scalar nature. 

In the quasiclassical limit, there is a canonical isomorphism 
$\varsigma_A:\mathsans{F}_A\mapsto \Fun(\widehat{E}^+[1])$
defined as follows. Redefine the fiber coordinates of $\widehat{E}^+[1]$ as 
\begin{equation}
\gamma^{\bar p}=\sigma^{\bar p}{}_{\bar a}\zeta^{\bar a}.
\label{acohom11}
\end{equation}
Then, for any operator $\mathcal{F}\in \mathsans{F}_A$ of the form
\eqref{acohla26}, $\varsigma_A({\mathcal{F}})\in \Fun(\widehat{E}^+[1])$ reads
\begin{equation}
\varsigma_A({\mathcal{F}})=\sum_{m,n}\frac{1}{m!n!}\phi_{\bar p_1\ldots \bar p_m;u_1\ldots u_n}\gamma^{\bar p_1}
\cdots \gamma^{\bar p_m}c^{u_1}\cdots c^{u_n}.
\label{acohla26/1}
\end{equation}
Under the isomorphism $\varsigma_A$, we have
\begin{equation}
\varsigma_A([Q^{(\nu)}{}_A,\mathcal{F}])=Q^{(\nu)}{}_{\hat E^+}\varsigma_A(\mathcal{F}),\vphantom{\Bigg]}
\label{acohla26/2}
\end{equation}
where $Q^{(\nu)}{}_{\hat E^+}$ is a certain homological vector field on $\widehat{E}^+[1]$, 
\begin{equation}
Q^{(\nu)}{}_{\hat E^+}{}^2=0.
\label{acohla26/3}
\end{equation}
Relation \eqref{acohla26/2} states that  
$\varsigma_A$ is a chain map of the complexes $(\mathsans{F}_A,[Q^{(\nu)}{}_A,\cdot\,])$,
$(\Fun(\widehat{E}^+[1]),Q^{(\nu)}{}_{\hat E^+})$. Therefore, in the quasiclassical limit, 
the chiral ring $\mathcal{R}^{(\nu)}{}_A$ is isomorphic to the cohomology of 
$(\Fun(\widehat{E}^+[1]),Q^{(\nu)}{}_{\hat E^+})$. Next, we are going to compute $Q^{(\nu)}{}_{\hat E^+}$
in such regime, having in mind that $Q^{(\nu)}{}_{\hat E^+}$ may receive complicated perturbative quantum 
corrections.

Using \eqref{hetersgm4/1},  we find that 
\begin{equation}
\tilde Q_{\hat E^+}=\rho^{\bar a}{}_{\bar p}\gamma^{\bar p}\partial_{z\bar a}
-\frac{1}{2}f^{\bar p}{}_{\bar q\bar r}\gamma^{\bar q}\gamma^{\bar r}\partial_{\gamma \bar p}, \vphantom{\Bigg]}
\label{acohom12}
\end{equation}
while $\Lambda_{\hat E^+}$ is still given by expression \eqref{acohom7}.
From \eqref{acohom11}, \eqref{acohom12}, we find easily
\begin{subequations}
\begin{align}
&\tilde Q_{\hat E^+}z^a=0, \qquad 
\tilde Q_{\hat E^+}z^{\bar a}=\rho_{\bar p}\gamma^{\bar p}
\vphantom{\Big]}
\label{acohla20} 
\\
&\tilde Q_{\hat E^+}\gamma^{\bar p}
=-\frac{1}{2}f^{\bar p}{}_{\bar q\bar r}\gamma^{\bar q}\gamma^{\bar r},
\vphantom{\Big]}
\label{acohla21} 
\\
&\tilde Q_{\hat E^+}c^u=0,
\vphantom{\Big]}
\label{acohla22} 
\end{align}
\label{acohla20-22}
\end{subequations} 
\!\!From \eqref{acohom7}, we obtain \hphantom{xxxxxxxxxxxxxxxxxxxxxxxxxxxxxxxx}
\vspace{-.09cm}
\begin{subequations}
\begin{align}
&\Lambda_{\hat E^+}z^a=0, \qquad \Lambda_{\hat E^+}z^{\bar a}=0,
\vphantom{\Big]}
\label{acohla23} 
\\
&\Lambda_{\hat E^+}\gamma^{\bar p}=0,
\vphantom{\Big]}
\label{acohla24} 
\\
&\Lambda_{\hat E^+}c^u=-\frac{1}{2}f^u{}_{vw}c^vc^w.
\vphantom{\Big]}
\label{acohla25} 
\end{align}
\label{acohla23-25}
\end{subequations}     
\!\!\!Comparing \eqref{atopsgm2} and \eqref{acohla20-22}, we find that
the action of the primary topological field variation operator $\delta_A$ on $\mathsans{F}_A$
is such that 
\begin{equation}
\varsigma_A(\delta_A\mathcal{F})=\tilde Q_{\hat E^+}\varsigma_A(\mathcal{F}),
\label{acohla36}
\end{equation}
for any operator $\mathcal{F}\in\mathsans{F}_A$. 
Likewise, comparing \eqref{aintflv1} and \eqref{acohla23-25}, we find that the
action of the secondary topological field varation operator $s_A$ 
is such that 
\eject\noindent
\begin{equation}
\varsigma_A(s_A\mathcal{F})=\Lambda_{\hat E^+}\varsigma_A(\mathcal{F}).
\label{acohla37}
\end{equation}

Consider the first BRST structure of the $A$ model.
In the quasiclassical limit, $Q^{(1)}{}_A=Q_A$ acts as $\delta_A$ (cf. subsect. \ref{subsec:topsgm7}, eq. 
\eqref{qopsbrstc1}). From \eqref{acohla26/2}, \eqref{acohla36}, in the same regime, we have 
therefore \hphantom{xxxxxxxxxxxxxxxx}
\begin{equation}
Q^{(1)}{}_{\hat E^+}=\tilde Q_{\hat E^+}.
\label{acohom26}
\end{equation}
We conclude that, {\it at the classical level, the chiral ring $\mathcal{R}^{(1)}{}_A$ 
is isomorphic to the cohomology of $(\Fun(\widehat{E}^+[1]),\tilde Q_{\hat E^+})$ and, hence, 
to the Dolbeault cohomology $H_{\bar\partial}{}^*(\wedge^*\ker_{\mathcal{E}}{}^*)$ or the 
sheaf cohomology $H^*(\mathcal{O}_{\wedge^*\ker_{\mathcal{E}}{}^*})$.} 
The first BRST structure is the one normally envisaged in twisted heterotic
sigma models. Here, we have recovered a well-known result about the heterotic chiral ring
\cite{Sharpe:2005fd}\,--\cite{Adams:2005tc}. 

Next, consider the second BRST structure. In the quasiclassical limit, $Q^{(2)}{}_A=Q_A+\Lambda_A$ 
acts as $\delta_A+s_A$ 
(cf. subsect. \ref{subsec:topsgm7}, eqs. \eqref{qopsbrstc1}, \eqref{qopsbrstc2}). 
From \eqref{acohla26/2}, \eqref{acohla36}, \eqref{acohla37}, 
in the same regime, we have then 
\begin{equation}
Q^{(2)}{}_{\hat E^+}=\tilde Q_{\hat E^+}+\Lambda_{\hat E^+}=Q_{\hat E^+},
\label{acohom27}
\end{equation}
where \eqref{acohom5} has been used. 
We conclude that, {\it at the classical level, the chiral ring $\mathcal{R}^{(2)}{}_A$ 
is isomorphic to the cohomology of $(\Fun(\widehat{E}^+[1]),Q_{\hat E^+})$, that is the 
Lie algebroid cohomology $H_{LA}{}^*(\widehat{E}^+)$}.

Next, we consider the problem of computing correlators of $A$ model chiral ring classes
at genus $\ell=0$ in the quasiclassical limit. This matter was discussed preliminarily in 
subsect.\ref{subsec:topsgm10}, which the reader is referred to. 
Recall that the quantum correlator $\langle\mathcal{O}\rangle$ of any chiral ring element 
$\mathcal{O}\in\mathcal{R}^{(\nu)}{}_A$ should be expressible as an integral over $M$ of a suitable 
top degree de Rham cohomology class $[\omega_{\mathcal{O}}]_{\mathrm{dR}}$ depending on $\mathcal{O}$.
As we have found above, with $\mathcal{O}$ there corresponds  
a cohomology class of the complex $(\Fun(\widehat{E}^+[1]),Q^{(\nu)}{}_{\hat E^+})$.
This indicates that there must exist a map 
$\widehat{\varphi}:\Fun(\widehat{E}^+[1])\mapsto\Fun(T_M[1])$
such that, for any chiral ring element $\mathcal{O}\in\mathcal{R}^{(\nu)}{}_A$, 
one has 
\eject\noindent
\begin{equation}
\langle\mathcal{O}\rangle=\int_{T_M[1]}\widehat{\varphi}\,\omega,
\label{acohom32}
\end{equation}
where $\omega\in \Fun(\widehat{E}^+[1])$ with $Q^{(\nu)}{}_{\hat E^+}\omega=0$
is a representative of $\mathcal{O}$. In \eqref{acohom32},  $\widehat{\varphi}\,\omega$ is to be thought of
as a non homogeneous form on $M$, whose top degree part is 
integrated on $M$.
The map $\widehat{\varphi}$ must have the following properties. First, 
\begin{equation}
\widehat{\varphi}\,Q^{(\nu)}{}_{\hat E^+}\omega=(-1)^{d-r}\dd\widehat{\varphi}\,\omega, \vphantom{\Bigg]}
\label{acohom29/2}
\end{equation}
for $\omega\in \Fun(\widehat{E}^+[1])$, where $\dd=\zeta^a\partial_a+\zeta^{\bar a}\partial_{\bar a}$ 
is the de Rham differential in supergeometric form. In virtue of \eqref{acohom29/2}, the right hand side of 
\eqref{acohom32} depends only on the $Q^{(\nu)}{}_{\hat E^+}$--cohomology class of $\omega$
and, so, is defined on $\mathcal{R}^{(\nu)}{}_A$. 
Second, 
\begin{equation}
\int_{T_M[1]}\widehat{\varphi}\,\omega=0,
\label{acohom29}
\end{equation}
whenever $\omega\in \Fun(\widehat{E}^+[1])$ does not have bidegree $(d,r)$. 
This ensures that the selection rule on the chiral ring correlators 
found earlier in subsect. \ref{subsec:topsgm10} is satisfied. 

Let \eqref{hetersgm36/1} hold. 
Then, the holomorphic line bundle $\wedge^{d}\mathcal{T}_{\mathcal{M}}{}^*
\otimes \wedge^{r}\ker\rho_{\mathcal{E}}$ is holo\-morphically trivial
and, so, it has a nowhere vanishing holomorphic section $\upsilon$.
$\upsilon$ induces an algebra morphism $\widehat{\varphi}_{\upsilon}
:\Fun(\widehat{E}^+[1])\mapsto \Fun(T_M[1])$ defined by 
\begin{equation}
\widehat{\varphi}_{\upsilon\,}\omega
=\frac{1}{d!r!}\upsilon_{a_1\ldots a_{d}}{}^{u_1\ldots u_{r}}
\zeta^{a_1}\cdots \zeta^{a_d}\partial_{cu_1}\cdots \partial_{cu_r}\omega,\vphantom{\Bigg]}
\label{acohom28}
\end{equation}
with $\omega\in\Fun(\widehat{E}^+[1])$. 
By the holomorphy of $\upsilon$,  
$\widehat{\varphi}_{\upsilon}$ fulfils \eqref{acohom29/2}, 
for the first BRST structure.
It does so also for the second BRST structure, provided that 
the unimodularity condition \eqref{hetersgm36/5} holds.
Moreover, $\widehat{\varphi}_{\upsilon}$ fulfils \eqref{acohom29} for either structures.  
In this way, we have justified the algebro--geometric conditions
\eqref{hetersgm36/1} and \eqref{hetersgm36/5} for the $A$ model introduced in subsect. 
\ref{subsec:topsgm10}.

\subsection{\normalsize \textcolor{blue}{The Lie algebroid $E^+$ and its cohomology}}\label{subsec:cohla6}

\hspace{.5cm} We are now going to show that the Lie algebroid with complex base $E^+$ has 
\eject\noindent
a structure analogous to that of the Lie algebroid $\widehat{E}^+$ 
constructed above and that its cohomology is intimately related to the BRST cohomology defining the 
type $B$ half--topological sigma model chiral ring.

Consider the complex vector bundle $E^+{}_\sigma=T_M{}^+\oplus\ker\rho_{E^+}$. 
We endow $E^+{}_\sigma$ with the anchor defined by \hphantom{xxxxxxxxxxxxxxxxxxxxxx}
\begin{equation}
\rho_{E^+{}_\sigma}(x\oplus u)=x
\label{bcohom1}
\end{equation}
with $x\in\Gamma(T_M{}^+)$, $u\in\Gamma(\ker\rho_{E^+})$, and with the Lie bracket defined by 
\begin{equation}
[x\oplus u,y\oplus v]_{E^+{}_\sigma}
=[x,y]_{T_M{}^+}\oplus\big([\sigma^cx,v]_{E^c}
-[\sigma^cy,u]_{E^c}+[u,v]_{E^c}\big) \vphantom{\Bigg]}
\label{bcohom2}
\end{equation}
with $x,y\in\Gamma(T_M{}^+)$, $u,v\in\Gamma(\ker\rho_{E^+})$.
Again, condition \eqref{hetersgm4/1} is required by the fulfillment of the Jacobi identity. 
$E^+{}_\sigma$ acquires in this way a structure of Lie algebroid with complex base, as is 
immediately verified.

The bundle map $\varpi_\sigma:E^+{}_\sigma\rightarrow E^c$
defined by $\varpi_\sigma(x\oplus u)=\sigma^cx+u$
for $x\in\Gamma(T_M{}^+)$, $u\in\Gamma(\ker\rho_{E^+})$
is a monomorphism of complex Lie algebroids. The image of $\varpi_\sigma$ 
in $E^c$ in nothing but the Lie algebroid $E^+$, as follows 
from relation \eqref{stconn1}. $\varpi_\sigma:E^+{}_\sigma\rightarrow E^+$
is then an isomorphism of Lie algebroids with complex base. Again, for simplicity,
we identify $E^+{}_\sigma$ and $E^+$ leaving the isomorphism $\varpi_\sigma$ understood.

Let us compute the homological vector field $Q_{E^+}$. 
As $E^+$ has a canonical holomorphic structure, there is a holomorphic
framing $\{e_i\}$. By the above findings, $\{\partial_a\}\cup\{e_u\}$ constitutes another 
distinguished framing of $E^+$. Let $\zeta^a$, $c^u$ be the corresponding fiber coordinates
of $E^+[1]$. Then, 
\begin{equation}
Q_{E^+}=\zeta^a\big(\partial_{za}-f^u{}_{iv}\sigma^i{}_ac^v\partial_{cu}\big)
-\frac{1}{2}f^u{}_{vw}c^vc^w\partial_{cu}. \vphantom{\Bigg]}
\label{bcohom3}
\end{equation}
It is straightforward to check that $Q_{E^+}$ is nilpotent
\begin{equation}
Q_{E^+}{}^2=0, 
\label{bcohom4}
\end{equation}
\eject\noindent
as required \footnote{$\vphantom{\bigg]}$ $Q_{E^+}$ extends to a homological vector field on $E^c[1]$.
Since the framing $\{\partial_a\}\cup\{e_u\}$ used in the construction of $Q_{E^+}$ 
is not holomorphic, the extension differs from the canonical extension 
$\mathcal{Q}_E$ defined above eq. \eqref{cohla17}, which employs the holomorphic framing
$\{e_i\}$ (cf.  footnote \ref{foot:1}). Indeed, one has $\mathcal{Q}_E=Q_{E^+}
-G^{\bar u}{}_{a\bar b}\zeta^a\zeta^{\bar b}\partial_{c\bar u}$.
So, $\mathcal{Q}_E$, $Q_{E^+}$ have the same restriction on $\Fun(E^+[1])$
but generally differ on $\Fun(E^c[1])$.}.

Next, we observe that $Q_{E^+}$ decomposes as
\begin{equation}
Q_{E^+}=\tilde Q_{E^+}+\Lambda_{E^+},
\label{bcohom5}
\end{equation}
where $\tilde Q_{E^+}$, $\Lambda_{E^+}$ are given by 
\begin{align}
&\tilde Q_{E^+}=\zeta^a\big(\partial_{za}-f^u{}_{iv}\sigma^i{}_ac^v\partial_{cu}\big),
\vphantom{\Big]}
\label{bcohom6}
\\
&\Lambda_{E^+}=-\frac{1}{2}f^u{}_{vw}c^vc^w\partial_{cu}.
\vphantom{\Big]}
\label{bcohom7}
\end{align}
$\tilde Q_{E^+}$, $\Lambda_{E^+}$ are nilpotent and anticommute
\begin{align}
&\tilde Q_{E^+}{}^2=0,
\vphantom{\Big]}
\label{bcohom8}
\\
&\Lambda_{E^+}{}^2=0,
\vphantom{\Big]}
\label{bcohom9}
\\
&\tilde Q_{E^+}\Lambda_{E^+}+\Lambda_{E^+}\tilde Q_{E^+}=0.
\vphantom{\Big]}
\label{bcohom10}
\end{align}
We note that \eqref{bcohom8} depends crucially on the validity of \eqref{hetersgm4/1}.

By \eqref{bcohom8}, $(\Fun(E^+[1]),\tilde Q_{ E^+})$ is a complex.
The algebra $\Fun(E^+[1])$ is isomorphic to the algebra of
$\wedge^*\mathcal{K}_{E^+\sigma}{}^*$--valued $(*,0)$ forms $\Omega^{*,0}(M,\wedge^*\mathcal{K}_{E^+\sigma}{}^*)$,
where the antiholomorphic vector bundle $\mathcal{K}_{E^+\sigma}$ was defined in subsect. \ref{subsec:topsgm10}.
Under the isomorphism, $\tilde Q_{ E^+}$ is identified with the $(1,0)$ connection
$\mathcal{D}^{\mathrm{ad}}_{E\sigma}$ with vanishing $(2,0)$ curvature that defines the antiholomorphic structure of
$\mathcal{K}_{E^+\sigma}$ and, so, with the holomorphic Dolbeault operator
$\partial_{\wedge^*\mathcal{K}_{E^+\sigma}{}^*}$. The cohomology of 
$(\Fun(\widehat{E}^+[1]),\tilde Q_{\hat E^+})$ is thus isomorphic
to the holomorphic Dolbeault cohomology $H_{\partial}{}^*(\wedge^*\mathcal{K}_{E^+\sigma}{}^*)$, that is 
the sheaf cohomology $H^*(\overline{\mathcal{O}}_{\wedge^*\mathcal{K}_{E^+\sigma}{}^*})$. 
This cohomology depends in general on the splitting $\sigma$, since $\mathcal{K}_{E^+\sigma}$ does.

We note that $\Fun(E^+[1])$ is  bigraded according to the polynomial degree
in $\zeta^a$, $c^u$ and that, 
by \eqref{bcohom8}--\eqref{bcohom10}, $(\Fun(E^+[1]),\tilde Q_{ E^+},\Lambda_{ E^+})$ is 
a double complex. By \eqref{bcohom5}, the total cohomology of this double complex is the cohomology 
of the complex $(\Fun(E^+[1]),Q_{ E^+})$, that is the Lie algebroid cohomology
$H_{LA}{}^*(E^+)$. Again, the latter can be computed using spectral sequence
methods. It is also manifestly independent from the splitting $\sigma$.

\subsection{\normalsize \textcolor{blue}{Lie algebroid cohomology of $E^+$ and $B$ model chiral ring}}
\label{subsec:cohla7}

\hspace{.5cm} Now, we are going to uncover the relation between the Lie algebroid 
cohomology of $E^+$ and the $B$ model chiral ring.
In outline, the analysis follows the same lines as that done for the $A$ model.

From subsect. \ref{subsec:topsgm8}, the chiral ring $\mathcal{R}^{(\nu)}{}_B$ of the 
$\nu$--th BRST structure of the type $B$ half--topological sigma model consists of 
operator $Q^{(\nu)}{}_B$--cohomology classes of vanishing scaling dimensions $h,\bar h$. 
Each class is represented by a scalar local operator $\mathcal{O}$ 
such that $[Q^{(\nu)}{}_B,\mathcal{O}]=0$ defined modulo an operator of the form 
$[Q^{(\nu)}{}_B,\mathcal{X}]$ with $\mathcal{X}$ any scalar local operator. 
Let $\mathsans{F}_B$ denote the algebra of the $B$ model 
scalar local operators. Then, $(\mathsans{F}_B,[Q^{(\nu)}{}_B,\cdot\,])$ is a complex
and the chiral ring $\mathcal{R}^{(\nu)}{}_B$ is its cohomology.

The operators $\mathcal{F}\in\mathsans{F}_B$ are all of the form
\begin{equation}
\mathcal{F}=\sum_{m,n}\frac{1}{m!n!}\phi_{p_1\ldots p_m;u_1\ldots u_n}(x)\chi^{p_1}\cdots\chi^{p_m}
\lambda^{u_1}\cdots \lambda^{u_n},
\label{bcohla26}
\end{equation}
the fields $\overline{\chi}^{\bar p}{}_{\bar z}$, $\lambda^*{}_{uz}$
as well as all the $\partial_z$, $\barpartial_{\bar z}$ derivatives of fields not contributing 
because of the scalar nature of $\mathcal{F}$. 

Analogously to the $A$ model, in the quasiclassical limit, 
there is a canonical isomorphism $\varsigma_B:\mathsans{F}_B\mapsto \Fun(E^+[1])$.
Define new fiber coordinates of $E^+[1]$ by 
\begin{equation}
\gamma^p=\sigma^p{}_a\zeta^a. \vphantom{\Bigg]}
\label{bcohom11}
\end{equation}
Then, for any operator $\mathcal{F}\in \mathsans{F}_B$ of the form
\eqref{bcohla26}, $\varsigma_B({\mathcal{F}})\in \Fun(E^+[1])$ is 
\eject\noindent
\begin{equation}
\varsigma_B({\mathcal{F}})=\sum_{m,n}\frac{1}{m!n!}\phi_{p_1\ldots p_m;u_1\ldots u_n}\gamma^{p_1}
\cdots \gamma^{p_m}c^{u_1}\cdots c^{u_n}.
\label{bcohla26/1}
\end{equation}
Again, analogously to the $A$ model, the isomorphism $\varsigma_B$ has the property that 
\begin{equation}
\varsigma_B([Q^{(\nu)}{}_B,\mathcal{F}])=Q^{(\nu)}{}_{E^+}\varsigma_B(\mathcal{F}),\vphantom{\Bigg]}
\label{bcohla26/2}
\end{equation}
where $Q^{(\nu)}{}_{E^+}$ is a certain homological vector field on $E^+[1]$, 
\begin{equation}
Q^{(\nu)}{}_{E^+}{}^2=0. \vphantom{\Bigg]}
\label{bcohla26/3}
\end{equation}
By \eqref{bcohla26/2}, $\varsigma_B$ is a chain map of the complexes $(\mathsans{F}_B,[Q^{(\nu)}{}_B,\cdot\,])$,
$(\Fun(E^+[1]),Q^{(\nu)}{}_{E^+})$ and, so, the quasiclassical chiral ring $\mathcal{R}^{(\nu)}{}_B$ is 
isomorphic to the cohomology of $(\Fun(E^+[1])$, $Q^{(\nu)}{}_{E^+})$. 
We are now going to compute $Q^{(\nu)}{}_{E^+}$ in that regime. 

Using \eqref{hetersgm4/1},  we find that 
\begin{equation}
\tilde Q_{E^+}=\rho^a{}_p\gamma^p(\partial_{za}-f^v{}_{iu}\sigma^i{}_ac^u\partial_{cv})
-\frac{1}{2}f^r{}_{pq}\gamma^p\gamma^q\partial_{\gamma r}. \vphantom{\Bigg]}
\label{bcohom12}
\end{equation}
while $\Lambda_{E^+}$ is still given by expression \eqref{bcohom7}.
From \eqref{bcohom11}, \eqref{bcohom12}, we find easily
\begin{subequations}
\begin{align}
&\tilde Q_{E^+}z^a=\rho_p\gamma^p, \qquad \tilde Q_{E^+}z^{\bar a}=0,
\vphantom{\Big]}
\label{bcohla20} 
\\
&\tilde Q_{E^+}\gamma^p=-\frac{1}{2}f^p{}_{qr}\gamma^q\gamma^r,
\vphantom{\Big]}
\label{bcohla21} 
\\
&\tilde Q_{E^+}c^u=-f^u{}_{iv}\sigma^i{}_a\rho^a{}_p\gamma^pc^v
\vphantom{\Big]}
\label{bcohla22} 
\end{align}
\label{bcohla20-22}
\end{subequations} 
\!\!while, from \eqref{bcohom7}, we obtain 
\begin{subequations}
\begin{align}
&\Lambda_{E^+}z^a=0, \qquad \Lambda_{E^+}z^{\bar a}=0,
\vphantom{\Big]}
\label{bcohla23} 
\\
&\Lambda_{E^+}\gamma^p=0,
\vphantom{\Big]}
\label{bcohla24} 
\\
&\Lambda_{E^+}c^u=-\frac{1}{2}f^u{}_{vw}c^vc^w.
\vphantom{\Big]}
\label{bcohla25} 
\end{align}
\label{bcohla23-25}
\end{subequations}      
\!\!\!By comparing \eqref{btopsgm2} with \eqref{bcohla20-22} and  
\eqref{bintflv1} with \eqref{bcohla23-25}, we find that
the actions of the 
topological field variation operators $\delta_B$, $s_B$ 
on $\mathsans{F}_B$ satisfy 
\begin{equation}
\varsigma_B(\delta_B\mathcal{F})=\tilde Q_{E^+}\varsigma_B(\mathcal{F}), \vphantom{\bigg]}
\label{bcohla36}
\end{equation}
and \hphantom{xxxxxxxxxxxxxxxxxxxxxxxxxxxxxxxxxxxx}
\begin{equation}
\varsigma_B(s_B\mathcal{F})=\Lambda_{E^+}\varsigma_B(\mathcal{F}).
\label{bcohla37}
\end{equation}
for any operator $\mathcal{F}\in\mathsans{F}_B$, extending the $A$ model findings. 

Consider the first BRST structure of the $B$ model.
In the quasiclassical limit, $Q^{(1)}{}_B=Q_B$ acts as $\delta_B$ (cf. subsect. \ref{subsec:topsgm7}, eq. 
\eqref{qopsbrstc1}). From \eqref{bcohla26/2}, \eqref{bcohla36}, in the same regime, we have 
therefore 
\vspace{-.1cm}
\begin{equation}
Q^{(1)}{}_{E^+}=\tilde Q_{E^+}. \vphantom{\Bigg]}
\label{bcohom26}
\end{equation}
It follows that, {\it at the classical level, the chiral ring $\mathcal{R}^{(1)}{}_B$ 
is isomorphic to the cohomology of $(\Fun(\widehat{E}^+[1]),\tilde Q_{E^+})$
and, hence, to the holomorphic Dolbeault cohomology $H_{\partial}{}^*(\wedge^*\mathcal{K}_{E^+\sigma}{}^*)$ or  
the sheaf cohomology $H^*(\overline{\mathcal{O}}_{\wedge^*\mathcal{K}_{E^+\sigma}{}^*})$.}

Next, consider the second BRST structure. In the quasiclassical limit, $Q^{(2)}{}_B=Q_B+\Lambda_B$ 
acts as $\delta_B+s_B$ 
(cf. subsect. \ref{subsec:topsgm7}, eqs. \eqref{qopsbrstc1}, \eqref{qopsbrstc2}). 
From \eqref{bcohla26/2}, \eqref{bcohla36}, \eqref{bcohla37}, 
in the same regime, we have then 
\begin{equation}
Q^{(2)}{}_{E^+}=\tilde Q_{E^+}+\Lambda_{E^+}=Q_{E^+}, \vphantom{\Bigg]}
\label{bcohom27}
\end{equation}
where \eqref{bcohom5} has been used. 
Consequently, {\it at the classical level, the chiral ring $\mathcal{R}^{(2)}{}_B$ 
is isomorphic to the cohomology of $(\Fun(E^+[1]),Q_{E^+})$, that is the 
Lie algebroid cohomology $H_{LA}{}^*(E^+)$}.

The problem of computing correlators of $B$ model chiral ring classes
at genus $\ell=0$ in the quasiclassical limit can be treated along the same lines as
the $A$ model. As we have found above, with any chiral ring class
$\mathcal{O}\in\mathcal{R}^{(\nu)}{}_B$ there 
corresponds a cohomology class of the complex $(\Fun(E^+[1]),Q^{(\nu)}{}_{E^+})$.
Thus, there must exist a map $\varphi:\Fun(E^+[1])\mapsto\Fun(T_M[1])$
such that, for 
$\mathcal{O}\in\mathcal{R}^{(\nu)}{}_B$, 
\begin{equation}
\langle\mathcal{O}\rangle=\int_{T_M[1]}\varphi\,\omega, \vphantom{\Bigg]}
\label{bcohom32}
\end{equation}
where $\omega\in \Fun(E^+[1])$ with $Q^{(\nu)}{}_{E^+}\omega=0$
is a representative of $\mathcal{O}$. The mapping 
\eject\noindent
$\varphi$ must have the properties that 
\begin{equation}
\varphi\,Q^{(\nu)}{}_{E^+}\omega=(-1)^{d-r}\dd\varphi\,\omega,
\label{bcohom29/2}
\end{equation}
for $\omega\in \Fun(E^+[1])$, to make the right hand side of 
\eqref{bcohom32} depend only on the $Q^{(\nu)}{}_{E^+}$--cohomology class of $\omega$
and, thus, defined on $\mathcal{R}^{(\nu)}{}_B$, and that
\begin{equation}
\int_{T_M[1]}\varphi\,\omega=0,
\label{bcohom29}
\end{equation}
whenever $\omega\in \Fun(E^+[1])$ does not have bidegree $(d,r)$,  
to ensure that the selection rule on the chiral ring correlators 
of subsect. \ref{subsec:topsgm10} is satisfied. 

Assume that \eqref{hetersgm36/2a}, \eqref{hetersgm36/2} hold. As 
the antiholomorphic line bundle $\wedge^{r}\mathcal{K}_{E^+\sigma}{}^*$ 
is antiholomorphically trivial, it admits a nowhere vanishing antiholomorphic section
$\upsilon$. Similarly, as the holomorphic line bundle $\wedge^{d}\mathcal{T}_{\mathcal{M}}{}^*$, 
is holomorphically trivial, it possesses a nowhere vanishing holomorphic sections $\Omega$. 
$\upsilon$, $\Omega$ induce an algebra morphism $\varphi_{\upsilon\Omega}
:\Fun(E^+[1])\mapsto \Fun(T_M[1])$ defined by 
\begin{equation}
\varphi_{\upsilon\Omega\,}\omega
=\frac{1}{d!r!}\overline{\Omega}_{\bar a_1\ldots \bar a_{d}}\upsilon^{u_1\ldots u_{r}}
\zeta^{\bar a_1}\cdots \zeta^{\bar a_d}\partial_{cu_1}\cdots \partial_{cu_r}\omega,
\label{bcohom28}
\end{equation}
with $\omega\in\Fun(E^+[1])$. 
By the antiholomorphy of $\upsilon$ and the holomorphy of $\Omega$  
$\varphi_{\upsilon\Omega}$ fulfils \eqref{bcohom29/2}, 
for the first BRST structure.
It does so also for the second BRST structure, provided, again, that 
the unimodularity condition \eqref{hetersgm36/5} holds.
Moreover, $\varphi_{\upsilon\Omega}$ fulfils \eqref{bcohom29} for either structures.  
In this way, as for the $A$ model, we have justified the algebro--geometric conditions
\eqref{hetersgm36/2a}, \eqref{hetersgm36/2} and \eqref{hetersgm36/5}  
for the $B$ model introduced in subsect. \ref{subsec:topsgm10}.

\subsection{\normalsize \textcolor{blue}{Lie algebroid chiral algebras}}
\label{subsec:cohla8}

In this final subsection, we make some remarks concerning the chiral algebras
of the half--topological sigma models. 

Consider first the chiral algebra $\mathcal{A}^{(\nu)}{}_A$ of the $A$ model. 
Since the local operators and the BRST charge can be described locally along $M$, 
it is possible to consider operators that are well--defined not throughout $M$, 
but only on local neighborhoods of $M$ \cite{Witten:2005px}. 
$Q^{(\nu)}{}_A$--cohomology classes of operators defined in a neighborhood have 
operator product expansions involving classes of operators 
defined in the same neighborhood and they can be restricted 
to smaller neighborhoods and patched on unions of neighborhoods
in a natural fashion. So, what we actually have is a 
sheaf of chiral algebras $\mathcal{C}^{(\nu)}{}_A$. 

An operator class $\mathcal{O}$ of $\mathcal{A}^{(\nu)}{}_A$ is characterized by 
its scaling dimensions $(h,0)$ 
and $R$/flavour charges $(q_R,q_L)$. A local operator $\mathcal{F}$ representing 
$\mathcal{O}$ must then have the same properties. $\mathcal{F}$ must so be a function of 
the fields $x^a$, $x^{\bar a}$, $\lambda^u$, $\lambda^*{}_{uz}$ and their 
$\partial_z$ derivatives and the field $\overline{\chi}^{\bar p}$ with $h$ $z$ indices, 
$q_R$ factors $\overline{\chi}^{\bar p}$ and $q_L$ $\lambda^u$ minus $\lambda^*{}_{uz}$ 
possibly differentiated factors. The fields $\chi^p{}_{\bar z}$ and $\barpartial_{\bar z}$ derivatives 
of fields cannot appear in the expression of $\mathcal{F}$ because
its antiholomorphic scaling dimension vanishes. 
The $\partial_z$ derivatives of $\overline{\chi}^{\bar p}$ and the fields
$l^u{}_{\bar z}$, $l^*{}_{uz}$ are not included as they can be eliminated 
using the field equations. 
It can be shown that the operators $\mathcal{F}$ of this form are in one--to--one 
correspondence with the elements of $\Omega^{0,q_R}(\mathcal{W}_{h,q_L})$, 
where $\mathcal{W}_{h,q_L}$ is a certain complicated holomorphic vector bundle 
depending on $h$, $q_L$  
constructed with $\mathcal{T}_{\mathcal{M}}$, $\ker_{\mathcal{E}}$ and their duals.

Consider now the first BRST structure of the $A$ model. In the quasiclassical limit, 
the BRST charge $Q^{(1)}{}_A$ acts on the local operators $\mathcal{F}$ of fixed scaling 
dimensions $(h,0)$ and flavour charge $q_L$ but varying $R$ charge $q_R$ as the Dolbeault operator 
$\barpartial_{\mathcal{W}_{h,q_L}}$ in many though not all cases,  
by reasons analogous to those found in the treatment of chiral ring 
$\mathcal{R}^{(1)}{}_A$ \footnote{$\vphantom{\Bigg]}$ In the case where
$E=T_M$ considered by Witten in \cite{Witten:2005px}, this is actually always true.}.
For the sake of the argument, let us leave aside this technical complication 
and pretend that $Q^{(1)}{}_A$ acts classically as $\barpartial_{\mathcal{W}_{h,q_L}}$. 
Since $\barpartial_{\mathcal{W}_{h,q_L}}$ obeys the $\barpartial$--Poincar\'e lemma, 
the cohomology of $\barpartial_{\mathcal{W}_{h,q_L}}$  is trivial on any 
local neighborhood of $M$. The operator $Q^{(1)}{}_A$--cohomology 
is therefore locally trivial classically on the local operators $\mathcal{F}$ 
of scaling dimensions $(h,0)$ and flavour charge $q_L$. Since $h$ and $q_L$ are arbitrary,
the operator $Q^{(1)}{}_A$--cohomology is locally trivial classically in general. 

It is believed that perturbative quantum corrections cannot create cohomology classes 
\cite{Witten:2005px}. Taking this for granted, it follows that 
the operator $Q^{(1)}{}_A$--cohomology is locally trivial also quantum mechanically in the 
perturbative regime we assume. By an argument totally analogous to that used to prove
the Dolbeault-\v Cech isomorphism in algebraic geometry, one then shows that 
the operator $Q^{(1)}{}_A$--cohomology, that is the chiral algebra $\mathcal{A}^{(1)}{}_A$,
is isomorphic to the \v Cech cohomology of the sheaf $\mathcal{A}^{(1)}{}_{A}{}^0$
of $R$ charge $0$ $Q^{(1)}{}_A$--closed operators. 
In \cite{Witten:2005px,Kapustin:2005pt}, Witten and Kapustin showed independently that,
for $E=T_M$ with $M$ Calabi--Yau,  
the \v Cech cohomology of $\mathcal{A}^{(1)}{}_{A}{}^0$ could be related to that of the sheaf of 
chiral differential operators introduced 
in \cite{Malikov:1998dw}, \cite{Gorbounov:1999}. This result was later generalized 
by Tan in \cite{Tan:2006qt}\,--\!\cite{Tan:2007bh} to heterotic sigma models with general gauge bundle.
It extends also to our type $A$ sigma model for the first BRST structure.

Does the above generalize to the type $A$ sigma model for the second BRST structure?
To be sure, not in a straightforward fashion. The 
argument just outlined for the first structure uses in an essential way the fact that, in the quasiclassical 
limit, $Q^{(1)}{}_A$ reduces to the appropriate Dolbeault operator $\barpartial$ and that,
by the $\barpartial$--Poincar\'e lemma,  the $\barpartial$--cohomology is locally trivial. 
For the second structure, one expects that $Q^{(2)}{}_A$, again quasiclassically, 
reduces instead to some generalization of the Lie algebroid differential $Q_{\hat E^+}$ 
studied in subsect. \ref{subsec:cohla4} by arguments analogous to those employed in the study 
of the chiral ring $\mathcal{R}^{(2)}{}_A$ in subsect. \ref{subsec:cohla5}. 
The point here is that there is no analog of the Poincar\'e lemma for $Q_{\hat E^+}$. 
Thus, the operator $Q^{(2)}{}_A$--cohomology is not locally trivial
in general and  
$\mathcal{A}^{(2)}{}_A$ does not have a \v Cech 
description the same way $\mathcal{A}^{(1)}{}_A$ does. 

The above analysis can be repeated along the same lines for the $B$ model with completely 
similar conclusions.

\vfill\eject

\section{\normalsize \textcolor{blue}{Conclusions and outlook}}\label{sec:outlook}

\subsection{\normalsize \textcolor{blue}{Summary of results}}
\label{subsec:outlook1}

\hspace{.5cm} 
In this paper, we have constructed a heterotic sigma model whose target space geometry
consists of a transitive Lie algebroid $E$ with complex structure on a Kaehler manifold
$M$ satisfying certain natural geometric conditions, the heterotic Lie algebroid sigma model.
We have then found that the model possesses a primary $(0,2)$ supersymmetry,
which it shares with the other heterotic sigma models \cite{Distler:1995mi}, 
and a novel secondary $(1,0)$
supersymmetry ensuing from the adjoint bundle $\ker\rho_E$, the kernel
of the anchor $\rho_E$ of $E$. 

This opens new possibilities for topological twisting.
The two twist prescriptions, considered in this paper, lead to 
half--topological sigma models of type $A$ and $B$,
as already studied in the literature. These sigma models, however, 
because of the higher amount of supersymmetry of the parent untwisted model
stemming from the Lie algebroid target space geometry, 
are characterized not only by the usual primary topological BRST operator but also by
a secondary Slavnov--like BRST operator anticommuting with the former. 
As a consequence, the sigma models possess two inequivalent BRST structures,
which share many, but not all, properties. 

Quantum mechanically, in the perturbative regime in which there are no world--sheet 
instanton contributions, the two half--topological sigma models constructed by us  
are conformally invariant at the level 
of BRST cohomology and are characterized by a holomorphic chiral algebra $\mathcal{A}$
and a chiral ring $\mathcal{R}$ with respect to both BRST structures,
on a par with the half--topological sigma models considered in earlier studies
\cite{Adams:2003zy}--\cite{Adams:2005tc}. The difference 
between the structures manifests itself with regard to the dependence of the chiral 
algebra correlators on the target space geometric data. 
The classical action of our sigma models contains also a Kaehler metric $g$ of 
$M$ and a splitting $\sigma$ of $E$. The chiral algebra correlators are independent 
from $g$ for both BRST structures while they are independent from $\sigma$  
for the second BRST structure only. The second structure is therefore 
``more topological'' than the first and, 
from a Lie algebroid theoretic point of view, also more natural.
Indeed, as we have seen, the second structure's chiral ring is directly 
related to the target space Lie algebroid cohomology in a way the 
first structure's one is not. 

Our conclusions rest on a considerable amount of educated guessing
and are therefore to a certain extent conjectural. We do not know
whether the two BRST structures coexist in the same quantum field theory
or, else, whether their pertain to two different quantum field theories
sharing the same quasiclassical limit. Happily, our conclusions apparently
do not hinge on the answer to this questions. In any case, as is well-known
\cite{Tan:2008mi,Tan:2008mc}, 
the inclusion of non perturbative world--sheet instanton effects can 
radically alter the perturbative picture that emerges from our analysis. 

\subsection{\normalsize \textcolor{blue}{Open problems}}
\label{subsec:outlook2}

\hspace{.5cm} There are several issues which we have not touched
and, related to these, problems still open. 

\vspace{.2cm}
{\it Lie algebroid sigma models with a regular but non transitive target}.

A Lie algebroid $E$ is regular if its anchor $\rho_E$ has locally constant rank \cite{Mackenzie1}. 
Every transitive Lie algebroid is regular, but not viceversa. 
The natural question arises whether it would be possible 
to generalize our analysis and produce a heterotic Lie algebroid sigma model with 
a regular non transitive target. Since our construction relies in an essential
way on the use of splittings, which exist only for transitive Lie algebroids, 
the task seems hopeless at first glance.  
However, as the base $M$ of a regular Lie algebroid
$E$ admits a foliation such that the restriction of $E$ to each leaf of $M$ 
is transitive, it is conceivable that a Lie algebroid sigma model with regular 
target may be built as a family of Lie algebroid sigma models with transitive 
target parametrized by the foliation of $M$ and connected in some way. 

\vspace{.2cm}
{\it Lie algebroid sigma models with $(2,2)$ supersymmetry}.

It would be desirable to have a Lie algebroid sigma model with $(2,2)$ supersymmetry.
For a target space geometry consisting of a transitive Lie algebroid $E$ with complex 
structure over a Kaehler base $M$, of the kind considered in this paper, in the same way as the $(0,2)$ 
heterotic sigma model arises by coupling fermionic degrees of freedom in $\ker\rho_{E^+}$ 
to the basic $(0,2)$ sigma model over $M$ in a way compatible with $(0,2)$ supersymmetry,
the $(2,2)$ sigma model should arise by coupling fermionic degrees of freedom in $\ker\rho_{E^c}$ 
to the basic $(2,2)$ sigma model over $M$ in a way compatible with $(2,2)$ supersymmetry. 
The closure of the $(2,2)$ supersymmetry algebra apparently requires the splitting $\sigma$ 
to be flat. While every transitive Lie algebroid admits a flat splitting locally, it generally does 
not globally \cite{Mackenzie1}. So, it is likely that such $(2,2)$ Lie algebroid sigma model, assuming 
that it does exist, will turn out to be only a very mild  generalization of the usual one.
Alternatively, one may try to define the $(2,2)$ sigma model locally in target space
and then try to glue the resulting field theories in a globally meaningful way, 
though it is difficult to fathom how this could be done.

\vspace{.2cm}
{\it Description of the chiral algebra}.

In this paper, we concentrated on the chiral rings of the two half--topological
Lie algebroid sigma models for each BRST structure and said comparatively little about 
their ambient chiral algebras (cf. subsect. \ref{subsec:cohla8}).
The point is that the condition $[Q^{(\nu)}{}_t,\mathcal{O}]=0$
is not very constraining for a local operator $\mathcal{O}$ of positive holomorphic scaling 
dimension. This is what renders chiral algebras very complicated objects which
resist any attempts at a simple description such as that of chiral rings.
Needless to say, much work remains to be done on this aspect of the theory.
 
\vspace{.2cm}
{\it Deformation of the target Lie algebroid structure}.

In general, infinitesimal deformations of the target space geometry of a sigma model 
result in the insertion  
of integrated vertex operators in correlators. For the sigma models studied in this 
paper, the target geometry consists of a transitive Lie algebroid $E$ with complex structure
over a manifold $M$ satisfying \eqref{transcmplx2} 
(cf. subsects. \ref{subsec:liealg2}, \ref{subsec:liecmplx2}). Its 
deformation involves that of $a)$ the Lie bracket structure $[\cdot,\cdot]_E$,
$b)$ the anchor structure $\rho_E$, $c)$ the fiber complex structure $J_E$
and $d)$ the base complex structure $J_{EM}$ subject to various constraints. 
Further, 
the deformations reducible to symmetry transformations 
must be modded out. This analysis requires a considerable amount of 
extra work and is left for future work. 

\vspace{1cm}

\textcolor{blue}{Acknowledgements}

We thank F. Bastianelli and G. Bonelli for useful discussions.
We warmly thank the Erwin Schroedinger Institute for Mathematical Physics (ESI) for the kind 
hospitality offered to us in July 2010.

\vfill\eject

\end{document}